\documentclass[a4paper,12pt]{article}
\usepackage[english]{babel}
\usepackage{amsmath, amssymb,graphicx}
\usepackage{jheppub}
\usepackage{comment}

\newcommand{\be}{\begin{equation}}
\newcommand{\ee}{\end{equation}}
\newcommand{\bea}{\begin{eqnarray}}
\newcommand{\eea}{\end{eqnarray}}
\newcommand{\nn}{\nonumber}

\title{Holographic Dual to Conical Defects:  \\I. Moving Massive Particle}
\author[a]{D.S. Ageev}
\author[a]{I.Ya. Aref'eva}
\author[b]{M.D. Tikhanovskaya}
\affiliation[a]{Steklov Mathematical Institute, Russian Academy of Sciences, Gubkin str. 8, 119991
Moscow, Russia}
\affiliation[b]{National Research Nuclear University "MEPhI" (Moscow Engineering Physics Institute), 115409 Moscow, Russia}

\emailAdd{ageev@mi.ras.ru}
\emailAdd{arefeva@mi.ras.ru}
\emailAdd{tikhanovskaya@mi.ras.ru}

\abstract{  We study  correlation functions of scalar operators
on the boundary of the $AdS_3$ space deformed by moving massive particles in the context of the AdS/CFT duality. 
To calculate two-point correlation functions we use the geodesic approximation   and the renormalized image method. 
We compare results of the renormalized image method with  direct calculations using  tracing  of winding  geodesics around the cone singularities, and show on examples that they are equivalent. We demonstrate that in the geodesic approximation
 the correlators exhibit a zone structure.  This structure substantially depends on the mass and velocity of the particle.}


\begin{document}
\maketitle

\newpage


\newpage
\section{Introduction}

The AdS/CFT, or more generally the gauge/gravity duality  \cite{Malda,GKP,Witten} is a powerful tool in the study of quantum systems in the strong coupling limit. Due to its flexibility there is a wide range of applications in heavy-ion collision \cite{IA,DeWolf,Solana}, condensed matter theory \cite{Herzog:2007ij,Hartnoll:2008vx,Hartnoll:2008kx}, thermalization of strongly coupling theories \cite{Balasubramanian:2011ur,Balasubramanian:2012tu,Lopez,Keranen:2011xs,Aref'eva:2015rwa}, entanglement entropy \cite{Ryu:2006bv,Ryu:2006ef,Nishioka:2009un} and quantum quenches \cite{0708.3750,Nozaki:2013wia}.

Two dimensional conformal field theory is the holographic dual of $AdS_3$ gravity. Three dimensional gravity  is topological and there are no propagating gravitons in this theory. Deformations of the three-dimensional gravity by point particles are only global \cite{Deser,Hooft}. This means, that locally the deformed space is still the $AdS_3$, but globally there are  wedges to cut out and glue
their  faces. In another words,   point particles induce  conical singularities. Scattering of point particles in the space with conical singularities was studied in \cite{Deser-Jackiw}. Classical and quantum  scalar  field theories  on a cone  have been  considered in several papers starting from \cite{QFT-cone} and scalar fields in the  flat space with defects have been studied in \cite{volovich}.  The cosmic  strings in the flat $M_4$ and  $AdS_4$ provide four-dimensional generalization \cite{CS,Kirsch:2004km,Bayona:2010sd}, while the cosmic membranes provide higher dimensional generalization of conical defect in the context of the TeV-gravity  \cite{CM}.

In is natural to ask a question about holographic dual to  $AdS_3$ with point particles \cite{Balasubramanian:1999zv}.  Correlators in the theory dual to  $AdS_3$ with a static particle  have been considered  within the geodesic approximation
\cite{Balasubramanian:1999zv,AB-TMF} and appearance of new excitations
in the boundary theory
has been noticed.
Then, the AdS/CFT correspondence for  the multi-boundary  $AdS_3$ orbifold has been studied   \cite{Balasubramanian}.     A new quantity, called entwinement,  in the dual CFT has been introduced in \cite{Balasubramanian:2014sra}, and it has been shown that it is related with  the conical defect geometry.  Correlators in the theory dual to the Gott time machine in the $AdS_3$ have been investigated
\cite{Arefeva:2015sza}.  A holographic dual model for defect conformal field theories
has been considered in \cite{Araujo:2015hna}.

In this paper we continue to study  boundary theories dual to  $AdS_3$ deformed by  massive moving point particles. To describe these deformations  it is convenient to consider  $AdS_3$  as an $SL(2,R)$ group manifold  \cite{Matschull:1998rv}. $AdS_3$
with a particle  is  a space that remains  after cutting out a special subset, called wedge, from $AdS_3$ spacetime, and then identifying the boundaries of this wedge in a
special  way \cite{Hooft, Matschull:1997du}.  The geodesics in this spacetime locally are the same as in the nondeformed $AdS_3$ and this drastically  simplifies the problem of constructing the boundary correlators in the geodesic approximation. In the geodesic approximation one has to find  all  geodesics connecting two given points on the boundary.  For one static particle one can find all  geodesics connecting two spacelike separated points explicitly in the Deser-Jackiw coordinates \cite{Balasubramanian:2014sra}. But the generalization of these coordinates to  multi-particle cases is not explicit \cite{Ciafaloni:1996qx} which makes the problem of analytical description of all geodesics rather complicated.

We study this problem using the cutting and gluing method that has been used  previously
in \cite{Balasubramanian:1999zv,AB-TMF,Balasubramanian:2014sra,Arefeva:2015sza}.  As in \cite{Arefeva:2015sza},  in this paper we have to use numerical simulations to take into account  all  geodesics  connecting two given points on the boundary in the present of  moving defects.

 The paper is organized as  follows. In Section 2, we remind the group structure of the $AdS_3$ and set the notations.
 In Section 3, the renormalized image  method is described. The relation of winding geodesics and
 imaged geodesics is clarified on several examples. In Section 4,  the zone structure of correlators  on the boundary of the $AdS_3$ deformed by moving particle is presented and discussed.

\section {Setup}
\subsection{$AdS_3$ space as a group manifold }
In this section we set the notations and the parametrization we use in this paper.
The $AdS_3$ is a hyperboloid, which in embedding coordinates $x_0$, $x_1$, $x_2$ and $x_3$ can be written as:
\be\label{det}
-x_0^2-x_3^2+x_1^2+x_2^2=-1.
\ee
We also use the barrel coordinates $(t, \chi, \phi)$:
\bea
\label{comp}
x_3=\cosh\chi \, \cos t,\\\nn
x_0=\cosh\chi \, \sin t,\\\nn
x_1=\sinh\chi \, \cos \phi,\\\nn
x_2=\sinh \chi \sin \phi,\eea
where $t$ is the time coordinate, $\chi$ is the radial coordinate and $\phi$ is the angular coordinate with period $2\pi$. The $AdS_3$ conformal boundary corresponds to $\chi \rightarrow \infty$.
In these coordinates the metric can be written out as:
\be\nn
ds^2=-\cosh^2\chi dt^2+d\chi^2+\sinh^2\chi d\phi^2.
\ee
Instead of $\chi$ and $\phi$ also we will use Poincare disc coordinates related with $\chi$ as $r=\tanh \chi/2$ and in these coordinates the metric has the form:
\be\nn
ds^2=-\left(\frac{1+r^2}{1-r^2}\right)^2 dt^2 + \left(\frac{2}{1-r^2}\right)^2(dr^2+r^2d\phi^2).
\ee

The $AdS_3$ also admits the representation as $SL(2,R)$ group of real 2x2 matrices:
\be\label{points}
{\bf x}= x_3 {\bf{1}}+ \sum_{\mu=0,1,2} \gamma_{\mu} x^\mu=\cosh \chi \, {\bf \Omega} (t) + \sinh{\chi} \, {\bf \Gamma}(\phi)=\left(
           \begin{array}{cc}
             x_3+x_2 & x_0+x_1 \\
             x_1-x_0 &  x_3-x_2\\
           \end{array}
         \right),
\ee
where
\begin{equation}
{\bf 1} =\left(
           \begin{array}{cc}
             1 & 0 \\
             0 & 1 \\
           \end{array}
         \right)
;~~~~
\gamma_0 = \left(
             \begin{array}{cc}
               0 & 1 \\
               -1 & 0 \\
             \end{array}
           \right)
;~~~~
\gamma_1 = \left(
             \begin{array}{cc}
               0 & 1 \\
               1 & 0 \\
             \end{array}
           \right)
;~~~~
\gamma_2 =\left(
            \begin{array}{cc}
              1 & 0 \\
              0 & -1 \\
            \end{array}
          \right),
\end{equation}
and
\bea\label{Omega}
{\bf \Omega} (t)=\cos t {\bf 1} +\sin t \gamma_0;\,\,\,\,\,\, {\bf \Gamma} (\phi) = \cos \phi \gamma_1 +\sin \phi \gamma_2.
\eea
The condition $\det {\bf x} = 1$ is equivalent to (\ref{det})

\subsection{Point particles in  $AdS_3$}

It is known, that the gravity in spacetime dimension 3 is almost trivial, in the sense
 of absence of propagating degrees of freedom. In works \cite{Deser,Hooft} it was shown, that point particle does not change the metric locally,  producing conical defect singularity. In this section we remind the structure of the $AdS_3$ deformed by  point particles.

\subsubsection{Static particle in $AdS_3$}
Let us recall  the Deser-Jackiw solution \cite{Deser}.
Consider the Einstein equation in the 3-dimensional spacetime with the cosmological constant which equals to $-1$ :

\be\label{Einst}
G^{\mu\nu} - g^{\mu\nu}=8\pi G T^{\mu\nu}.
\ee

The ansatz for the metric $ds_{DJ}^2$ supported by the time independent point-like source is:
\bea\label{ansatz}\nn
ds_{DJ}^2&=&-N^2(R)dt^2+ \Phi(R)(dR^2+R^2d\tilde{\phi}^2),\\\nn
T^{00}&=&\frac{m}{\sqrt{-g}}N(R)\delta(R),
\eea
where functions $\Phi(R)$ and $N(R)$ are:

 \begin{eqnarray}\label{DJsol}\nn
  \Phi(R) = \frac{4A^2}{\Lambda R^2 (\left(R/R_0)^{A}+(R/R_0)^{-A} \right)^2},\,\,\,\,\,\,\,\,\,\,
  N(R) = \frac{(\left(R/R_0)^{A}-(R/R_0)^{-A} \right)}{(\left(R/R_0)^{A}+(R/R_0)^{-A} \right)}
\end{eqnarray}
Parameter $A$  connects with the mass of the particle as $ A=1-4 G m$. After the change of variables:
\bea\nn
\sinh\chi=\frac{1}{2} \left(\left(\frac{R}{R_0}\right)^{ A }+\left(\frac{R}{R_0}\right)^{-A }\right), \,\,\,\,\,\,
   \phi=A \tilde{\phi},
\eea
we get the $AdS_3$ metric in the barrel coordinates with a different angular coordinate range of values,

\bea
\label{barrA}\nn
ds^2=-\cosh^2\chi dt^2+d\chi^2+\sinh^2\chi d\phi^2,\,\,\,\,\,\,
\phi\in \left(0,2 \pi A \right).
\eea

\begin{figure}
\centering
\begin{picture}(200,200)
\put(0,0){\includegraphics[width=6cm]{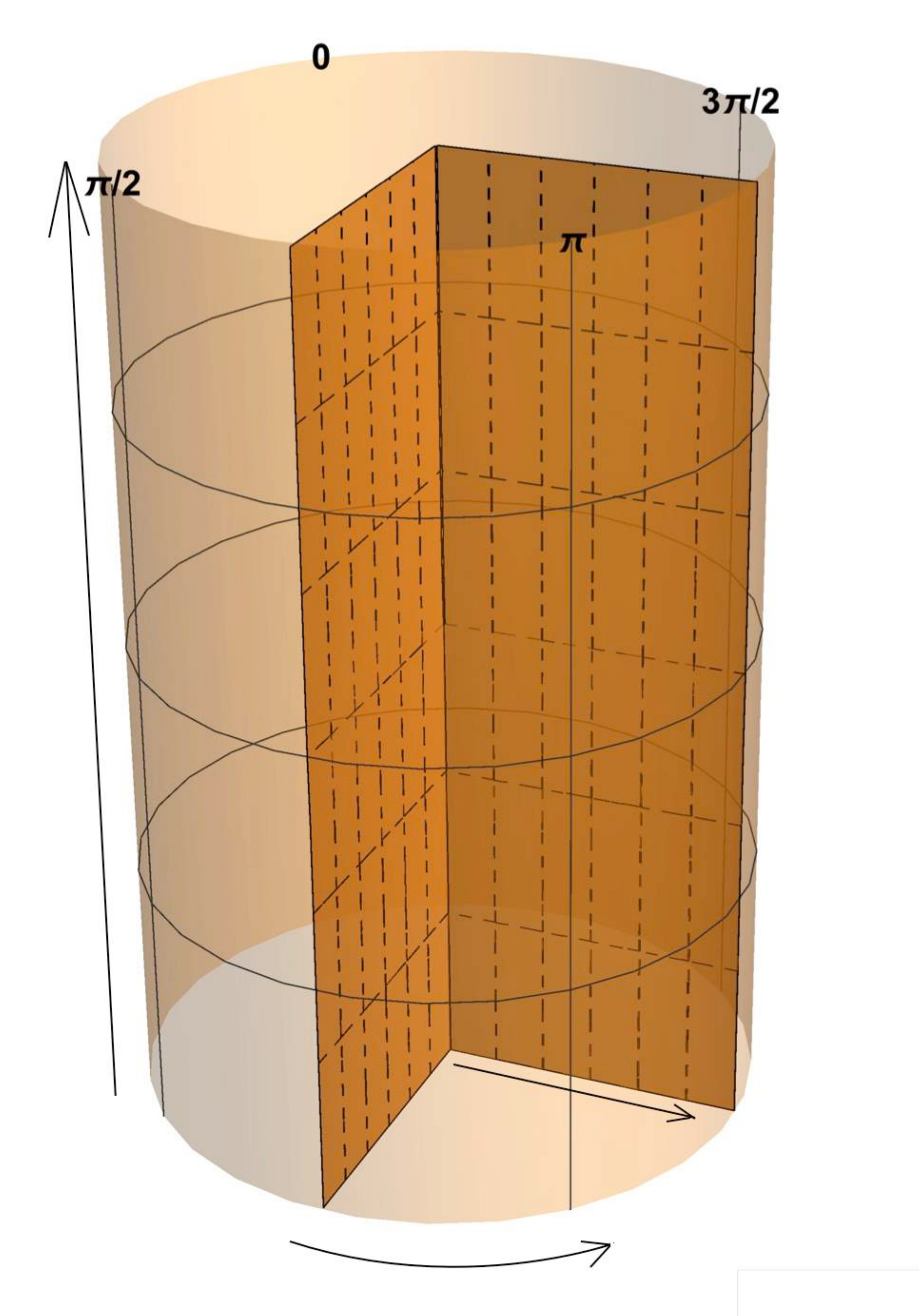}}
\put(10,100){$\tau$}
\put(90,0){$\phi$}
\put(100,30){$\chi$}
\end{picture}
\caption{The $AdS_3$ deformed by the static particle. Two constant angle surfaces incident from the origin of the $AdS$ are the faces of the wedge to cut out and identify.} \label{Fig:staticwedge}
\end{figure}
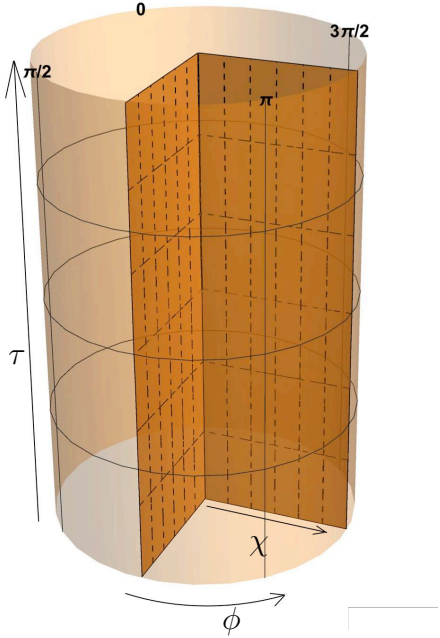

Let us now consider the static particle case from the group language. Resting in the center of the $AdS_3$ static particle cuts out the wedge that  can be described by two faces that are some constant angle surfaces (see Fig.\ref{Fig:staticwedge}). These two faces are identified in the constant $t$ sections. In matrix notation the first face of the wedge is:
\be
\label{1-st-face-m}
{\bf x}_{1-st \, face}=\cosh\chi \,{\bf \Omega}(t)+\sinh \chi \,{\bf \Gamma}(-\alpha/2).\ee
The face is parameterized by two values: $\alpha$ and $t$,  $\alpha$ is proportional to the mass of the particle and $t$ is time coordinate.  The second face of the wedge can be obtained by rotation of  the first face by the angle $\alpha$. Writing out rotation:
\be\nn
 {\bf x}_{rot}= u_{rot}^{-1} \, {\bf x}\,  u_{rot},\,\,\,\,\,\,
  u_{rot}={\bf \Omega}(-\alpha/2),\ee
 we get the second face:
\be
\label{rel-1-2-faces}
{\bf x} _{2-nd\, face}={\bf \Omega}(\alpha/2)\,\cdot\,{\bf x} _{1-st\, face}\,\cdot\,{\bf \Omega}(-\alpha/2),\ee
where ${\bf \Omega}(\alpha/2)$ is given by (\ref{Omega}).
\subsubsection{Moving massive particle in the $AdS_3$}

To consider a massive moving particle and get it's group language description one can consider a static particle and  boost it. The massive particle moves along the periodic worldline oscillating in the bulk of the $AdS_3$. The constant angle faces of the wedge to be identified become some surfaces that one can get by
boosting the wedge of the static particle. These faces are glued as in the static case along the constant time slices and symmetrically with respect to the boost direction, but now they exhibit some nontrivial isometry due to nontrivial holonomy induced by the  moving particle.

To obtain the faces of the wedge of moving massive particle we make the boost, that in the matrix notation has the form:
\bea\label{movp}
{ \bold u} = \cosh(\xi/2)\, {\bf 1} - \sinh(\xi/2) \, \gamma_2\label{boost-u}=\cosh(\xi/2)\, {\bf \Omega}(0) - \sinh(\xi/2) \,{\bf \Gamma}(\pi/2),
\eea
i.e. we apply (\ref{movp}) to the faces of the wedge (\ref{1-st-face-m}) and (\ref{rel-1-2-faces}) and get:
\bea
{\bf x}_{1\,-\,mov\,-\,face}={ \bold u}^{-1}{\bf x}_{1\,-\,face}{ \bold u},\,\,\,\,
{\bf x}_{2\,-\,mov\,-\,face}={ \bold u}^{-1}{\bf x}_{2\,-\,face}{ \bold u}.
\eea
From (\ref{rel-1-2-faces}) we find the isometry map identifying these two wedges:
\bea
\label{2-st-face-mm}
{\bf x}_{2-nd \,mov\,\, face}
&=&{\bf \Omega}_{{ \bold u}}(\alpha/2,\xi/2)\,\cdot\,{\bf x} _{1-st\, mov\,\,face}\,
\cdot\,{\bf \Omega}_{{ \bold u}}(-\alpha/2,\xi/2),\eea
where
\be\nn
{\bf \Omega}_{{ \bold u}}(-\alpha/2,\xi/2) \equiv{\bf \Omega}_{{ \bold u}}={ \bold u}^{-1}(\xi/2)
{\bf \Omega}\left(-\alpha/{2}\right){ \bold u}(\xi/2).\ee
Finally the isometry induced by the presence of the moving massive particle in $\text{AdS}_3$ is:
\be\label{isom}
{\bf x}^*={\bf \Omega}_{{ \bold u}}^{-1}{\bf x}{\bf \Omega}_{{ \bold u}}=\left(
           \begin{array}{cc}
             x_3^{*}+x_2^{*} & x_0^{*}+x_1^{*} \\
             x_1^{*}-x_0^{*} &  x_3^{*}-x_2^{*}\\
           \end{array}
         \right),\ee
where {\bf x} is the $AdS_3$ point defined as (\ref{points}).

Rewriting  (\ref{points}) in an explicit form using barrel coordinates:
\be\nn
{\bf x}=\left(
\begin{array}{cc}
 \cos t \cosh  \chi+\sin  \phi  \sinh  \chi & \cosh  \chi
   \sin t+\cos  \phi  \sinh  \chi \\
- \cosh  \chi \sin t+\cos  \phi  \sinh  \chi & \cos t \cosh
    \chi-\sin  \phi  \sinh  \chi \\
\end{array}
\right),\ee

and \eqref{isom} has the form:

\be\nn
{\bf x}^*=\left(
\begin{array}{cc}
 \cos t^{*} \cosh  \chi^{*}+\sin  \phi^{*}  \sinh  \chi^{*} & \cosh  \chi^{*}
   \sin t^{*}+\cos  \phi^{*}  \sinh  \chi^{*} \\
- \cosh  \chi^{*} \sin t^{*}+\cos  \phi^{*}  \sinh  \chi^{*} & \cos t^{*} \cosh
    \chi^{*}-\sin  \phi^{*}  \sinh  \chi^{*} \\
\end{array}
\right).
\ee

 After some algebra we get an explicit coordinate expression for isometry as:
\bea\nn
\tan t^*&=& \mathcal{B}_{\xi}(\alpha)\sec t  \tanh \chi \cos \phi +\tan t \left(1+2\sinh^2\xi\sin^2\frac{\alpha}{2}\right),\\ 
\tan\phi^*&=&-2 \mathcal{\xi(\alpha)}^{-1} \tan \phi,
\label{im3}\eea
   \bea\label{chi}\nn
\cosh\chi^*&=&\cosh\chi[ \left(\mathcal{B}_{\xi}(\alpha)\tanh \chi \cos \phi +  \sin t (1+2\sinh^2\xi\sin^2\frac{\alpha}{2})\right)^2+ \cos ^2t ]^{\frac{1}{2}},
    \eea
where
\bea\label{B-F}
\mathcal{B}_{\xi}(\alpha)&=&\sinh \xi\left( \sin \alpha  \tan \phi
 -2\cosh\xi \sin^2\frac{\alpha}{2}\right),\\\nn
\mathcal{\xi(\alpha)}&=&\cosh \xi  (2 \sin \alpha \tan \phi
   -\cos \alpha +\cos \phi )\\\nn&+&\sec \phi \cos (\alpha +\phi
   )+\cos \alpha \cosh2 \xi-2 \sinh ^2\xi.
\eea

From \eqref{im3}  taking the limit $\chi\rightarrow\infty$ we get the expression for isometry near the boundary of the $AdS_3$:
 \bea\label{nb-mov-iso}
 \tan t_b^* &=&\mathcal{B}_{\xi}(\alpha)\sec t_b \cos \phi_b+\tan t_b \left(1+2\sinh^2\xi\sin^2\frac{\alpha}{2}\right),\\\nn
\tan \phi_b^* &=& -{2\mathcal{\xi(\alpha)}^{-1}\tan \phi_b}.\eea
The expression for the radial coordinate  $\chi$ after the isometry near the boundary is:
\bea\label{expchi}
\text{e}^{\chi_{nb}^*} &= & \text{e}^{\chi_{nb}} \sqrt{\mathcal{A}}\,,\eea
where
\bea
   \mathcal{A}&=& \left(\mathcal{B}_{\xi}(\alpha)\cos \phi_b+\sin t_b (1+2\sinh^2\xi\sin^2\frac{\alpha}{2})\right)^2 + \cos ^2t_b.  \nn
\eea

Now we derive equations defining the wedge faces. As it has been mentioned above, to fix the wedges we must find  points that are constant in time and symmetrical in angle under the isometry,
 i.e. they are fixed by conditions $t^*=t$ and $\phi^{*}=-\phi$. So, solving equation $\tan t = \tan t^*$ we get the expression for the wedge face:
 \bea
\label{section-mm}
 \tanh \chi
&=&
 \frac{2\sin t\sinh \xi \sin^2 \frac{  \alpha}{2}}{   2\cosh \xi  \sin ^2 \frac{\alpha}{2} \cos \phi \pm \sin  \alpha \sin \phi
   }. \eea
It is useful to change the variable as $r=\tanh(\chi/2)$
and get $r$ as function of $\phi$ and $t$ for two wedges:
  \bea \label{faces}
r(\phi,t)&=& \tanh \left(\frac{1}{2} \text{arctanh}  \frac{2\sin t\sinh \xi \sin^2 \frac{ \alpha}{2}}{   2\cosh \xi  \sin^2  \frac{\alpha}{2} \cos \phi \pm \sin  \alpha \sin \phi
   } \right).\eea
   The intersection of two surfaces determined by (\ref{faces}) gives a fixed point of the isometry (or equally the massive particle worldline):
   \bea\nn
   r(t)&=&\frac{1-\sqrt{1-\tanh ^2\xi \sin^2t}}{\tanh \xi \sin t}.
      \eea
      
\begin{figure}
    \centering
     \includegraphics[width=6cm]{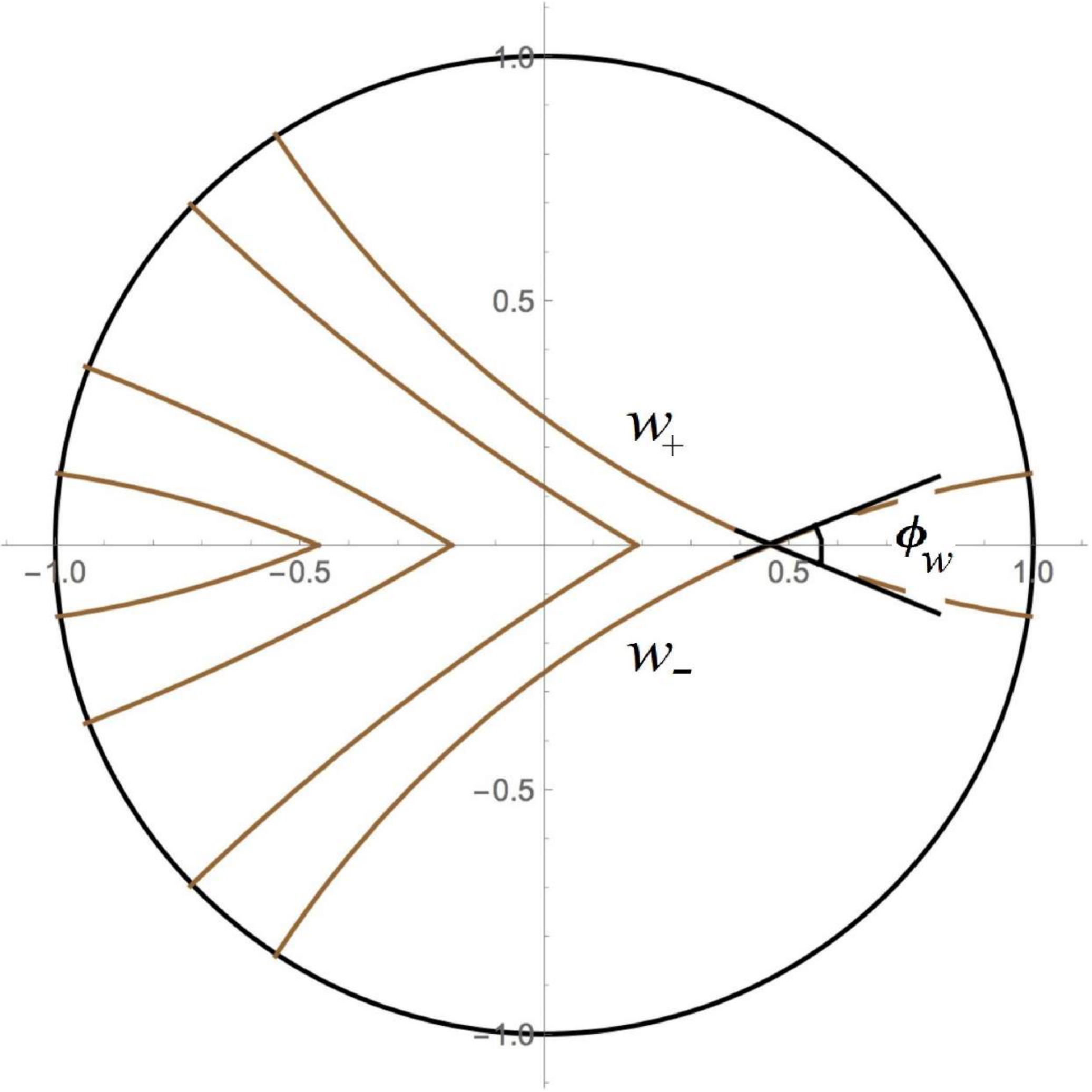}  {\bf A}$\,\,\,\,\,\,\,$
     \includegraphics[width=6cm]{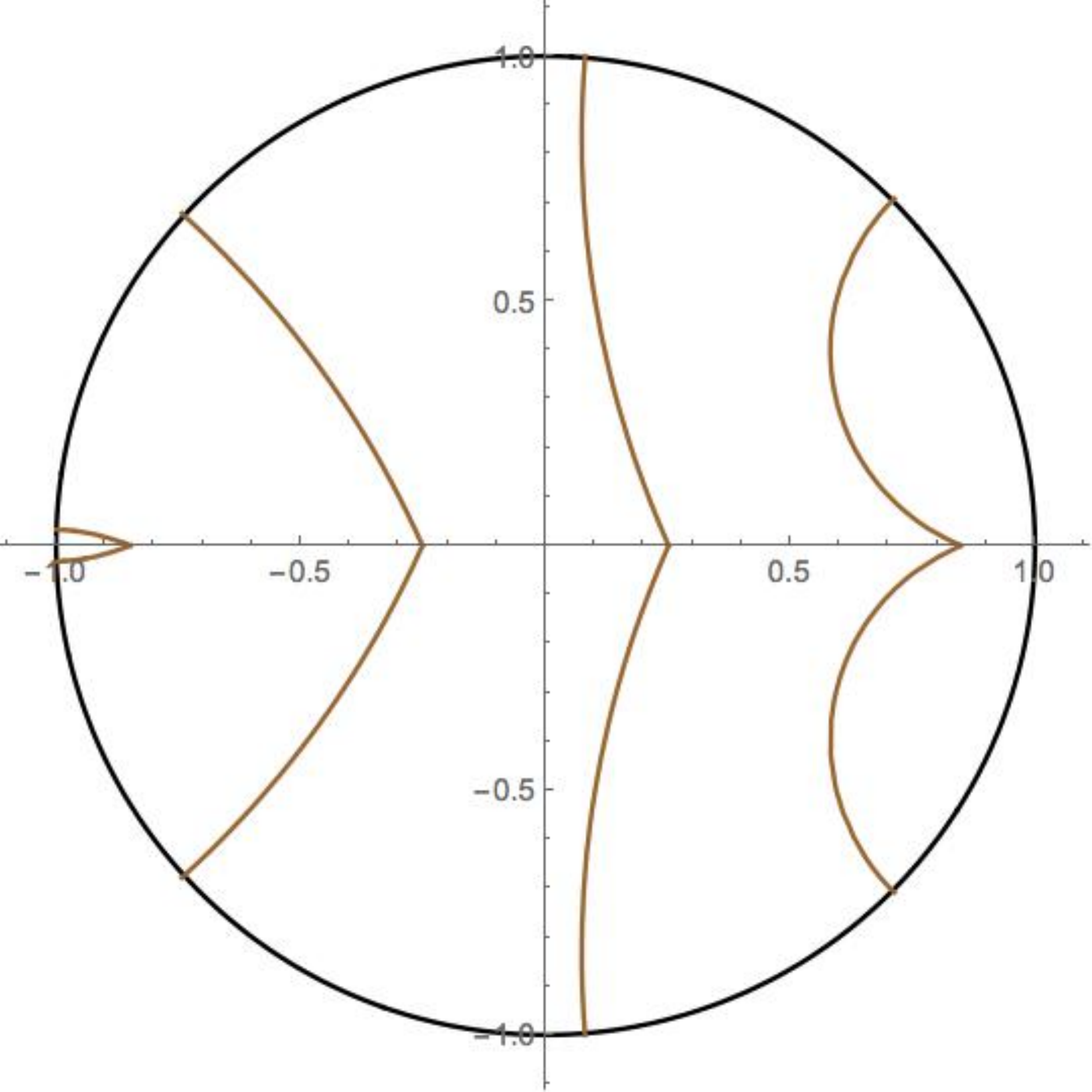} {\bf B}\\$\,$\\
       \caption{Constant time slices of the $AdS_3$ with  a moving massive particle for different times. The black circle represents the boundary of the $AdS_3$ space and brown curves are sections of the faces of the wedge in different time moments: $t=-\pi/2, -0.5, 0.5$ and $\pi/2$ from the left to the right for the each plot. The  section at $t=\pi/2$ of the wedge faces  is indicated by $w_\pm$ and  $\varphi$
       is the angle between $w_\pm$ at the crossing point, the location of the particle at $t=\pi/2$. We take parameter values to be $\xi=1$, $\alpha=\pi/4$ (A) and to be $\xi=2.5$, $\alpha=\pi/4$ (B).  }\label{2d}
\end{figure}

The massive particle moves from the left to right and vice versa periodically (with period $T=2\pi$).  Note that if $\xi \rightarrow \infty$ we will obtain the case of massless moving particle $r=\tan(t/2)$ that coincides with formulas for the massless particle in paper \cite{Matschull:1998rv}.
For constant time $t$ slices the wedge faces are some curves intersecting at the particle position.
The angle $\phi_w$ between these two curves at the intersection point can be expressed as:
\be\nn
\varphi=\arctan\left(\frac{4 \sin ^2\frac{\alpha
   }{2} \sin \alpha  \cosh
   \xi  \sqrt{1-\sin ^2 t \tanh
   ^2\xi }}{\sin ^2\alpha -4
   \sin ^4\frac{\alpha
   }{2} \left(1+2\sinh^2\xi\sin^2\frac{\alpha}{2}\right)}\right).
\ee
From this formula we can see, that the angle $\phi_w$  is maximal at $t=\pm\frac{\pi}{2}$ if $\alpha>\pi$ and at $t=0$ if $\alpha<\pi$ (see Fig.\ref{2d}).

\begin{figure}
    \centering
     \includegraphics[width=6cm]{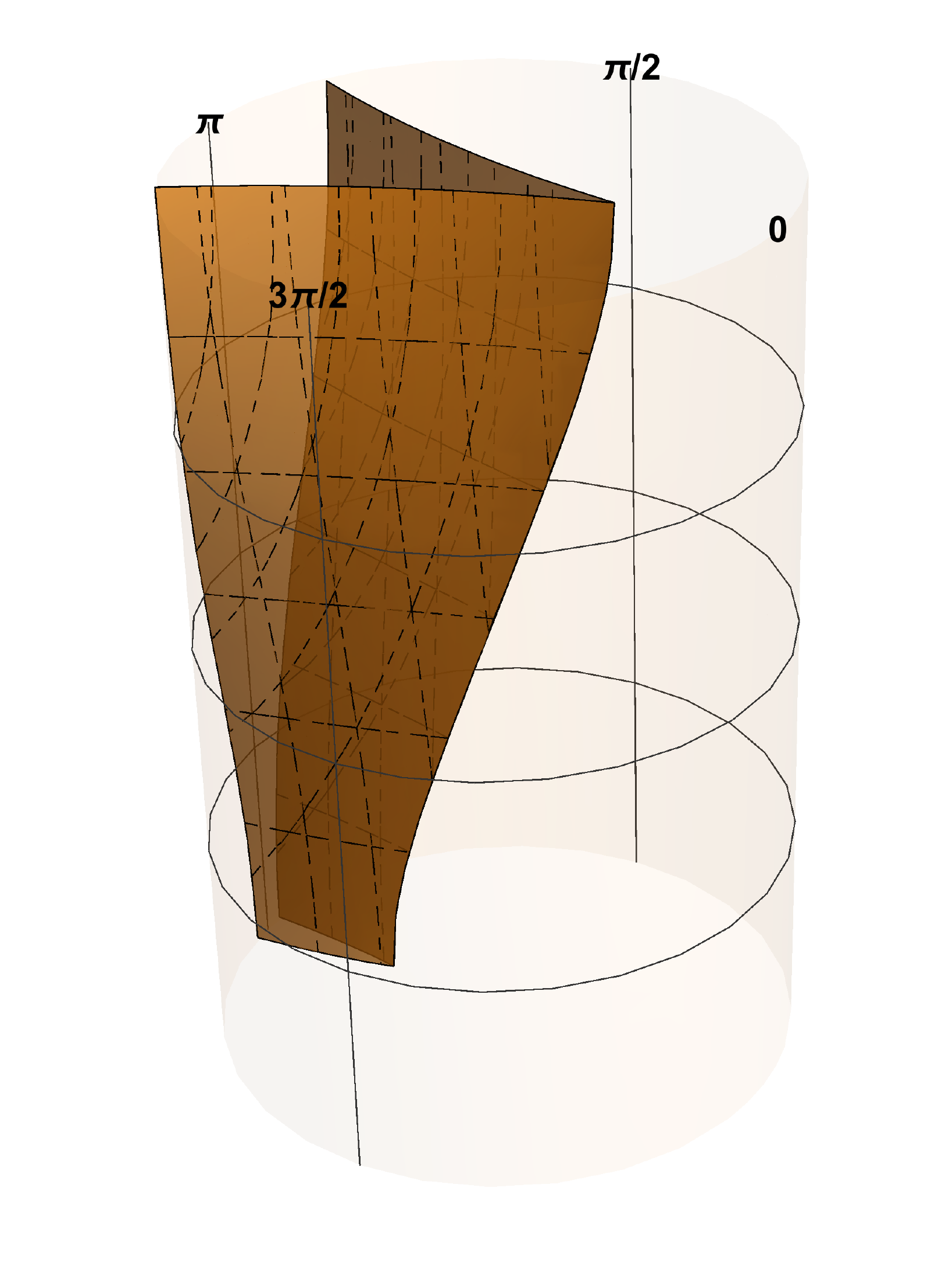}  {\bf A}$\,\,\,\,\,\,\,\,\,\,\,\,\,\,\,\,\,\,\,$
     \includegraphics[width=6.2cm]{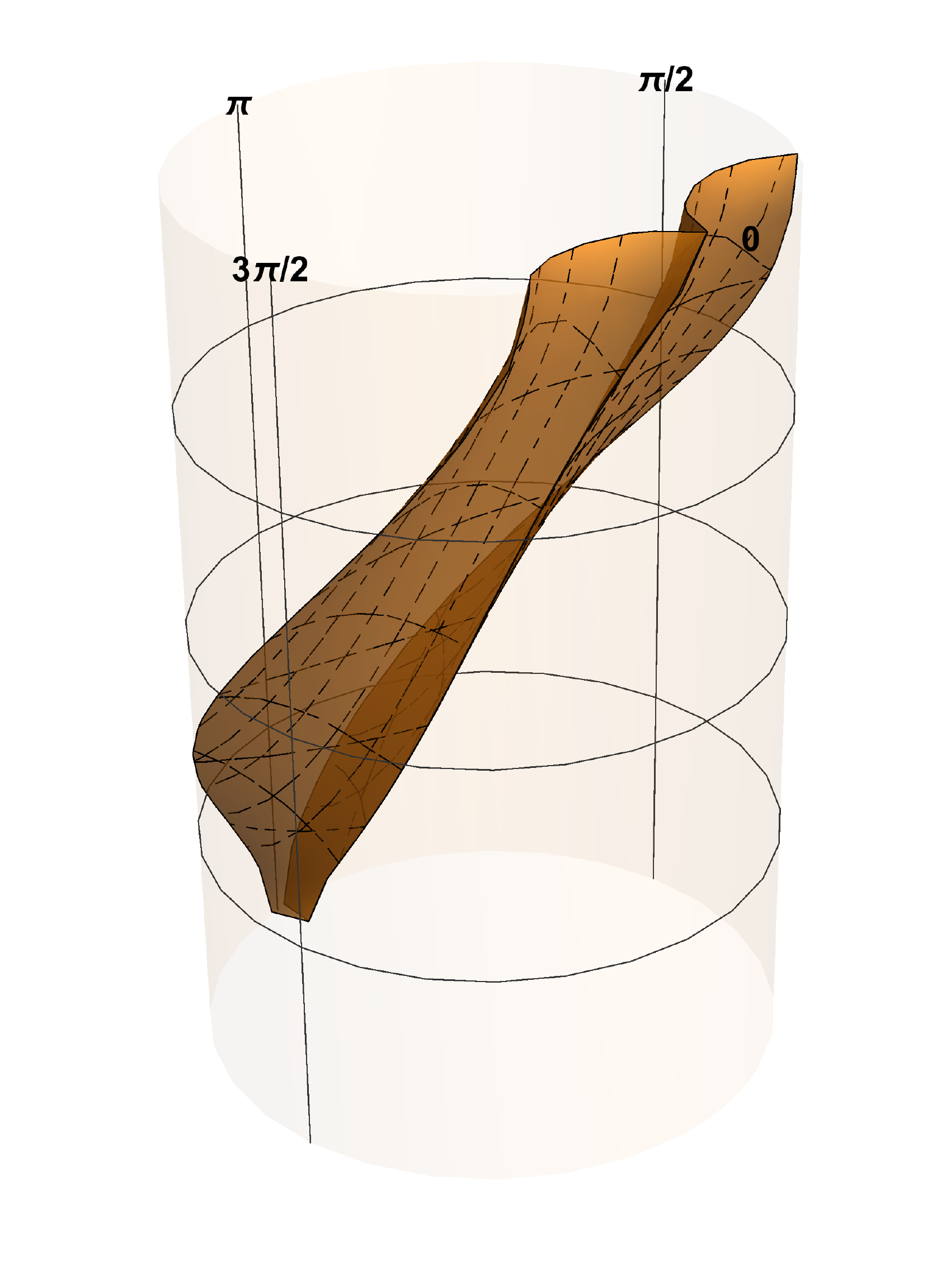} {\bf B}\\$\,$\\
       \caption{ The plot of the wedge of moving massive particle with certain parameters $\xi$ and $\alpha$ values. Here $-\pi/2<t<\pi/2$.  A. Here we take parameter values to be $\xi=1$, $\alpha=\pi/8$.  B. Here we take parameter values to be $\xi=2.5$, $\alpha=\pi/4$}\label{3d}
\end{figure}

Three dimensional plots of the wedge are presented in Fig.\ref{3d}.

\subsection{Correlation functions on the boundary and geodesics in the $AdS_3$.}\label{AdS-corr}

 In the AdS/CFT correspondence, to calculate the two-point correlation function of a scalar operator $\Phi_\Delta $ with large  conformal weight $\Delta$ on the $AdS_3$ boundary,  one can use the geodesic
 approximation \cite{Balasubramanian:1999zv}. In this approximation one defines the correlator:

  \be\label{main-g}
G_\Delta(\phi_a,t_a;\phi_b,t_b)=
e^{-\Delta {\cal L}_{ren}(\phi_a,t_a;\phi_b,t_b)}.
 \ee
Here $a$ and $b$ are two points on the boundary of the $AdS_3$ with coordinates   $(\phi_a,t_a)$ and $(\phi_b,t_b)$.  For the  spacelike  separated points $a$ and $b$,
 ${\cal L}_{ren}(\phi_a,t_a;\phi_b,t_b)$ is the renormalized length of the geodesic connecting these points \cite{Balasubramanian:1999zv,AB-TMF}. If $a$ and  $b$ are timelike separated points, there is no geodesic between them. In the paper we restrict ourselves to the geodesics between spacelike-separated points. The contribution of  timelike separated points from the geodesic prescription is considered in the next paper.

\subsubsection{Spacelike separated points}\label{SL-sep}
Now we remind how the geodesic approximation works in the $AdS_3$  global coordinates.
We consider  two spacelike separated boundary points $(\phi_a,t_a)$ and $(\phi_b,t_b)$.
We assume for definiteness  $\phi_b>\phi_a$.  The   geodesic curve in the bulk
$(\phi,t,r)=(\phi(\lambda),t(\lambda),r(\lambda))$, connecting these points is given by :
\bea\label{geod-lambda11m}
\phi_{ab}(\lambda)&=&\arctan\left(\tan \frac{D_s[\phi_a,\phi_b]}{2}\cdot
\tanh \lambda\right)+\Sigma_s[\phi_a,\phi_b],\\
\label{geod-lambda111m}
t_{ab}(\lambda)&=&\arctan \left(\tan \frac{D[t_a,t_b]}{2}\cdot\tanh\lambda\right)+\Sigma[t_a,t_b]\eea
and
\bea\nn
r(\lambda)&=&   \frac
   {\sqrt{\cos (2
   D[t_a,t_b]+\cosh (2  \lambda)}-\sqrt{\cos (2
   D[t_a,t_b]-\cos (2 D_s[\phi_a,\phi_b]))}}
   {\sqrt{\cos (2
   D_s[\phi_a,\phi_b] )+\cosh (2 \lambda)}}.\eea
Here the parametrization is taken so that the parameter value $\lambda=-\infty$ corresponds
to  the point $(\phi_a,t_a)$ on the boundary and $\lambda=+\infty$
corresponds to  the point $(\phi_b,t_b)$. $D_s[\phi_a,\phi_b]$, $D[t_a,t_b]$, $\Sigma_s[\phi_a,\phi_b]$ and $\Sigma[t_a,t_b]$ are defined as
\bea\label{DSl}
\phi_b-\phi_a<\pi:\,\,\,\,\,\,D_s[\phi_a,\phi_b]&=&\phi_b-\phi_a,\\\nn
\Sigma_s[\phi_a,\phi_b]&=&\frac{\phi_b+\phi_a}{2};\\\nn
\phi_b-\phi_a>\pi:\,\,\,\,\,\,D_s[\phi_a,\phi_b]&=&\phi_b-\phi_a-2\pi,\\\nn
\label{Sigma}\Sigma_s[\phi_a,\phi_b]&=&\frac{\phi_b+\phi_a}{2}+\pi;
\\\nn
D[t_a,t_b]&=&t_b-t_a,\\\nn
\Sigma[t_a,t_b]&=&\frac{t_b+t_a}{2}.
\eea

Note, that in \cite{Arefeva:2015sza} another parametrization for this geodesic has been used.

 It is easy to calculate the geodesic length ${\cal L}_{AdS}$ between two points  on the curve \eqref{geod-lambda11m}-\eqref{geod-lambda111m} using
 \bea\label{Lads}
\cosh  {\cal L}_{AdS}&=&-\frac{1}{2}\left[\det({\bf x}_a-{\bf x}_b)-2\right]\\&=&\nn -(x_{(a)},x_{(b)})=x_{0,a}x_{0,b}+x_{3,a}x_{3,b}-x_{1,a}x_{1,b}-x_{2,a}x_{2,b},
 \label{prod}\eea
  where $x_{i,a}$ and $x_{i,b}$, $i=0,1,2,3$ are embedding coordinates (\ref{comp}) of the endpoints.
   Writing down \eqref{Lads} explicitly  in coordinates $(\phi,t,\chi)$ we  express the geodesic length between points $a=(\phi_a,t_a,\chi_a)$ and $b=(\phi_b,t_b,\chi_b)$ as:
\bea\label{Lads-coord}
    &-&\cosh {\cal L}(a;b)=\\\nn&=&-\cosh\chi_a \, \sin t_a\cosh\chi_b \, \sin t_b
-\cosh\chi_a \, \cos t_a\cosh\chi _b\, \cos t_b \\\nn&+&
\sinh\chi _a\, \cos \phi_a \sinh\chi _b\, \cos \phi _b+
\sinh \chi_a \sin \phi_a \sinh \chi_b \sin \phi_b.
   \eea
  When points $a$ and $b$ go to the boundary (i.e. $\chi_{a,b} \to \infty$) one gets:

\bea
\label{non-bdr}
{\cal L}_{reg}(a;b)&=&\ln \left[ \left( \cos (t_a- t_b ) -\cos (\phi_a  - \phi _b)\right)\frac{e^{\chi_a+\chi_b}}{2}\right]\\
&=&\ln \left[ 2(\cos (t_a- t_b ) -\cos (\phi_a  - \phi _b))\right]+\delta _a+\delta_b,\nn
\eea
where $\delta _a=\chi_a-\ln 2$ and $\delta _b=\chi_b-\ln 2$.

 Removing the divergent parts $\delta _a$, $\delta _b$ in \eqref{non-bdr} we get the renormalized geodesic length for the spacelike geodesic connecting two points on the boundary:
 \be
 \label{ren-2}
 {\cal L}_{ren}(t_a,\phi_a;t_b,\phi_b)=\ln[2\left(\cos(t_a-t_b)-\cos(\phi_a-\phi_b)\right)],
 \ee
 From this formula and \eqref{main-g} the two-point function on the $AdS_3$ boundary  is:

 \be\label{G0}
G_{\Delta, AdS}\left(\phi_a,t_a,\phi_b,t_b\right)=\left(\frac{1}{2(\cos (t_a-t_b)-\cos(\phi_a-\phi_b))}\right)^\Delta.
 \ee

\subsubsection{Timelike separated points}\label{TL-sep}
Mentioned above, there are no continous geodesics in the $AdS_3$ connecting two timelike separated points on the  boundary. So, to use the geodesic approximation in calculation of two-point functions for timelike separated points, we use the prescription proposed in \cite{AB-TMF,Arefeva:2015sza}. This prescription is related with the prescription that has been early proposed in \cite{Balasubramanian:2012tu} in the Poincare  patch.  According to this prescription to calculate the  correlator for  timelike separated points one has to relate these
 points by a quasigeodesic, that consists of two pieces of the spacelike geodesics with a discontinuity at the Poincare horizon. The explicit formulae are
 \bea\label{t-geod-phi-lambda12m}
 \phi_{ab}(\lambda)&=&\arctan\left(\tan \frac{D[\phi_a,\phi_b]}{2}\cdot
\coth \lambda\right)+\Sigma[\phi_a,\phi_b],\\\nn
\\
\label{t-geod-lambda111m}
t_{ab}(\lambda)&=&\arctan \left(\tan \frac{D[t_a,t_b]}{2}\cdot\coth\lambda\right)+\Sigma[t_a,t_b],\eea
and
\bea
r(\lambda)=   \frac
   {\sqrt{\cos (2
   D[t_a,t_b])+\cosh (2  \lambda)}-\sqrt{\cos (2
   D[t_a,t_b])-\cos (2 D[\phi_a,\phi_b]))}}
   {\sqrt{\cos (2
   D[\phi_a,\phi_b]))+\cosh (2 \lambda)}},
  \label{tr-geod-lambda11m} \eea
 where $D[t_a,t_b]$ and $\Sigma[t_a,t_b]$ are defined as (\ref{DSl})

\begin{figure}[htbp]
\begin{center}
\includegraphics[width=5cm]{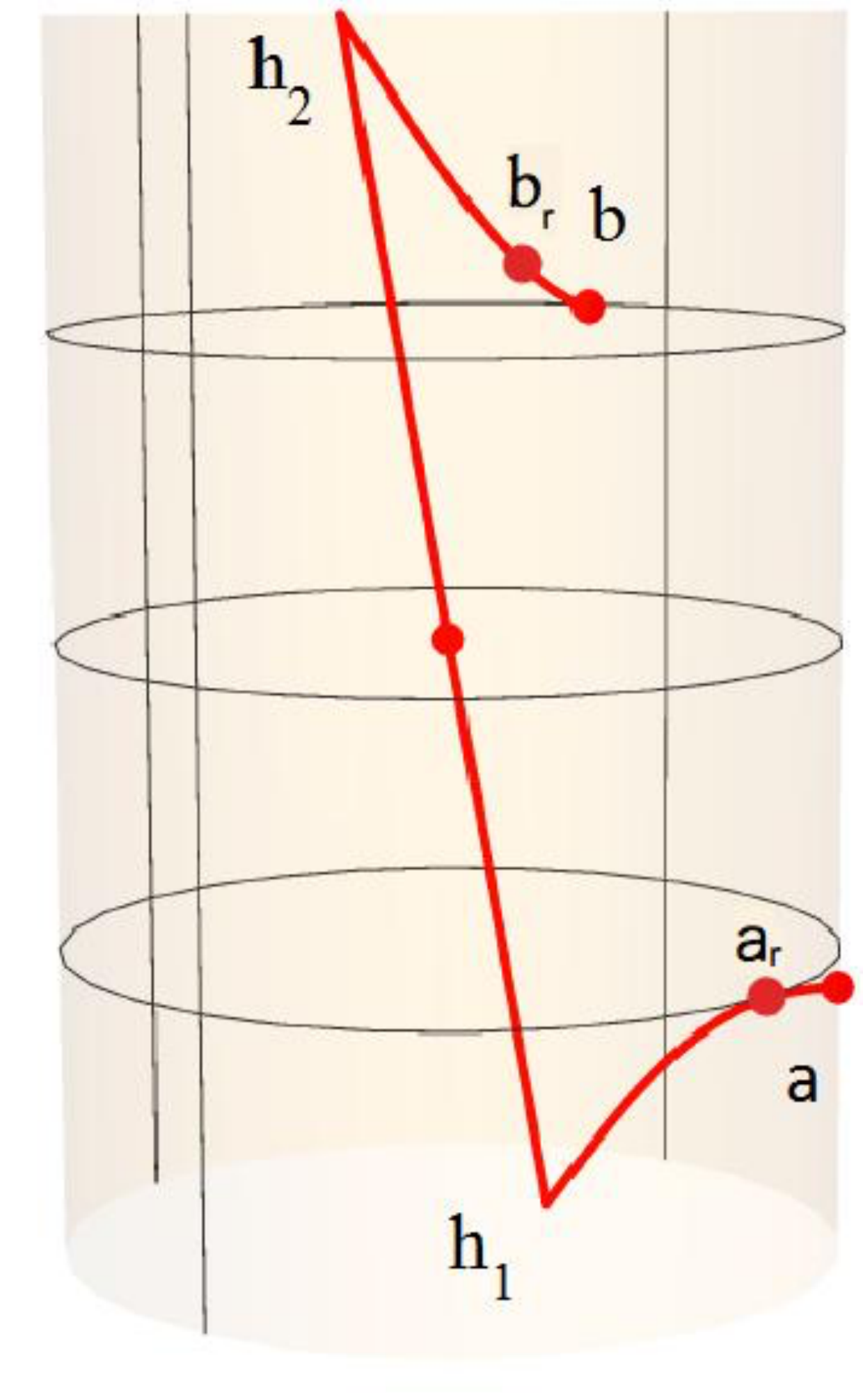}
\caption{The plot of the quasigeodesic connecting two points  $a$ and $b$ on the boundary. Points $a$ and $b$ correspond to affine parameter
limits $\lambda\to \mp\infty$, respectively.
Points $a_r$ and $b_r$ are the nearest points in the bulk corresponding to  finite $\mp \lambda$ parameters. The curve $ah_1$  is  the spacelike geodesic connecting the point $a$ at the boundary  and the point $h_1$ at the Poincare horizon. The curve $ah_2$ is the spacelike geodesic connecting the point  $b$ at the boundary and the point $h_2$ at the Poincare horizon.
}
\label{Fig:jupm}
\end{center}
\end{figure}

In Fig.\ref{Fig:jupm} we plot the quasigeodesic corresponding to the boundary  points $a$ and $b$
with coordinates $(\phi_a,t_a)$ and $(\phi_b,t_b)$, respectively.

 For simplicity we consider the case of symmetric points. In this case the boundary points are taken to be $(\delta \phi,\delta t)$, $(-\delta \phi,-\delta t)$,
 where
 \be
 \label{delta}
 \delta \phi=\frac{\phi_a-\phi_b}{2},\,\,\,\,\,\,\,\,  \delta t=\frac{t_a-t_b}{2}.
 \ee
The part of the quasigeodesic that starts at the point $a$ at $\lambda=-\infty$  reaches the Poincare horizon at the point with coordinates $(  \phi_{h_1}, t_{h_1}, \chi_{h_1})$.
  These coordinates correspond to the values of the right hand side of formulae (\ref{t-geod-phi-lambda12m})-(\ref{tr-geod-lambda11m})  at $\lambda \to-0$:
\bea\nn
t_{h_1}=-\frac{\pi}{2},\,\,\,\,\,
  \phi_{h_1}=-\frac{\pi}{2},\,\,\,\,\,
\chi_{h_1}={\mbox{arcsinh}}\frac{|\sin \delta \phi|}{\sqrt{\sin^2 \delta t-\sin^2 \delta \phi}}.
\eea
The part of the quasigeodesic that reaches  the point $b$ at $\lambda=+\infty$  starts from the Poincare horizon at the point with coordinates $(  \phi_{h_2}, t_{h_2}, \chi_{h_2})$, that correspond  to the values of the right hand side of formulae (\ref{t-geod-phi-lambda12m})-(\ref{tr-geod-lambda11m})  at $\lambda \to+0$:
\bea\nn
t_{h_2}=\frac{\pi}{2},\,\,\,\,
  \phi_{h_2}=\frac{\pi}{2},\,\,\,\,
\chi_{h_2}={\mbox{arcsinh}}\frac{|\sin \delta \phi|}{\sqrt{\sin^2 \delta t-\sin^2 \delta \phi}}.
\eea

Taking $b=h_1$  and $a=a_r$  in formula \eqref{Lads-coord} we get the geodesic length between $a_r$, the  point near the boundary (i.e. $\chi$ is large) with coordinates  $(-\delta\phi,-\delta t,\chi)$ and the point $h_1$ with coordinates  $(\phi_{h_1},t_{h_1},\chi_{h_1})$:
\bea\label{quasi-half}
{\cal L}(a_r;h_1)=\ln\left(2\sqrt{\sin^2\delta t-\sin^2\delta
\phi\,}\right)\, +\delta_{a_r} +...,\,\,\,\,\,\,
\delta_{a_r}=\chi_{a_r}-\ln 2,
\eea
where dots mean subleading terms when $\chi\to \infty$.
Subtracting the linear on $\chi $ term we get the renormalized geodesic length between the point $a$ on the boundary and point $h_1$:
\be\label{quasi-half-ren}
{\cal L}_{ren}(a;h_1)=\ln \left(2\sqrt{\sin^2\delta t-\sin^2\delta
\phi\,}\right). \ee

In a similar way   taking $b=h_2$  and $a=b_r$  in formula \eqref{Lads-coord} and subtracting
the divergent term $\delta_{b_r}=\chi_{b_r}-\ln 2$ we get
the renormalized geodesic length between the point $b$ on the boundary and the point
$h_2$ on the Poincare horizon:
\be\label{quasi-half-ren2}
{\cal L}_{ren}(b;h_2)=\ln\left(2\sqrt{\sin^2\delta t-\sin^2\delta
\phi \,}\right). \ee
Summing (\ref{quasi-half-ren}) and  (\ref{quasi-half-ren2}) we get
\bea\label{quasiren}\nn
&\,&{\cal L}_{quasi,ren}(t_a,\phi_a;t_b,\phi_b)= {\cal L}_{ren}(a;h_1)+{\cal L}_{ren}(b;h_2)\\&=&2\ln (2\sqrt{\sin^2\delta t-\sin^2\delta
\phi})=\ln(2(-\cos(t_a-t_b)+\cos(\phi_a-\phi_b))),
\eea
Combining \eqref{quasiren} with the (\ref{main-g}) we get the answer for the two-point correlation function
for timelike separated points
 \be\label{GO-t}
G_{\Delta, AdS}\left(\phi_a,t_a,\phi_b,t_b\right)=\left(\frac{1}{2(-\cos (t_a-t_b)+\cos(\phi_a-\phi_b))}\right)^\Delta.
 \ee
Comparing   (\ref{G0}) and (\ref{GO-t}) we can write

\be\label{GOG}
G_{\Delta, AdS}\left(\phi_a,t_a,\phi_b,t_b\right)=\left(\frac{1}{2|\cos (t_a-t_b)-\cos(\phi_a-\phi_b)|}\right)^\Delta.
 \ee
\subsubsection{Reflection symmetry}
\begin{figure}[htbp]
\begin{center}
\includegraphics[width=4cm]{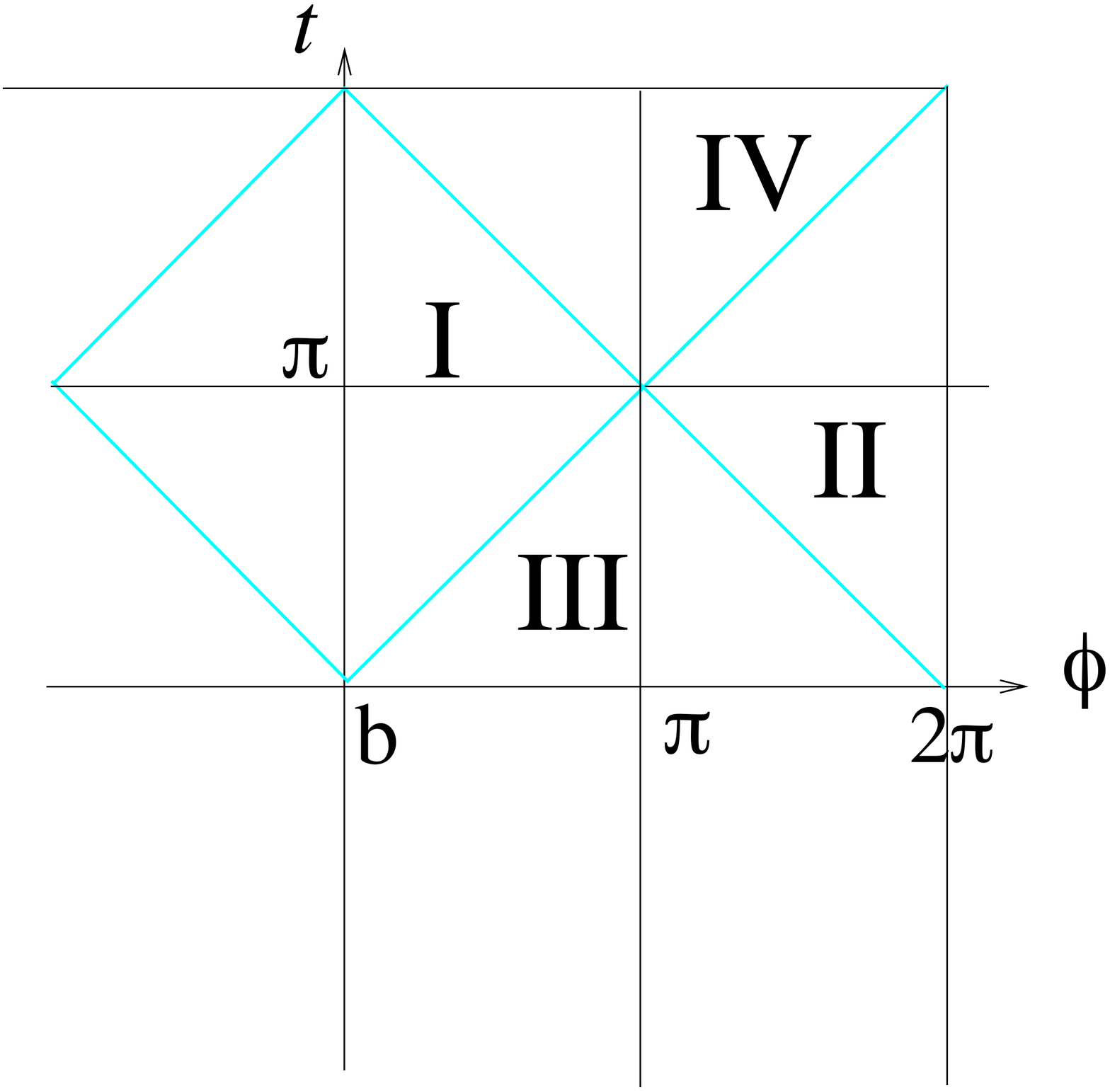}A\,\,\,\,\,
\includegraphics[width=4cm]{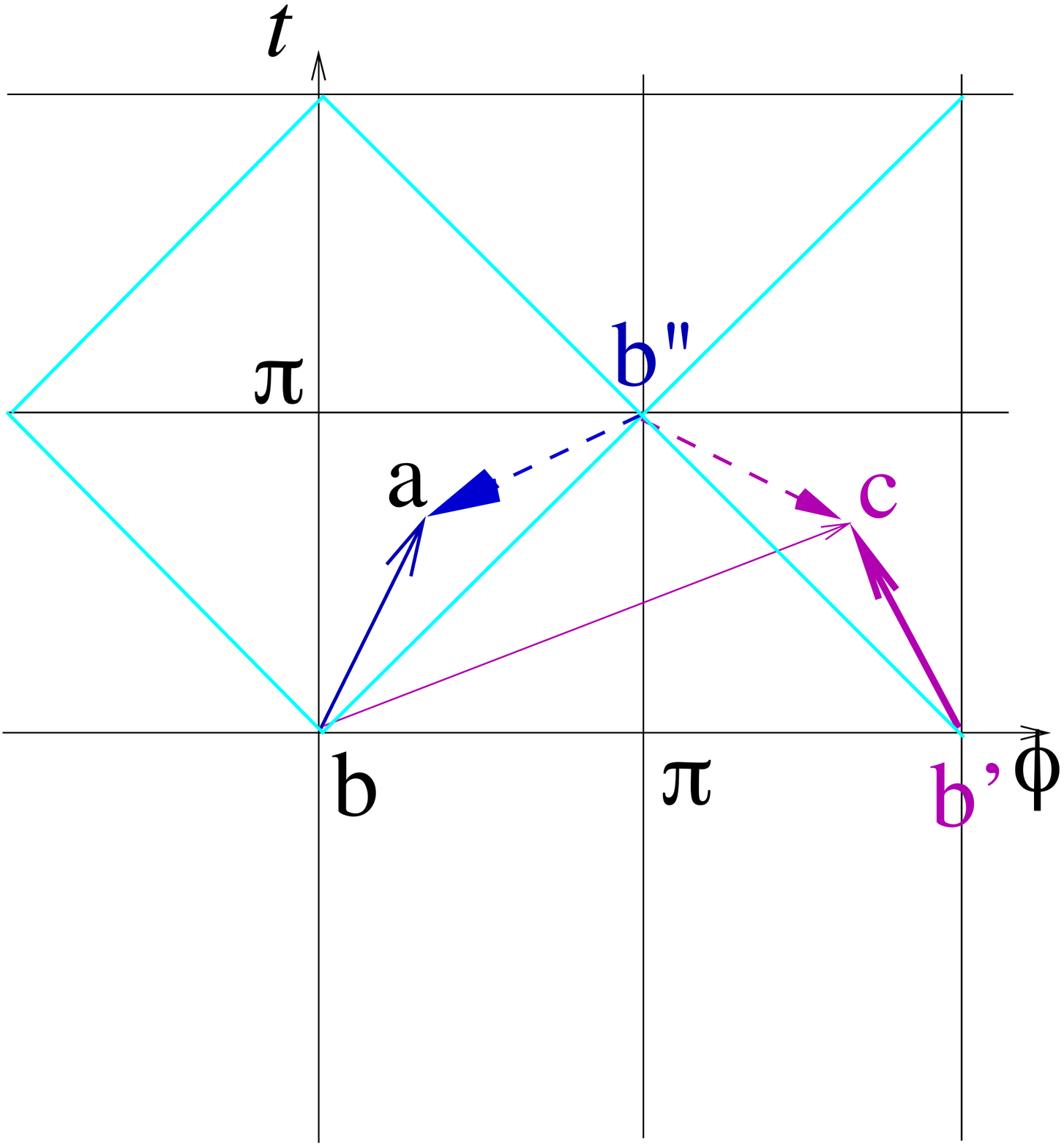}B\,\,\,\,\,
\includegraphics[width=4cm]{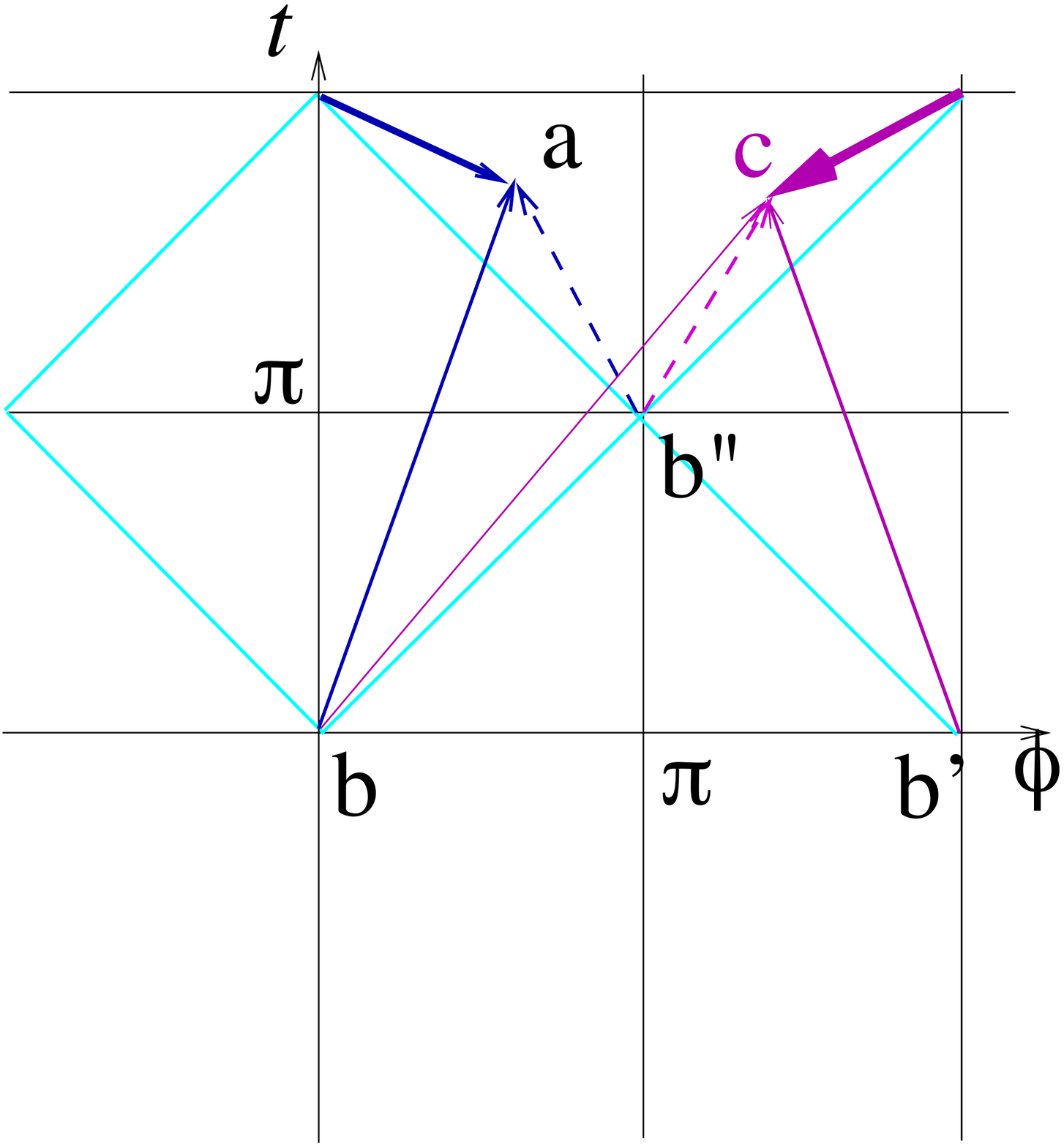}C
\caption{A. The plot shows different regions  I, II, III and IV.  Regions I and II are timelike regions with respect to the "trigonometrical"
interval $-\cos (t_b-t_c)+\cos(\phi _b-\phi_c)$ for $t_b=0,\,\,\phi_b=0$. Regions III and IV are spacelike regions with respect to the "trigonometrical"
interval.   B. The plot of the reflection transformation in the parts of  the region I (blue vectors) $ \vec{ba}\rightarrow \vec{b^{\prime \prime }a}$, in the
  the region II (magenta vectors) $ \vec{b^{\prime}c}\rightarrow \vec{b^{\prime \prime }c}$, here  $0<t_a<\pi$. C. The plot of the  reflection transformation $ \vec{ba}\rightarrow \vec{b^{\prime \prime }a}$ and $ \vec{bc}\rightarrow \vec{b^{\prime \prime }c}$ in the region IV.
}
\label{Fig:reflection}
\end{center}
\end{figure}

As has been noted in \cite{Arefeva:2015sza} the correlator (\ref{GOG}) possesses the reflection symmetry, 
\be
\nn
G_{\Delta,AdS}(\phi_a,t_a;\phi_b,t_b)=G_{\Delta,AdS}(\phi_a,t_a;\phi_b+\pi,t_b+\pi),
\ee
i.e. this  correlator
is invariant under a shift on $\pi$ of both arguments $t$ and $\phi$ simultaneously.
The transformation $(t,\phi)\to (t+\pi,\phi+\pi)$ is the reflection transformation. Under this transformation
the timelike interval $ab$ transforms to the spacelike one $ab''$ and vice versa (see Fig.\ref{Fig:reflection}).

\subsubsection{Remarks about the Wightman, causal and  retarded correlators}

Let us note the relation of the function (\ref{GOG}) defined via the geodesics approximation with the causal, retarded and Wightman correlators. The Wightman correlators are obtained \cite{OS,Luscher:1974ez,haag}  by the $i\epsilon $ prescription from the  CFT
correlators on the Euclidean cylinder \be\label{eucl}
G_{E}(\phi_a,\tau_a;,\phi_b,\tau_b)=\left(\frac{1}{2(\cosh(\tau_a-\tau_b)-\cos(\phi_a-\phi_b))}\right)^{\Delta}.
\ee
and can be written as
\bea\nn
G_W( \phi_a,t_a;\phi_b,t_b)& =&\langle O(\phi_a,t_a)O(\phi_b,t_b)\rangle\\&=&\left(\frac{1}{2(\cos (t_a-t_b-i\epsilon) - \cos (\phi_a-\phi_b))}\right)^\Delta
\label{extoom-S}
\eea
In the sense of distributions \cite{GS} we can present (\ref{extoom-S}) as
\bea\label{extoom-cc-m}
&\,&G_{W}(\phi_a,t_a;\phi_b,t_b)
=G_{\Delta,AdS}(\phi_a,t_a;\phi_b,t_b)\\\nn
&\cdot&\left\{\Psi_W(t_a;t_b)\,\theta(-\cos (t_a-t_b)+ \cos (\phi_a-\phi_b)]+\theta(\cos (t_a-t_b)  - \cos (\phi_a-\phi_b))\right\}\nn
\eea
where $G_{\Delta,AdS}$ is defined as (\ref{GOG}) and $\Psi_W$ function is:
\bea\nn
\Psi_W(t _a;t_b)&=&e^{-i\,\pi\,\Delta \,\,{\mbox {sign}}(\sin (t _a-t_b))}.
\eea

Using (\ref{extoom-S}) one gets the representation for the causal  correlator of the  conformal fields on the cylinder:
\bea\nn
G_{c}(\phi_a,t_a;\phi_b,t_b)&=&\langle TO(\phi_a, t_a)O(\phi_b,t_b)\rangle_{c}
\\ &=&\left(
\frac{1}{2(\cos (t_a-t_b-i\epsilon( t_a-t_b)) - \cos (\phi_a-\phi_b))}\right)^\Delta,
\label{Gc}
\eea
that can be written as
\bea\label{causal}
&\,&G_c(\phi_a,t_a;\phi_b,t_b)=G_{\Delta,AdS}(\phi_a,t_a;\phi_b,t_b)\\
&\cdot&\left\{\Psi_c(t_a;t_b)\,\theta(-\cos (t_a-t_b)+ \cos (\phi_a-\phi_b))+\theta(\cos (t_a-t_b)  - \cos (\phi_a-\phi_b))\right\},\nn
\eea
where
\bea\nn\\
\Psi_c(t _a;t_b)&=&\nn
e^{-i\,\pi\,\Delta \,\,{\mbox {sign}}(\sin (t _a-t_b))}\,\theta(t_a-t_b)+e^{i\pi\,\Delta \,{\mbox {sign}}(\sin (t _a-t_b))}\,\theta(-t_a+t_b).\\\nn
\,\label{qqq}
\eea

 The retarded correlator can be represented as
  \bea
  \label{ret}
G_{ret}(\phi_a,t_a;\phi_b,t_b)&\equiv &\theta(t_a-t_b)\langle [O(\phi_a,t_a),O(\phi_b,t_b)]\rangle,
\eea
and then we have
\bea\nn
&G&_{ret}(\phi_a,t_a;\phi_b,t_b)=\\&=&G_{\Delta,AdS}(\phi_a,t_a;\phi_b,t_b)\Psi_{ret}(t_a;t_b)\cdot\,\theta(t_a-t_b)
\theta(-\cos (t_a -t_b) +\cos (\phi_a-\phi_b)),\nn
\eea
where
\bea\nn
\Psi_{ret}(t_a;t_b)=-2i\sin( \pi \Delta \,{\mbox{sign}}(\sin (t_a-t_b))) .\eea

The above formula can be written in the universal way
\bea\label{extoom-cc-m}\nn
G_{A,\Delta}(\phi_a,t_a;\phi_b,t_b)
&=&G_{\Delta,AdS}(\phi_a,t_a;\phi_b,t_b)\cdot \Phi_{A,\Delta}(\phi_a,t_a;\phi_b,t_b),
\eea
where the subscript $A$ means Wightman $(W)$, causal $(c)$ or retarded $(r)$ and
\bea
&\Phi&_{B,\Delta}(\phi_a,t_a;\phi_b,t_b)=\label{Phi-1}\\&=&\Psi_{B,\Delta}(t_a;t_b)\cdot\,\theta(-\cos (t_a-t_b)+ \cos (\phi_a-\phi_b))+\theta(\cos (t_a-t_b)  - \cos (\phi_a-\phi_b))\nn\\
&\Phi&_{ret,\Delta}(\phi_a,t_a;\phi_b,t_b)=\label{Phi-2}\\&=&\Psi_{ret,\Delta}(t_a;t_b)\cdot\,\theta(t_a-t_b)\theta(-\cos (t_a -t_b) +\cos (\phi_a-\phi_b)),\nn
\eea
and where subscript $B$ stands for Wightman $(W)$ or causal $(c)$.

For an integer $\Delta$ the functions defined by \eqref{extoom-cc-m} and \eqref{causal} coincide
and the factor  $\Psi_{ret}$ is zero.

\section{Image method and winding geodesics}

\subsection{Image method on the living space }

 When the $AdS_3$ is deformed by the point particle, formulae (\ref{extoom-S}), (\ref{Gc})
 and (\ref{ret}) have to be modified. In particular,
 \bea
\label{gen}
&\,&<T O(\phi_a,t_a) O(\phi_b,t_b) >_{l.s.}= \\\nn&=& \left(\frac{1}{2(\cos(t_a-t_b +i(t_a-t_b)\epsilon) -
\cos(\phi_a-\phi_b))}\right)^\Delta\Theta_{0}(\phi_a,t_a;\phi_b,t_b)\\&+&\sum_{n\in  Z} \left(\frac{1}{2(\cos(t^*_{a,n}-t_b +i(t^*_{a,n}-t_b)\epsilon) -
\cos(\phi^*_{a,n}-\phi_b))}\right)^\Delta\Theta_{n}(\phi^*_{a,n},t^*_{a,n};\phi_b,t_b)Z_n(\phi^*_{a,n},t^*_{a,n};\phi_b,t_b)\nn\\
&+& \sum_{n\in  Z} \left(\frac{1}{2(\cos( t^\#_{a,n} -t_b+i(t^{\#}_{a,n} -t_b)\epsilon) -
\cos( \phi^*_{a,n} -\phi_b))}\right)^\Delta\bar{\Theta}_{n}( \phi^\#_{a,n}, t^\#_{a,n};\phi_b,t_b) \bar{Z_n}( \phi^\#_{a,n},t^{\#}_{a,n};\phi_b,t_b)\nn
\eea
Here  the subscript l.s. in the LHS of (\ref{gen}) means a living space of the boundary of the $AdS_3$ with  static or moving defects and
\be\nn
(\phi_a,t_a)^{*\,n}=(\phi^*_{a,n},t^*_{a,n}),
\ee
are coordinates of the image points obtained by n-times applications of the isometry *-transformation (\ref{isom}).  The $\#$-transformation is defined so that
\be\nn
\label{d-tr}
(\phi_a^\#\,t_a^\#)^{*}=( \phi_a, t_a),  \ee
and we also use notations
\be
\label{d-trn}
(\phi_a,t_a)^{\#\,n}=(\phi^\#_{a,n},t^\#_{a,n}).
\ee
In comparison with the usual image formula for Green functions, see for example eq.(4.1.35) in \cite{Skenderis:2008dg}, we put in (\ref{gen}) the extra factors:
\bea\nn
\Theta_{n}(\phi^*_{a,n},t^*_{a,n};\phi_b,t_b), \,\,\,\,\,\, \Theta_{n}( \phi^\#_{a,n}, t^\#_{a,n};\phi_b,t_b),\\\nn
Z_n(\phi^*_{a,n},t^*_{a,n};\phi_b,t_b), \,\,\,\,\,\, \bar{Z_n}( \phi^\#_{a,n},t^{\#}_{1,n};\phi_b,t_b),
\eea
The first two factors take values 1 or 0, depending on a particular image contribution, see more explanations below. Factors $Z$ and $Z_{n}$ are related to renormalizations, see also below Sect.\ref{renorm}.

According to (\ref{causal}) we have
\bea
\label{gen-explicit}
&\,&<T O(\phi_a,t_a) O(\phi_b,t_b) >_{l.s.}= \\\nn&=&\Phi_{c,\Delta}(\phi_a,t_a;\phi_b ,t_b) G_{\Delta,AdS} (\phi_a,t_a;\phi_b ,t_b)\,\Theta_{0}(\phi_a,t_a;\phi_b,t_b)\\&+&\sum^{n_{max}}_{n\in  Z}\Phi_{c,\Delta}(\phi^*_{a,n},t^*_{a,n};\phi_b,t_b)\,G_{\Delta,AdS}(\phi^*_{a,n},t^*_{a,n};\phi_b,t_b)\,\Theta_{n}(\phi^*_{a,n},t^*_{a,n};\phi_b,t_b)\,Z_n(\phi^*_{a,n},t^*_{a,n};\phi_b,t_b)\nn\\
&+& \sum^{\bar n_{max}}_{n\in  Z} \Phi_{c,\Delta}( \phi^\#_{a,n}, t^\#_{a,n};\phi_b,t_b)\,G_{\Delta,AdS}
( \phi^\#_{a,n}, t^\#_{a,n};\phi_b,t_b)\bar{\Theta}_{n}( \phi^\#_{a,n}, t^\#_{a,n};\phi_b,t_b)\, \bar{Z}_n( \phi^\#_{a,n},t^*_{a,n};\phi_b,t_b)\nn
\eea
where $G_{\Delta,AdS}(\phi_a,t_a;\phi_b ,t_b)$ is given by (\ref{GOG}) and $\Phi_{c,\Delta}(\phi_a,t_a;\phi_b,t_b)$ by (\ref{Phi-1}).

The presence of the $\Phi$-factors in summands in (\ref{gen-explicit}) is related to the
change of the causal relation between two points on the boundary under the isometry transformation \eqref{im3}. This is illustrated in Fig.\ref{Fig:Change-signature-d} and  Fig.\ref{Fig:Change-signature-s}. 

In Fig.\ref{Fig:Change-signature-d} a schematic plot of geodesics connecting the  points $a_i,\,i=1,2$ with  $b=(0,0)$
and  $b^\#$ is presented. The coordinates of the point  $b^\#$ is defined by the transformation
(\ref{d-trn}). We see that originally spacelike separated points can keep their causal relation  after the  $*$  and $\#$ transformations and also can change their causal relation.

In Fig.\ref{Fig:Change-signature-s} a schematic plot of geodesics connecting the  points $a_i,\,i=1,2$ with  $b=(0,0)$
and  $b^*$ is presented. The coordinates of the point  $b^*$ is defined by the transformation
(\ref{isom}). We see that originally spacelike separated points can keep their causal relation as well, after the isometry $*$ transformation,  can become timelike separated.

In this paper we ignore the contribution from timelike separated points, so we ignore $\Phi$-factors and define
\bea
\label{gen-ignore}
G_{l.s.} (\phi_a,t_a,\phi_b,t_b)&=& G_{\Delta,AdS} (\phi_a,t_a;\phi_b ,t_b)\,\Theta_{0}(\phi_a,t_a;\phi_b,t_b)\\&+&\sum_{n}\,G_{\Delta,AdS}(\phi^*_{a,n},t^*_{a,n};\phi_b,t_b)\,Z_n(\phi^*_{a,n},t^*_{a,n};\phi_b,t_b)\,\Theta_{n}(\phi^*_{a,n},t^*_{a,n};\phi_b,t_b)\nn\\
&+& \sum_{n}\,G_{\Delta,AdS}
( \phi^\#_{a,n}, t^\#_{a,n};\phi_b,t_b), \bar{Z}_n( \phi^\#_{a,n},t^*_{a,n};\phi_b,t_b)\bar{\Theta}_{n}( \phi^\#_{a,n}, t^\#_{a,n};\phi_b,t_b)\nn
\eea
\begin{figure}[h]
   \centering
\begin{picture}(180,90)
\put(-110,0){\includegraphics[width=6cm]{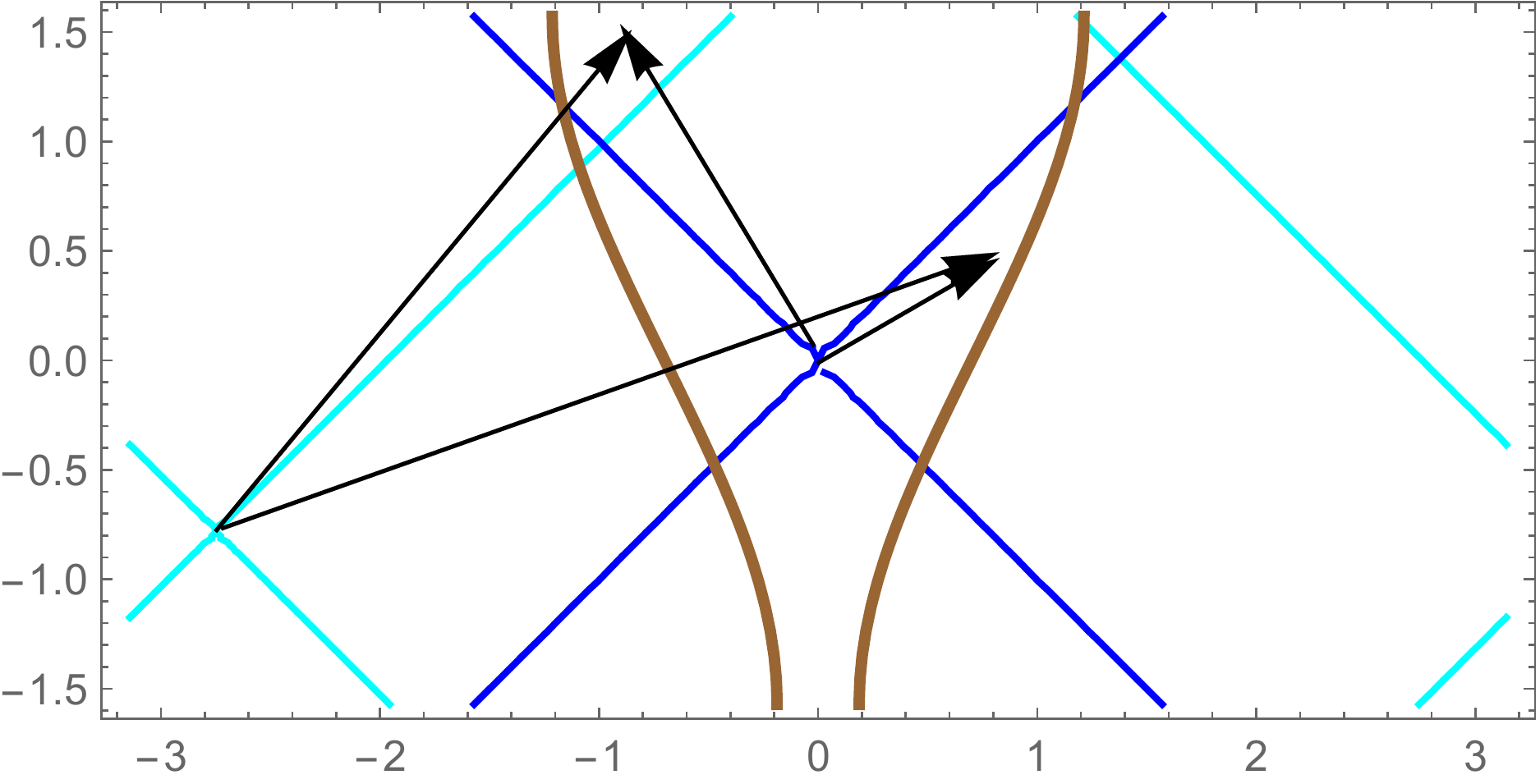}}
\put(-35,80){$a_1$}
\put(-20,35){$b$}
\put(5,55){$ a_2$}
\put(-90,35){$ b^{\#}$}
\put(130,0){\includegraphics[width=6cm]{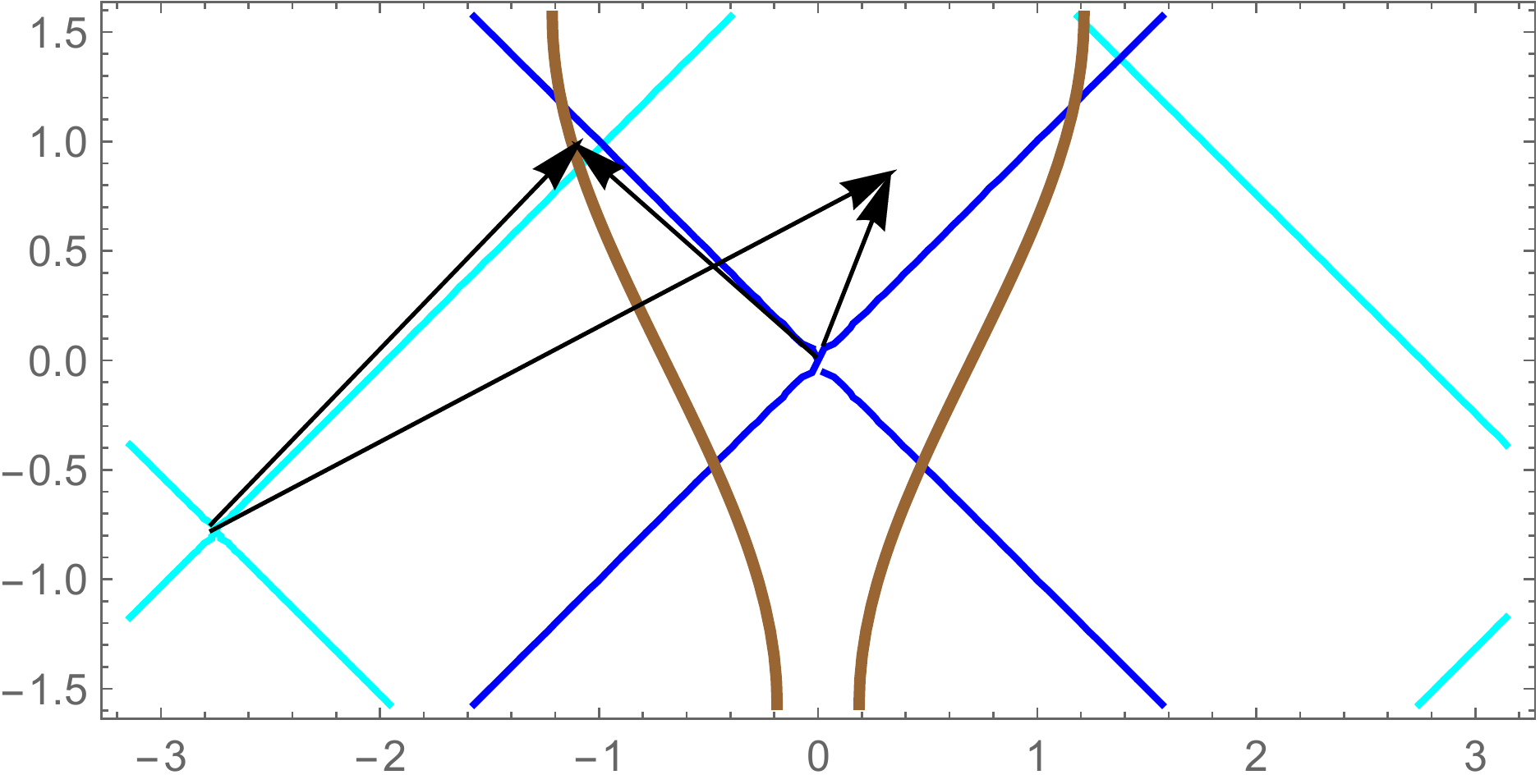}}
\put(200,75){$a_1$}
\put(218,35){$b$}
\put(230,69){$ a_2$}
\put(150,35){$ b^{\#}$}
\end{picture}\\
\caption{The schematic plot of geodesics connecting the  points $a_i,\,i=1,2$ with points  $b=(0,0)$
and  $b^\#$. A. The points $b$ and $a_1$  are timelike separated and points $b^\#$ and $a_1$
are also timelike separated. The points $b$ and $a_2$  are spacelike separated and also points $b^\#$ and $a_2$
are  spacelike separated.  B. The points $b$ and $a_1$  are spacelike separated while points $b^\#$ and $a_1$
are also timelike separated. The points $b$ and $a_2$  are timelike separated while points $b^\#$ and $a_2$
are  spacelike separated.
 Here $\alpha=1,\,\,\xi=1$.
} \label{Fig:Change-signature-d}\end{figure}
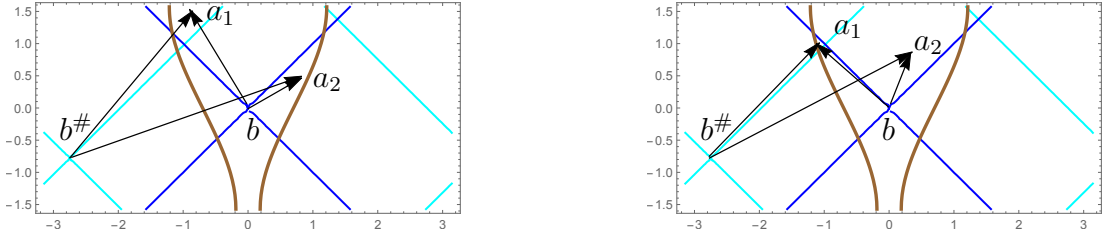
\begin{figure}[h]
   \centering
\begin{picture}(180,90)
\put(-110,0){\includegraphics[width=6cm]{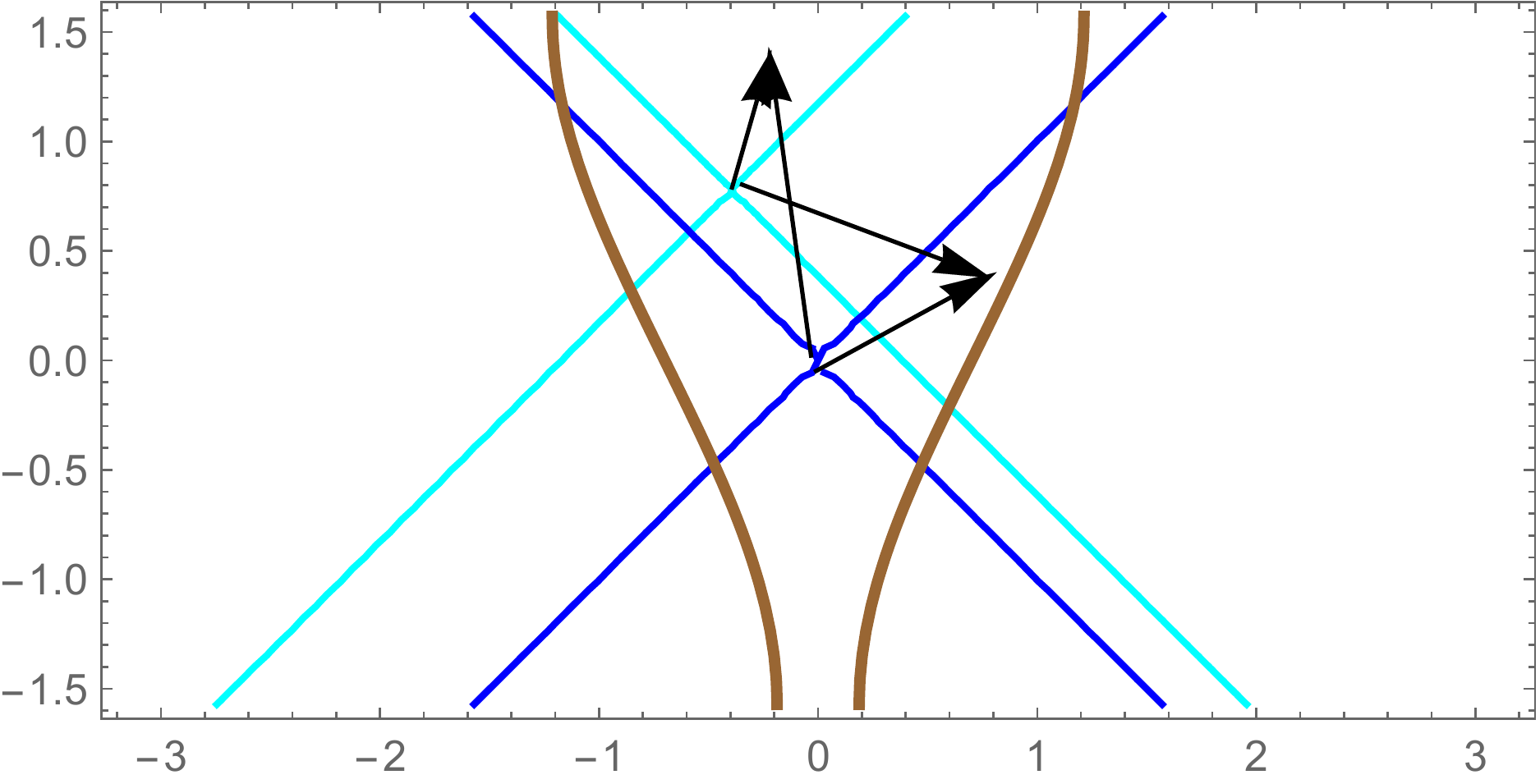}}
\put(-32,83){$a_1$}
\put(-20,35){$b$}
\put(5,55){$ a_2$}
\put(-40,65){$ b^{*}$}
\put(130,0){\includegraphics[width=6cm]{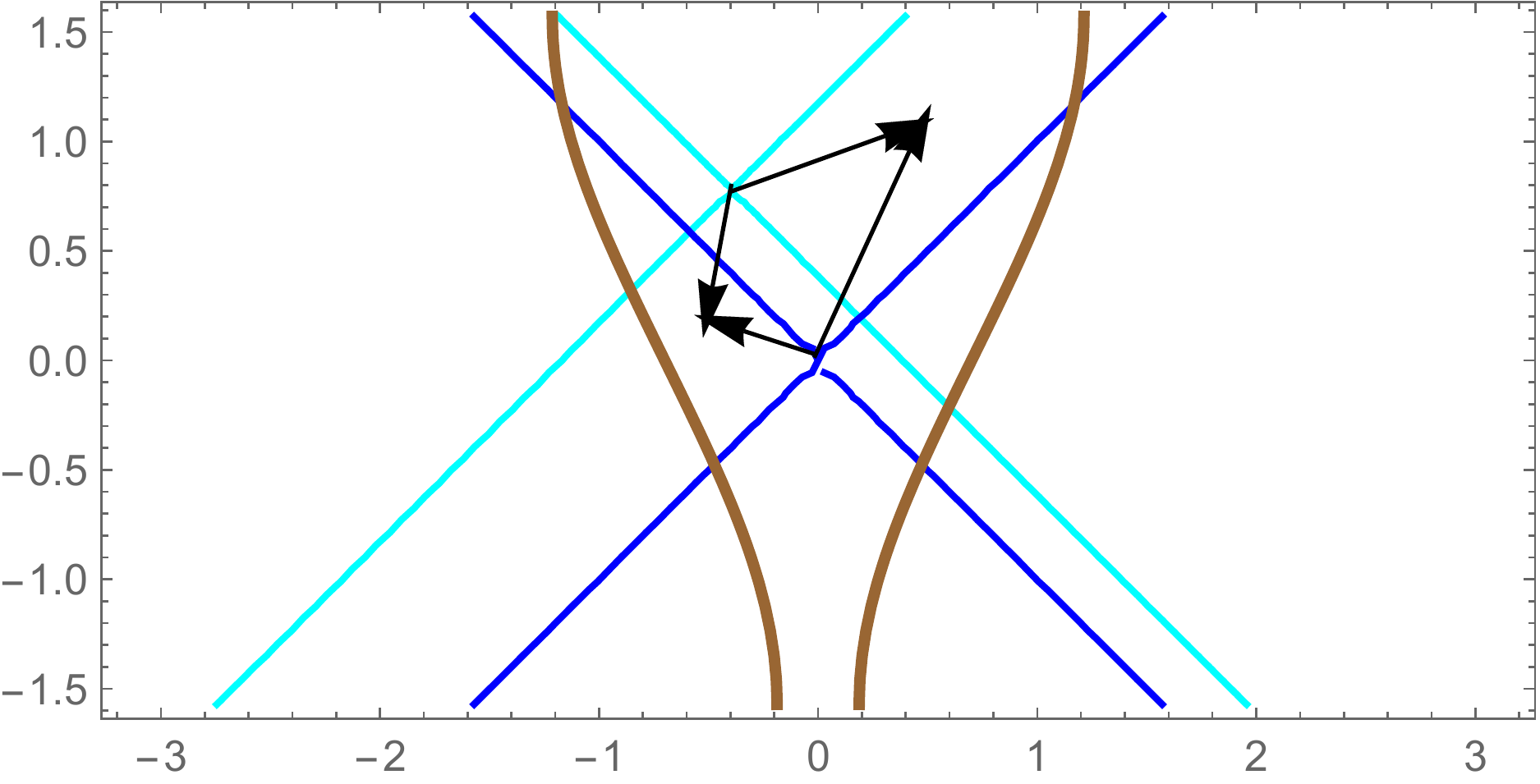}}
\put(199,45){$a_1$}
\put(218,35){$b$}
\put(235,75){$ a_2$}
\put(200,65){$ b^*$}
\end{picture}\\
\caption{The schematic plot of geodesics connecting the  points $a_i,\,i=1,2$ with  $b=(0,0)$
and  $b^*$. A. The points $b$ and $a_1$  are timelike separated and  points $b^*$ and $a_1$
are also  timelike separated. The points $b$ and $a_2$  are spacelike separated and  points $b^*$ and $a_2$
are  also spacelike separated.  B. The points $b$ and $a_1$  are spacelike separated while points $b*$ and $a_1$
are  timelike separated. The points $b$ and $a_2$  are timelike separated while points $b^*$ and $a_2$
are  spacelike separated.
 Here $\alpha=1,\,\,\xi=1$.
} \label{Fig:Change-signature-s}\end{figure}
Absorbing   $Z$-factors into the definition of $G_{\Delta,ren,n}(t^*_{a,n},\phi^*_{a,n};t_b,\phi_b)$
\bea\label{G-ren-definition}
G_{\Delta,ren,n}(\phi^*_{a,n},t^*_{a,n};\phi_b,t_b)&\equiv &G_{\Delta,AdS}(\phi^*_{a,n},t^*_{a,n};\phi_b,t_b)Z_n(\phi^*_{a,n},t^*_{a,n};\phi_b,t_b)\\\nn
G_{\Delta,ren,n}
( \phi^\#_{a,n}, t^\#_{a,n};\phi_b,t_b)&\equiv &G_{\Delta,AdS}
( \phi^\#_{a,n}, t^\#_{a,n};\phi_b,t_b)\, \bar{Z}_n( \phi^\#_{a,n},t^{\#}_{1,n};\phi_b,t_b)\eea
we get
\bea
\label{gen-ignore-2}
G_{l.s.} (\phi_a,t_a,\phi_b,t_b)&=& G_{\Delta,AdS} (\phi_a,t_a;\phi_b ,t_b)\,\Theta_{0}(\phi_a,t_a;\phi_b,t_b)\\&+&\sum_{n}\,G_{\Delta,ren,n}(\phi^*_{a,n},t^*_{a,n};\phi_b,t_b)\,\Theta_{n}(\phi^*_{a,n},t^*_{a,n};\phi_b,t_b)\nn\\
&+& \sum_{n}\,G_{\Delta,ren,n}
( \phi^\#_{a,n}, t^\#_{a,n};\phi_b,t_b)\,\bar{\Theta}_{n}( \phi^\#_{a,n}, t^\#_{a,n};\phi_b,t_b),\nn
\eea
and in one keeps the isometry invariance in the renormalization prescription then the following properties take place:
\bea\nn
G_{\Delta,ren,n}(\phi^*_{a,n},t^*_{a,n};\phi_b,t_b)&=&G_{\Delta,ren,n}(\phi^*_{a,n-1},t^*_{a,n-1}; \phi^\#_{b,1}, t^\#_{b,1})=G_{\Delta,ren,n}(\phi_a,t_a; \phi^\#_{b,n}, t^\#_{b,n})\\\nn
G_{\Delta,ren,n}(\phi^\#_{a,n},t^\#_{a,n};\phi_b,t_b)&=&G_{\Delta,ren,n}(\phi^\#_{a,n-1},t^\#_{a,n-1}; \phi^{*}_{b,1}, t^{*}_{b,1})=G_{\Delta,ren,n}(\phi_a,t_a; \phi^*_{b,n}, t^*_{b,n}).
\eea

In formulae \eqref{gen-ignore} and \eqref{gen-ignore-2} we do not specify ranges of summation over $n$. We clarify ranges of summation for static defect in Sect.\ref{Sec::static} and for moving defect in Sect.\ref{Sec::moving}.
As has been noted in Sect.\ref{AdS-corr} the function $G_{\Delta,AdS} (\phi_a,t_a;\phi_b ,t_b)$ is related with the
renormalized geodesic lengths. The function $G_{\Delta,ren,n}(\phi^*_{a,n},t^*_{a,n};\phi_b,t_b)$ also is related with  renormalized geodesic  lengths
for the cases when geodesics  cross the wedge, see Sect.\ref{renorm}
Therefore, we can shortly write
   \bea\label{main-g-sum}
G_{l.s.}(\phi_a,t_a;\phi_b,t_b)=
\sum e^{-\Delta {\cal L}_{ren}(\phi_a,t_a;\phi_b,t_b)}.
 \eea
Here the sum is over all geodesics connecting the points $a$ and $b$ with coordinates   $(\phi_a,t_a), (\phi_b,t_b)$ that belong to the living space  of the
$AdS_3$ with a wedge. The presence of $\Theta$-functions is implicitly assumed
and summation only over geodesics with $\Theta=1$ is relevant.
   These geodesic configurations can be different for different  points choice and characteristics of moving particles. 
   
   In   Sect.\ref{Sec::moving} we consider  representation (\ref{main-g-sum})
 for the $AdS_3$ with one moving defect.  To make our presentation more clear we start from
 one static defect, Sect.\ref{Sec::static}.
\subsection{Static defect.}\label{Sec::static}
In this section we formulate the renormalized image method of counting and calculation of  geodesic contributions in the right hand side of formula (\ref{main-g-sum}) for the $AdS_3$ with one static defect. This case has been considered in the previous papers \cite{Balasubramanian:1999zv} for spacelike separated points and in  \cite{AB-TMF,Arefeva:2015sza} for timelike separated points. We start from this case to set the notations and to describe our general method in a simpler case.
\subsubsection{Equal-time points.}\label{Sec:Equal-time points}
Now we formulate the image method for the case of one  static defect  and for  equal-time points.
Our prescription for calculation gives:

\bea\label{two-point-1}
G_{l.s}\left(\phi_a,t_a,\phi_b,t_a\right)&=&G_{\Delta,AdS}\left(\phi_a,t_a,\phi_b,t_a\right)\\&+&
\sum _{n=1}^{n=n_{max}}G_{\Delta,ren,n}\left(\phi_a,t_a,\phi_{b^{\star\,n}},t_a\right) +
\sum _{n=1}^{n=\bar n_{max}}G_{\Delta,ren,n}\left(\phi_a,t_a,\phi_{ b^{\# n}},t_a\right).\nn\eea
Let us make a few comment about this formula. According to this formula to calculate the correlator
we have to take into account:
\begin{itemize}
\item contribution from the basic geodesic connecting points $a$ and $b$;
\item  contributions from the geodesics connecting $a$ and all imaginary points $b^{\star\,n}$ of $b$,
 \be \nn
 b^{*\,n}\equiv b^{\underbrace{*...*}_{n}}
 \ee
that lie in the  right from $a$ half  circle,
 i.e.
 \be\label{right}
|\phi_a-\phi _{b^{*\,n}}|<\pi,\,\,\,\,\,n\leq n_{max}\ee

 \item  contributions from the geodesics connecting $A$ and all imaginary points $b^{\#\,n}$ of $b$,
\be\nn
b^{\#\,n}= b^{\underbrace{\#...\#}_{n}}
\ee
 lying in the left from $a$ half circle
 i.e.
 \be
 \label{left}
 |\phi_a-\phi _{ b^{\#\,n}}|<\pi,\,\,\,\,\,n\leq \bar n_{max}.
\ee

\end{itemize}
More explicitly our prescription has the form:
\bea\label{two-point-1m}
&\,&G_{l.s.}\left(\phi_a,t_a,\phi_b,t_a\right)
=G_{\Delta,AdS}\left(\phi_a,t_a,\phi_b,t_a\right)\\\nn&+&
\sum _{n=1}^{n=n_{max}}G_{\Delta,AdS}\left(\phi_a,t_a,\phi_{b}+n\bar\alpha,t_a\right) +
\sum _{n=1}^{n=\bar n_{max}}G_{\Delta,AdS}\left(\phi_a,t_a,\phi_{b}-n\bar\alpha,t_a\right),
\eea
where $\bar{\alpha}=2\pi-\alpha$, $n_{max}$ and $\bar{n}_{max}$ are given by \eqref{right} and \eqref{left}.
In the case $\alpha < \pi$ we have only one image point and the presence of the contribution of the additional geodesic depends on position of the points $a$ and $b$, see
\cite{AB-TMF}. For the case $\alpha>\pi$ we can get several terms in (\ref{two-point-1m}).

\begin{figure}[htbp]
\begin{center}
\includegraphics[width=5.5cm]{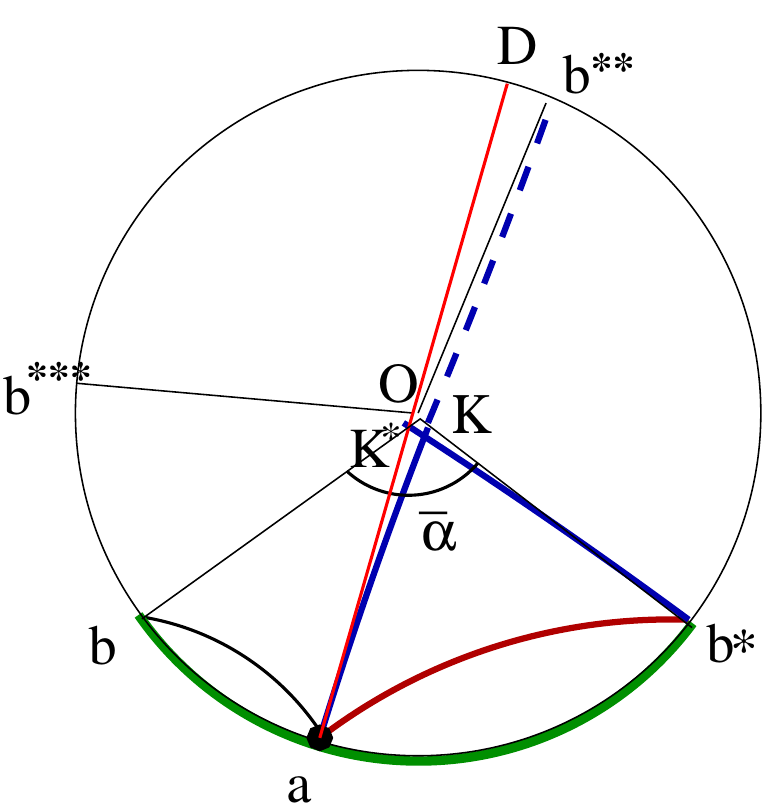}A\,\,\,\,\,\,\,
\includegraphics[width=5.5cm]{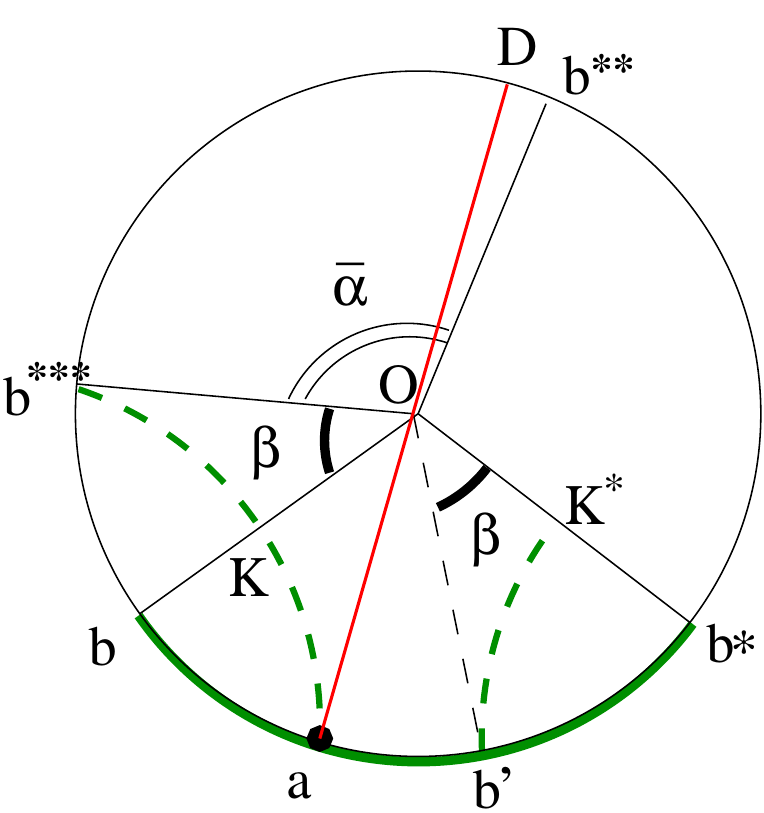}B\\
\caption{
 Plots of geodesics  connecting  the point $a$ with 3 image points $b^{*}$,  $b^{**}$ and  $b^{***}$ (A). The contribution of the $ab^{***}$ corresponds to the connection of the points $a$ and $b'$ (B).}\label{Fig:right3}
\end{center}
\end{figure}

\begin{figure}[htbp]
\begin{center}
\includegraphics[width=5.5cm]{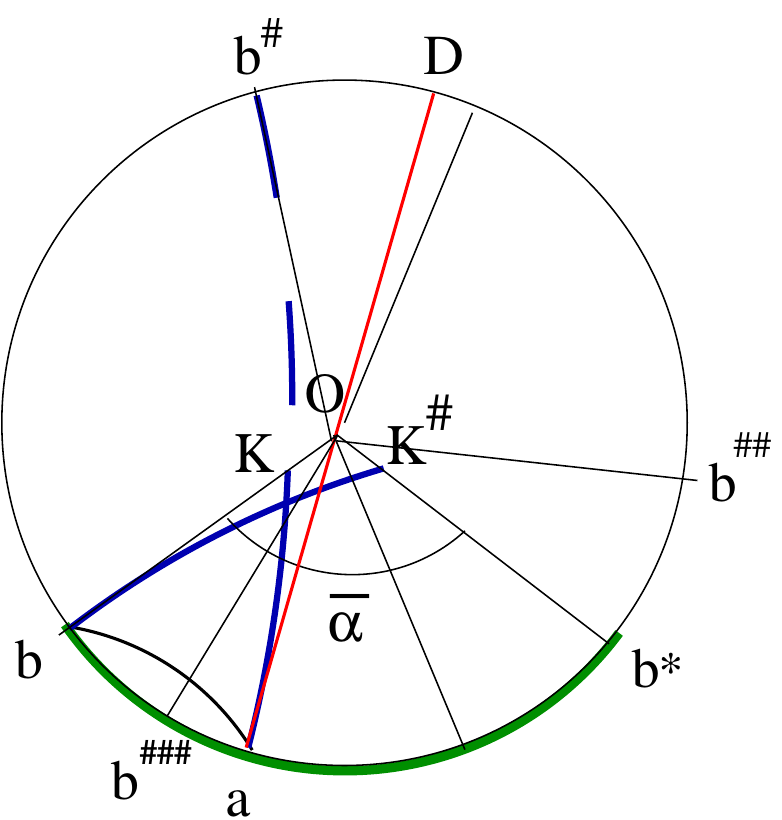}A\,\,\,\,\,\,\,
\includegraphics[width=5.5cm]{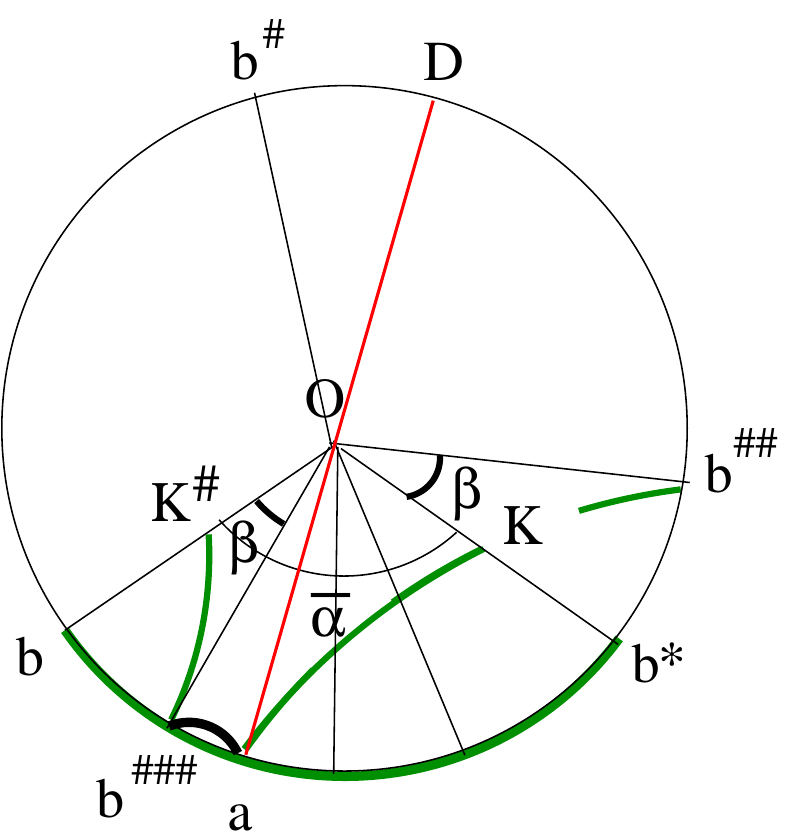}B\\
\caption{
Plots of geodesics  connecting  the point $a$ with 3 image points $b^{\#}$,  $b^{\#\#}$ and  $b^{\#\#\#}$ (A). Imaginary points  $b^{\#\#}$ and $b^{\#\#\#}$ are out of the left  semi circle indicated by the red line (B).}\label{Fig:left}
\end{center}
\end{figure}

In Fig.\ref{Fig:right3}.A the contributions of additional geodesics are shown. The geodesics connect the point $a$ with 3 image points obtained by the counterclockwise rotation of the point $b$ on the angle $\bar\alpha$,  $2\bar\alpha$ and  $3\bar\alpha$, respectively. Only  3 geodesics  $ab$, $ab^*$ and $ab^{**}$ contribute for the given position of the points $a$ and $b$. The geodesic $ab^{***}$ does not contribute since $b^{***}$ is out of the right semi circle indicated by the red line. In the Fig.\ref{Fig:left}.A  contributions of additional geodesics are shown. The geodesics connect the point $a$ and the  imaginary point $b^\#$, obtained by the clockwise rotation on the angle $\bar\alpha$ of the point $b$, can be represented as a sum of two geodesics . The contributions can be represented as the sum of two geodesics.  The first geodesic connects the point $a$ with a point on the wedge, the point $K$,  and the second one connects the point $K^\#$, the image of the point $K$, with the point $b$. One can seen that the geodesics $ab^{\#\#}$  and $ab^{\#\#\#}$ do not contribute since its length corresponds to a connection  of the point $a$ with the point $b'$ that is not the image of $b$.
Fig.\ref{Fig:right3}.B and  Fig.\ref{Fig:left}.B show the role of restrictions (\ref{right}) and
 (\ref{left}). 

\subsubsection{Proof of periodicity}
To proof that (\ref{two-point-1m}) defines the correlator on  the circle, we have to check that
\be\label{period}
G_{l.s.}\left(\phi_a,t_a,\phi_b,t_a\right)=G_{l.s.}\left(\phi_a,t_a,\phi_b+\bar\alpha,t_a\right).\ee

\begin{figure}[htbp]
\begin{center}
\includegraphics[width=5.5cm]{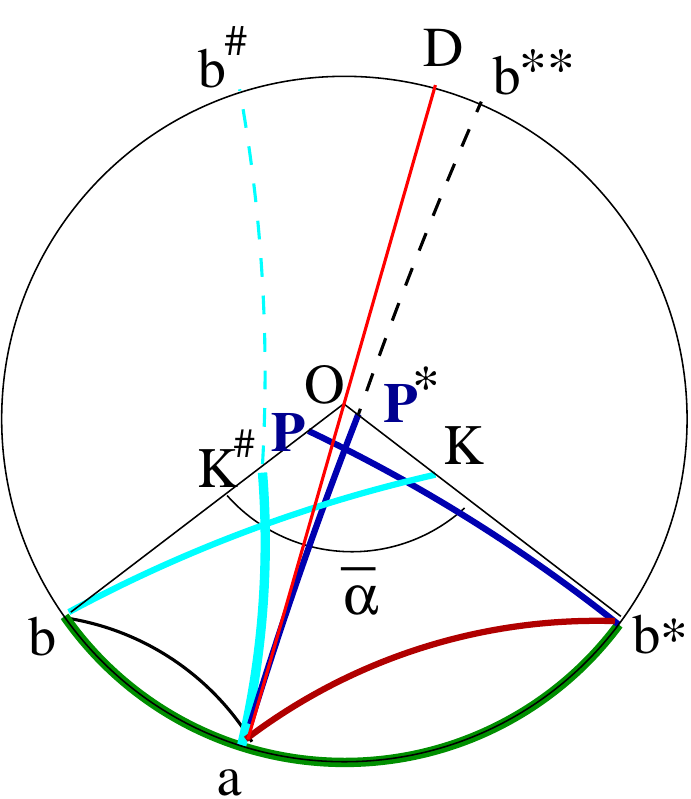}A\,\,\,\,\,\,\,
\,\,\,\,\,\,\,
\,\,\,\,\,\,\,
\includegraphics[width=5.5cm]{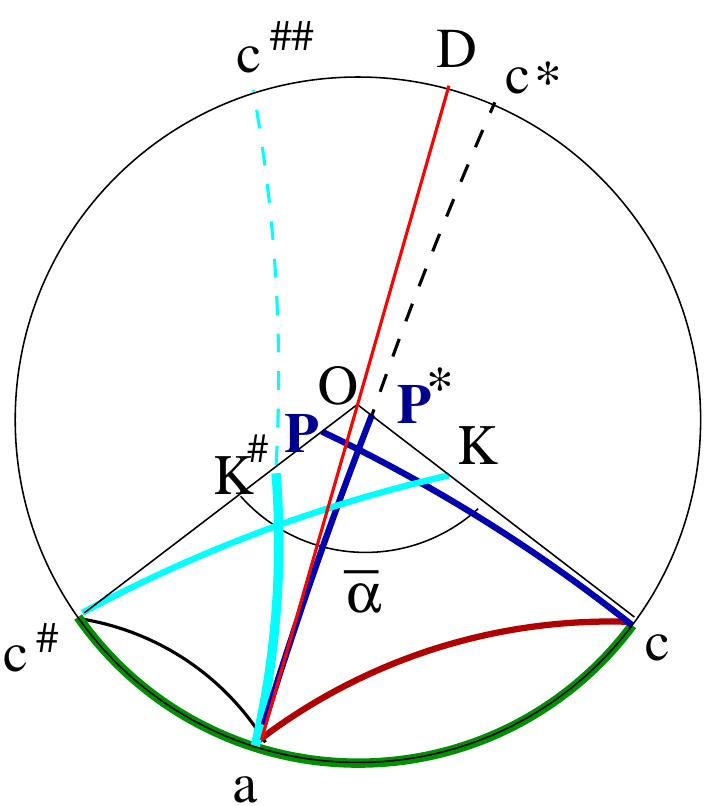}B\\
\caption{
A. Plots of geodesics connecting points $a$ and $b$, $b^{\#}$, $b^{**}$.
B. Plots of geodesics connecting points $a$ and $c$ (where $\phi_c=\phi_b+\bar\alpha)$, $c^{\#}$, $c^{*}$, $c^{\#\#}$.}\label{Fig:period}
\end{center}
\end{figure}

 Let us consider a particular case presented in  Fig.\ref{Fig:period}.A. For this case there are the  following contributions to the LHS of (\ref{period}): the contribution from the geodesic connecting the points  $a$ and $b$ (the basic geodesic), then the  contributions from geodesics connecting the point $a$ with the image points $b^{*}$, $b^{**}$ and $b^{\#}$, i.e. in (\ref{right}), (\ref{left}) $n_{max}=2$, $\bar{n}_{max}=1$.

   To calculate the RHS of
\eqref{period} we note that after the shift $\phi_b\to\phi_b+\bar\alpha$ according to our prescription there are the following contributions.
 There is the contribution from the basic geodesic between points pair $(a,c)$,  here we denote the point with coordinates $(\phi_b+\bar\alpha,t_a)$ as $c$ (see Fig.\ref{Fig:period} B). There are also contributions from  the image geodesics between points pairs $(a,c^{*})$ that is the same as $(a,b^{**})$, $(a,c^{\#})$ that is the same as $(a,b)$ and $(a,c^{\#\#})$ that is the same as $(a,b^{\#})$, i.e. $n_{max}$ and $\bar{n}_{max}$ are changed  so that $n_{max}=1$, $\bar{n}_{max}=2$ (see Fig.\ref{Fig:period} B). Therefore the changes of $n_{max}$ and $\bar{n}_{max}$ after the shift on the period preserves the set of contributing geodesics, that  makes $G_{l.s.}$ periodic.

\subsubsection{Spacelike separated points}
\begin{figure}[h]
   \centering
\begin{picture}(90,170)
\put(-110,0){\includegraphics[width=10cm]{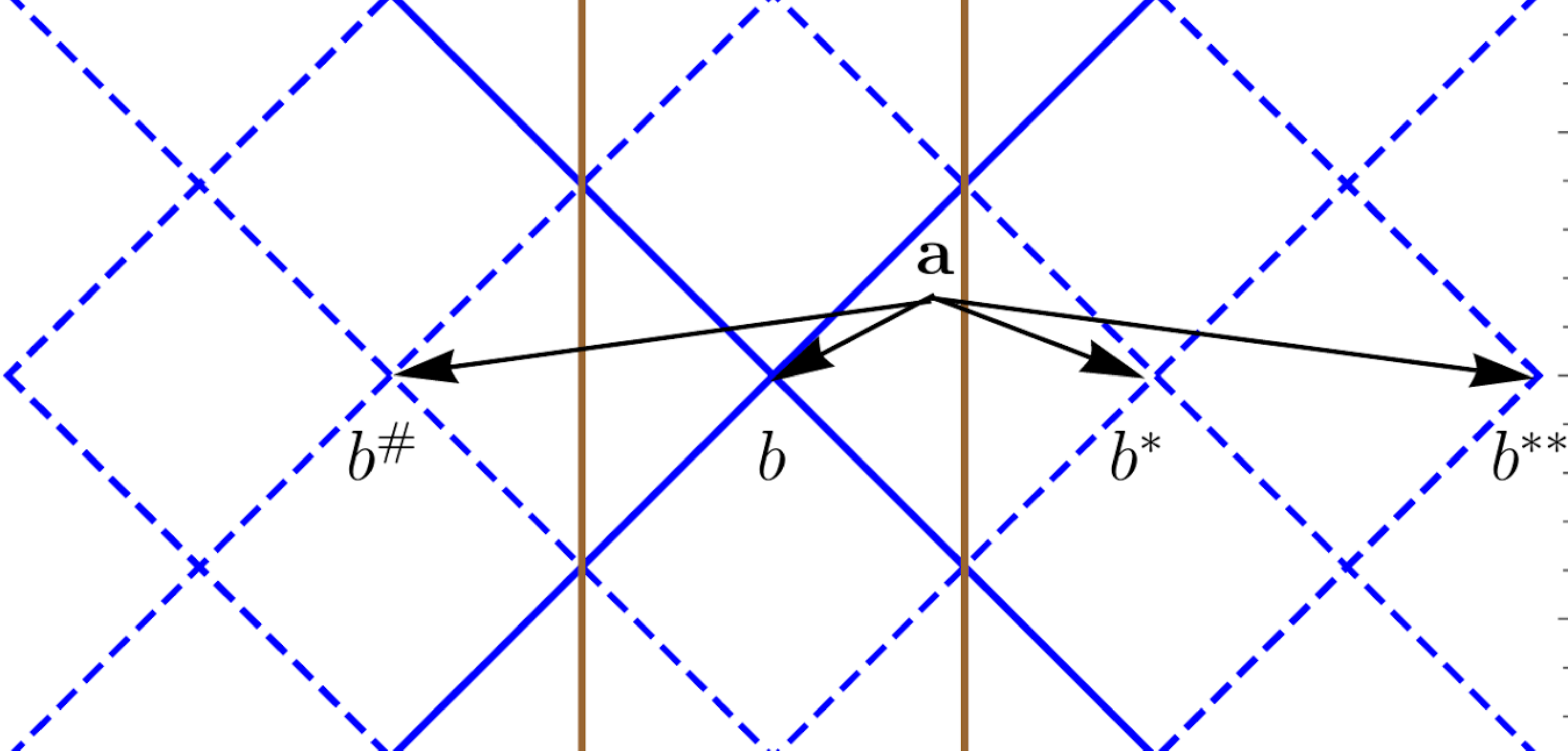}}
\put(-140,75){$(-\pi,0)$}
\put(0,65){$(0,0)$}
\put(172,75){$(\pi,0)$}
\end{picture}\\$\,$\
\caption{The schematic plot of geodesics connecting the spacelike separated points $a$ and $b=(0,0)$  and the point $a$ with  images of the point $b$: $b^*=(0,\bar\alpha)$, $b^{**}=(0,2\bar\alpha)$, $b^\#=(0,-\bar\alpha)$.
 Here $\alpha={3\pi}/{2}$.
} \label{Zones0}\end{figure}
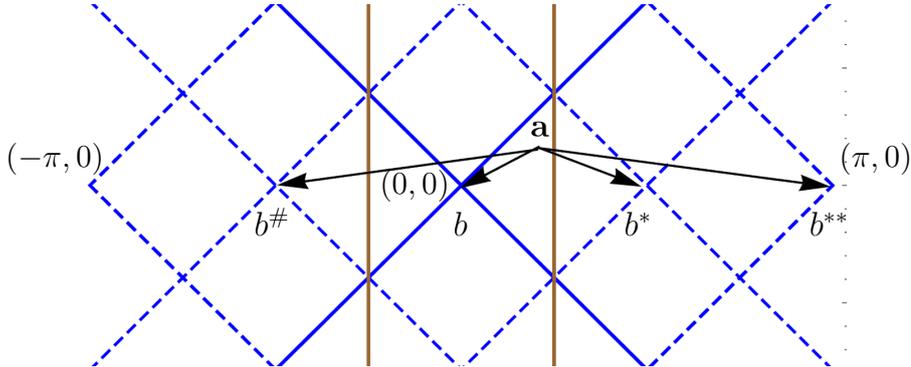

Our rule of construction the correlators for  two spacelike separated points in the presence of the static defect is the same as for equal-time points:

\bea\label{two-point-2}
G_{l.s.}\left(\phi_a,t_a,\phi_b,t_b\right)
&=&G_{\Delta,AdS}\left(\phi_a,t_a,\phi_b,t_b\right)\\&+&
\sum _{n=1}^{n=n_{max}}G_{\Delta,AdS}\left(\phi_a,t_a,\phi_{b}+n\bar\alpha,t_b\right) +
\sum _{n=1}^{n=\bar n_{max}}G_{\Delta,AdS}\left(\phi_a,t_a,\phi_{b}-n\bar\alpha,t_b\right),\nn\eea
where $n_{max}$ and $\bar n_{max}$ are found from restrictions (\ref{right}) and (\ref{left}).
A schematic picture for different contributions  to the right hand side of (\ref{two-point-2}) is presented in Fig.\ref{Zones0}.

\subsubsection{Universal formula and isometry invariance}
If we have the invariance of $G_{ren}$ under the isometry related with the defect
\be
\label{invar}
G_{ren}\left(\phi_a,t_a,\phi_{b}-\bar\alpha,t_b\right)=G_{l.s.}\left(\phi_a+\bar\alpha,t_a,\phi_{b},t_b\right),\ee
then we can rewrite formula (\ref{two-point-2}) in the form
\bea\label{two-point-4}
G_{ren}\left(\phi_a,t_a,\phi_b,t_b\right)
&=&G_{\Delta,AdS}\left(\phi_a,t_a,\phi_b,t_b\right)\Theta_{0}(\phi_a,t_a,\phi_b,t_b)\\&+&
\sum _{n=1}^{n=n_{max}}G_{\Delta,AdS}\left(\phi_a,t_a,\phi_{b}+n\bar \alpha,t_b\right) +
\sum _{n=1}^{n=\bar n_{max}}G_{\Delta,AdS}\left(\phi_a+n\bar\alpha,t_a,\phi_{b},t_b\right).\nn\eea

\subsection{ Moving particle.}\label{Sec::moving}

In this section we consider the two-point correlator of operators on the boundary in the presence of moving defect in the bulk.\footnote{If we consider on massive particle, then we can make Lorentz transformation to switch to a reference frame, where the particle is static. Then the problem that is under consideration in Sec\ref{Sec::moving} reduces to problem from Sec.\ref{Sec::static}. Nevertheless, we consider massive moving in laboratory frame, taking into account future generalization to the multiple particles case \cite{AA}.} Now we have a fixed direction that specifies  the defect movement. We choose the coordinate system according to Fig.\ref{2d}.The isometry is given by formulae (\ref{nb-mov-iso}). 

For moving massive particle the analog of formula \eqref{gen-ignore-2} is

\bea\label{two-point-6}
G_{l.s.}\left(\phi_a,t_a,\phi_b,t_b\right)&=&G_{\Delta,AdS}\left(\phi_a,t_a,\phi_b,t_b\right)\Theta_{0}(\phi_a,t_a;\phi_b,t_b;\alpha,\xi)\\\nn&+&
\sum _{n=1}^{n=n_{max}}G_{\Delta,ren,n}\left(\phi_a^{\#n},t_a^{\#n},\phi_b,t_b\right)\Theta_{cr}(\phi_a^{\#n},t_a^{\#n};\phi_b,t_b;\alpha,\xi) \\
&+&\sum _{n=1}^{n=\bar n_{max}}G_{\Delta,ren,n}\left(\phi_a^{*n},t_a^{*n},\phi_{b},t_b\right)\Theta_{cr}(\phi_a^{*n},t_a^{*n};\phi_{b},t_{b};\alpha,\xi),\nn\eea
where $n_{max}$ and $\bar n_{max}$ are given by \eqref{right} and \eqref{left}, i.e. they are as in the static case. The renormalizations defined $G_{\Delta, ren, n}$ are assumed to respect isometry (\ref{nb-mov-iso}) and we elaborate on this issue in the Sect.\ref{renorm}. In \eqref{two-point-6} we introduce functions $\Theta_{0}(\phi_a,t_a;\phi_{b},t_{b};\alpha,\xi)$ and  $\Theta_{cr}(\phi_a,t_a;\phi_{b},t_{b};\alpha,\xi)$ defined below.

The function $\Theta_{0}$ is defined as:

\begin{itemize}
\item $\Theta_{0}(\phi_a,t_a;\phi_b,t_b;\alpha,\xi)=1$ if geodesic connecting points $(\phi_a,t_a)$ and $(\phi_b,t_b)$ does not cross the wedge;
\item $\Theta_{0}(\phi_a,t_a;\phi_b,t_b;\alpha,\xi)=0$ if geodesic connecting points $(\phi_a,t_a)$ and $(\phi_b,t_b)$ crosses the wedge.
\end{itemize}

$\Theta_{cr}$ ("cr" means "crossing") is defined as following:
\begin{itemize}
\item$ \Theta_{cr}(\phi_a,t_a;\phi_b,t_b;\alpha,\xi)=1$ if geodesic connecting $(\phi_a,t_a)$ and $(\phi_b,t_b)$ crosses the face $w_{-}$ of the wedge;
\item $\Theta_{cr}(\phi_a,t_a;\phi_b,t_b;\alpha,\xi)=0$ if geodesic connecting $(\phi_a,t_a)$ and $(\phi_b,t_b)$ does not cross the face $w_{-}$ of the wedge (see Fig.\ref{2d}). 
\end{itemize}

\begin{figure}[htbp]
\begin{center}
\includegraphics[width=4cm]{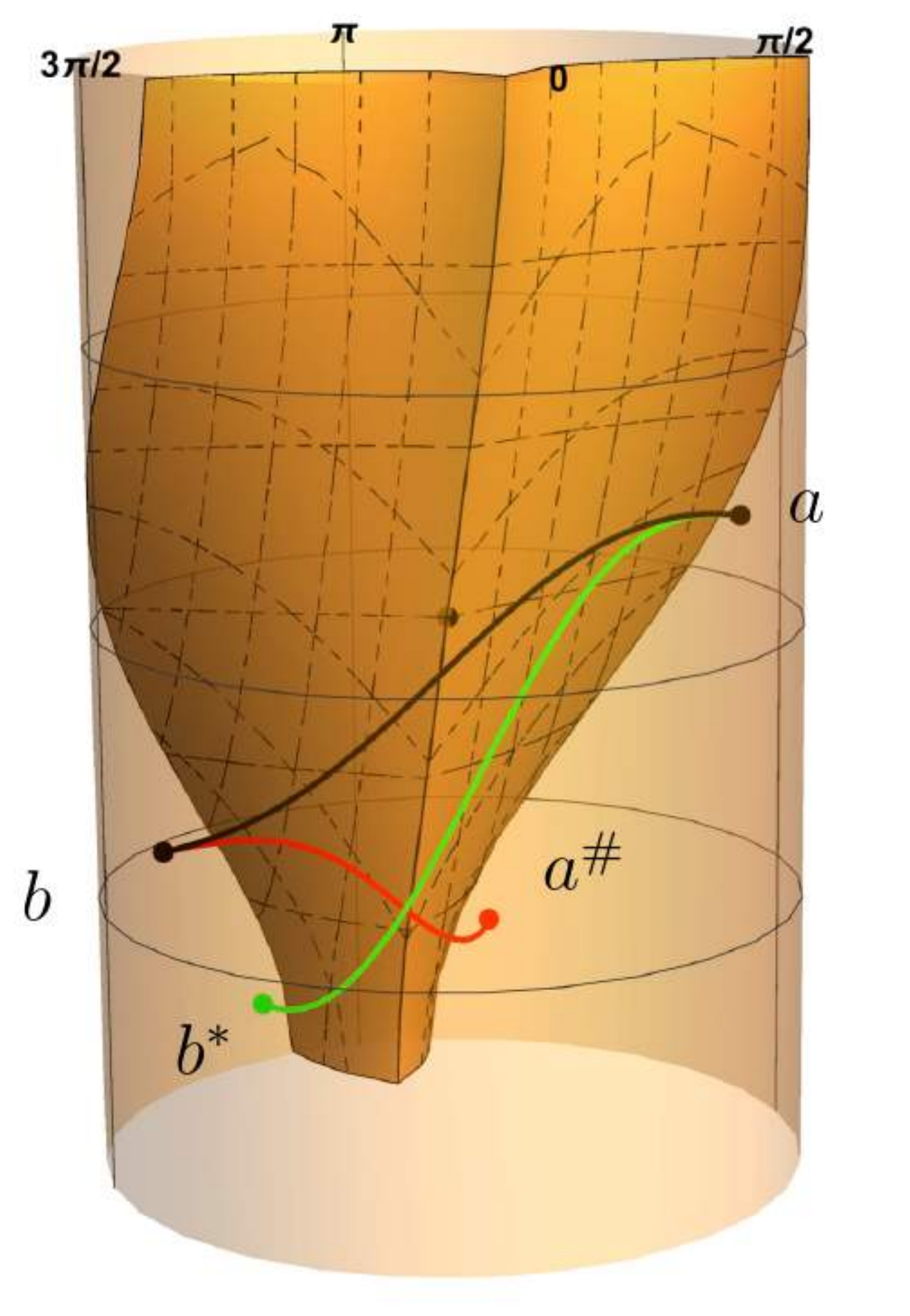}A\,\,\,\,\,
\includegraphics[width=4cm]{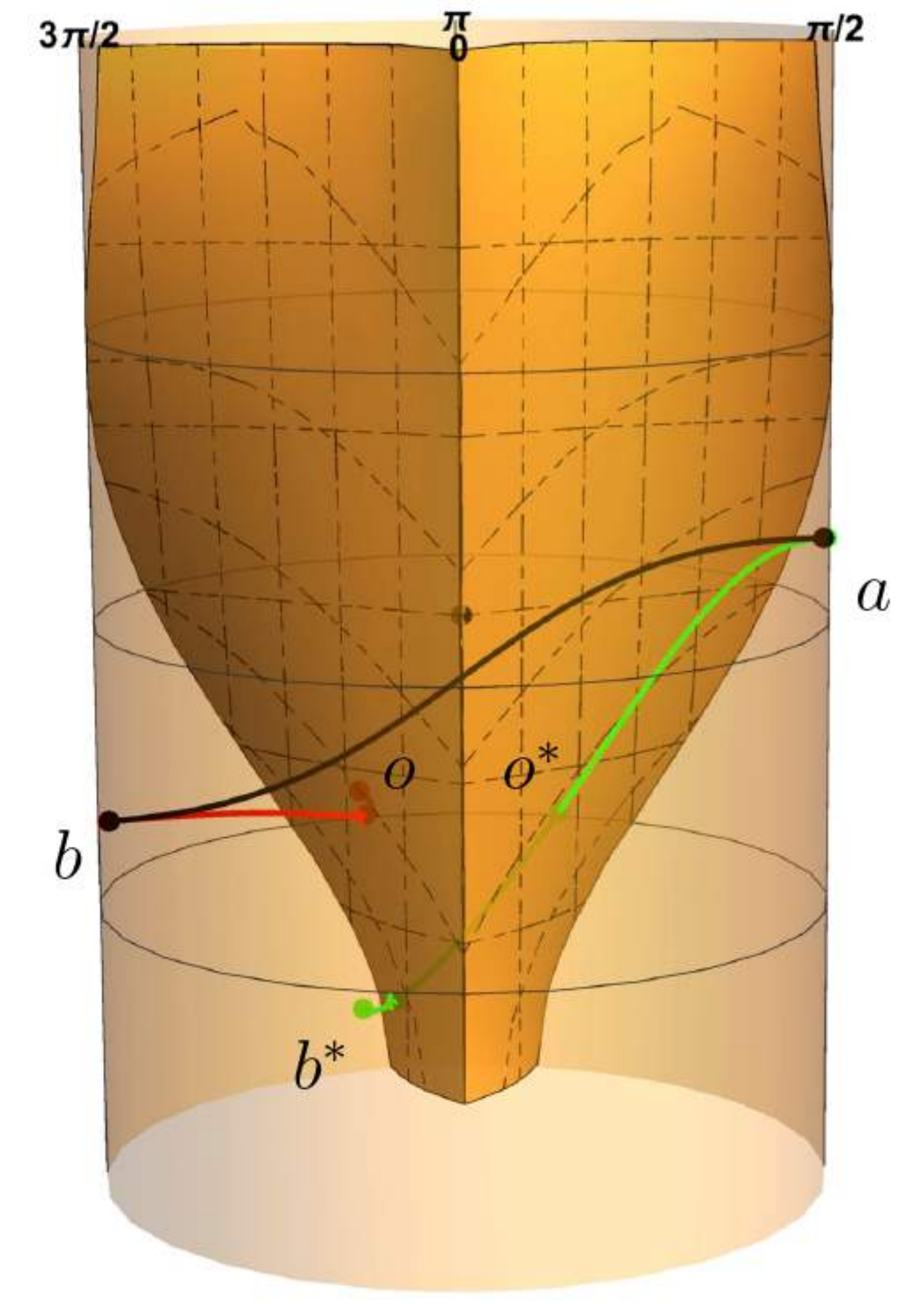}B\,\,\,\,\,
\includegraphics[width=4cm]{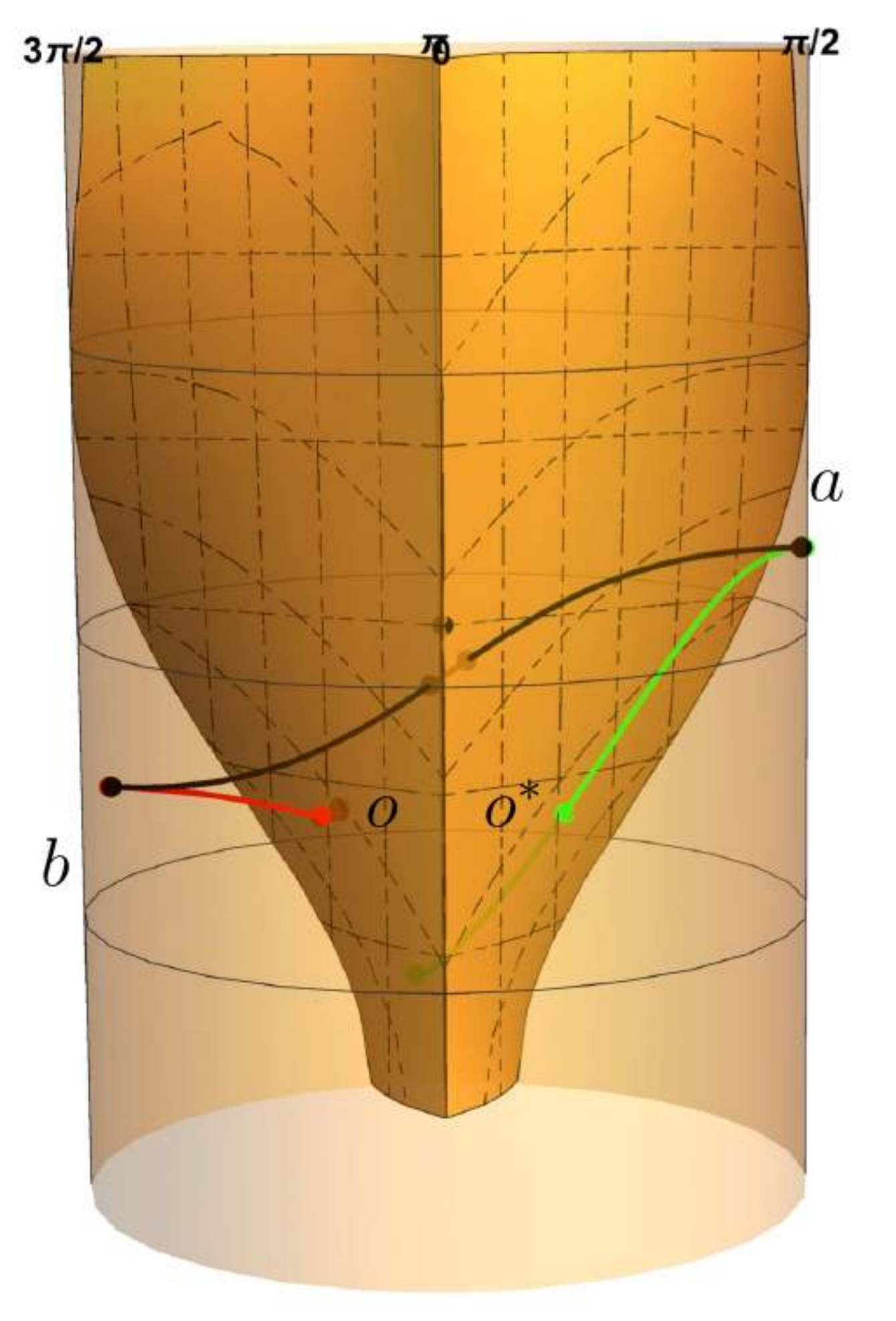}C
 \caption{The plot of the wedge and geodesic configurations for the  light particle. The black curves are the basic geodesics between points $a$ and $b$. The red and green curves are the image geodesics between points $(a^{\#}, b)$ and $(a,b^*)$ respectively. A. Boundary points are taken to be $\phi_a=5.1$, $t_a=-0.5$, $\phi_b=0.6$, $t_b=0.4$, parameter values are $\alpha = \pi/2$ and $\xi=1.3$. B. Boundary points are taken to be  $\phi_a=5$, $t_a=-0.5$, $\phi_b=1.6$, $t_b=0.2$, parameter values are $\alpha = \pi/2$ and $\xi=1.3$. C. Boundary points are taken to be  $\phi_a=4.4$, $t_a=-0.5$, $\phi_b=1.6$, $t_b=0.2$, parameter values are $\alpha = \pi/2$ and $\xi=1.3$\label{geodconf}}
 \end{center}
\end{figure}

\subsubsection{Light moving massive particle}\label{Sec:light}
For the case  $\alpha<\pi$ there are only two terms in (\ref{two-point-6}),
 one contribution comes from the "basic" geodesic, another two  come from its images:
  \bea\label{cor*}
\nn G_{\alpha,\xi}(\phi_a,t_a,\phi_b,t_b) &=&G_{\Delta,AdS}\left(\phi_a,t_a,\phi_b,t_b\right)\Theta_{0}(\phi_a,t_a;\phi_b,t_b;\alpha,\xi)\\ &+&
G_{\Delta,ren,1}{(\phi_{a^{*}},t_{a^{*}},\phi_b,t_b)}
\,\Theta_{cr}(\phi_{a^{*}},t_{a^{*}};\phi_{b},t_{b};\alpha,\xi)\nn\\
&+&
G_{\Delta,ren,1}{(\phi_a,t_a,\phi_{b^{*}},t_{b^*})}
\,\Theta_{cr}(\phi_a,t_a;\phi_{b^{*}},t_{b^{*}};\alpha,\xi).
 \eea

Finding support of functions $\Theta_{0}$ and $\Theta_{cr}$ numerically we get different possibilities for the geodesic structure. We call the basic geodesic the geodesic that connects two points $a$ and $b$ on the boundary without crossing the wedge. We call winding or image geodesic the one that starting from the boundary meets the wedge at a point, comes out from it at the image point and reaches the boundary.  !!!!!! In particular case when the point $a$ is  on one of the faces of the wedge (for example on the bottom side) then the imaginary point $a^*$ is on the upper side. The winding geodesic is $a^*b$. 
  
There are  three different combinations of image and basic geodesics contributing in the correlator on the boundary of the $AdS_3$ space deformed by the massive light particle:
\begin{itemize}
\item The basic geodesic contributes and the winding one does not.
\item The basic geodesic does not contribute, and the winding one does.
\item Both types of geodesics contribute in the correlator.
\end{itemize}
In Fig.\ref{geodconf}  we plot the different cases of geodesic configurations contributing to the propagator for certain values of $\alpha$ and $\xi$.
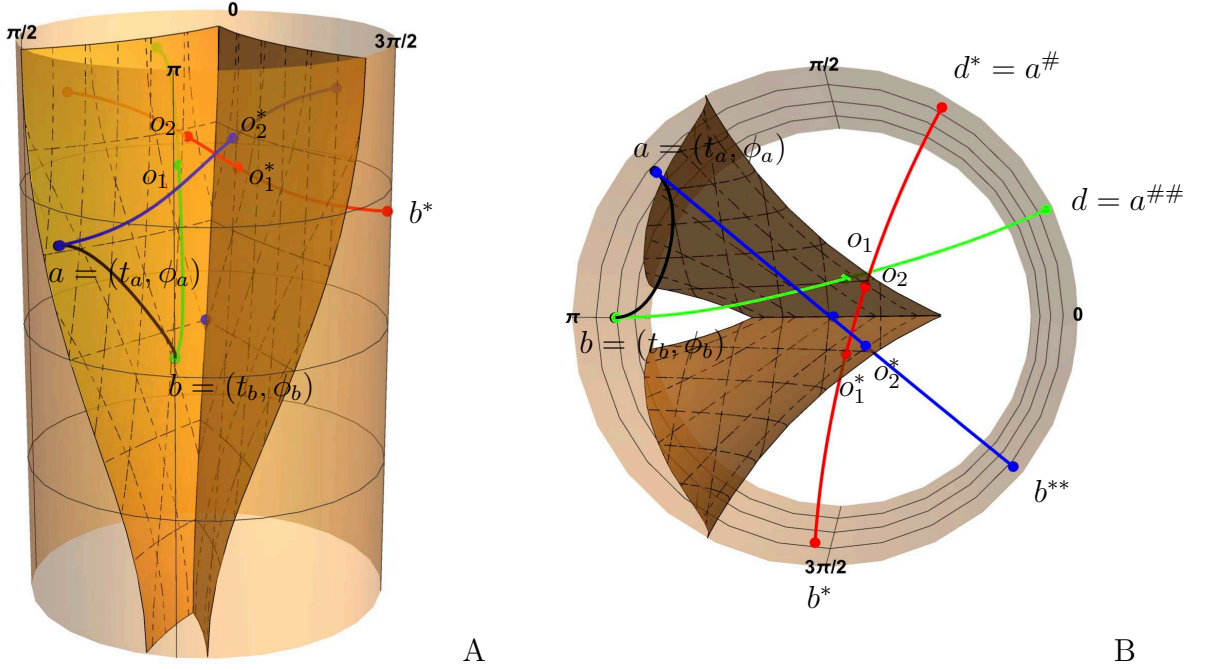
\begin{figure}
\centering
  \begin{picture}(200,100)
\put(0,40){\includegraphics[width=6cm]{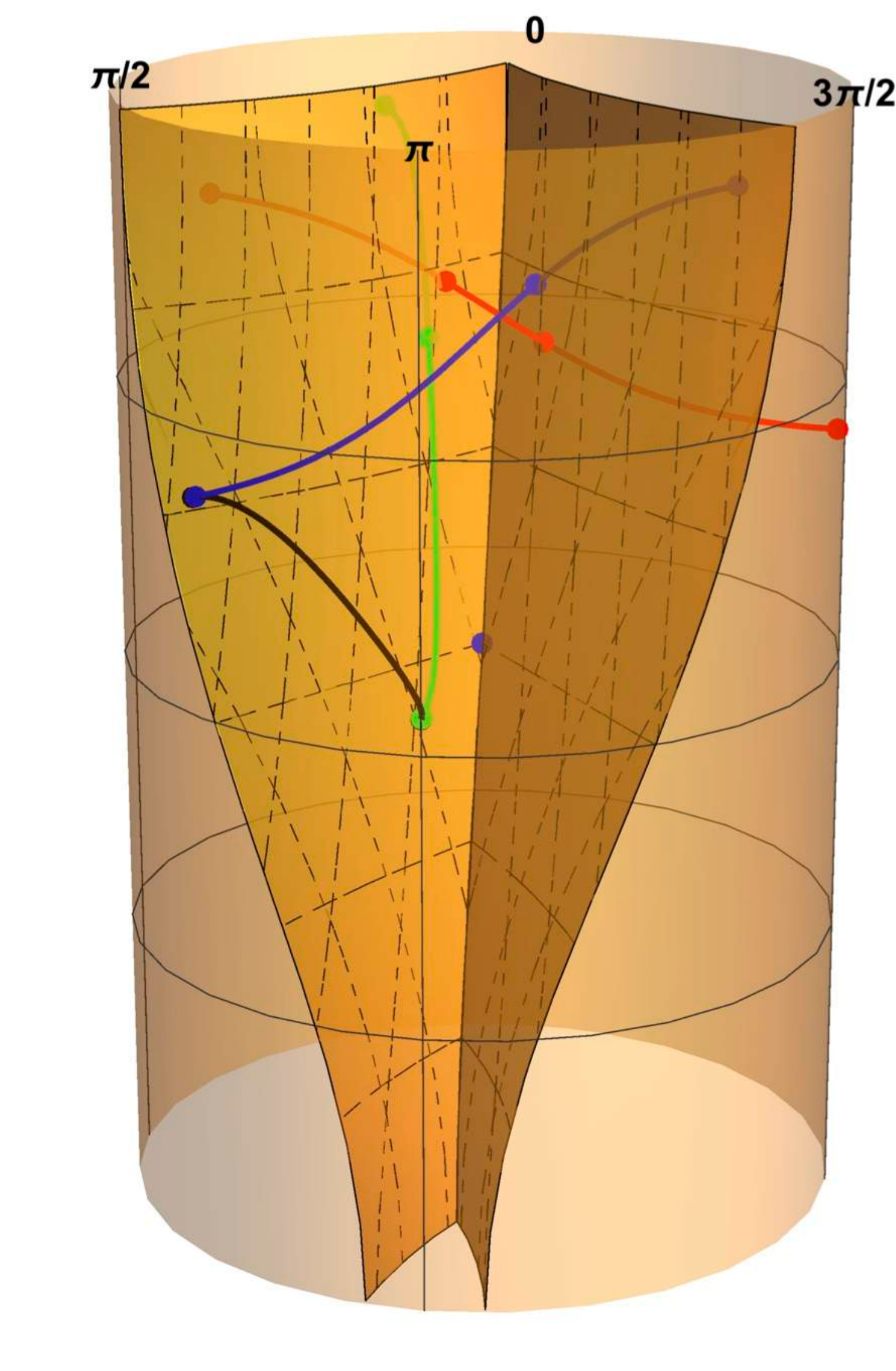}}$\,\,\,$\,\,\,\,\,\,
\put(15,193){$a=(t_a,\phi _a)$}
\put(60,150){$b=(t_b,\phi _b)$}
\put(150,215){$b^*$}
\put(90,230){$o_1^{*}$}
\put(53,250){$o_2$}
\put(50,230){$o_1$}
\put(87,250){$o_2^{*}$}
\put(170,50){A}
\end{picture}
\centering
 \begin{picture}(220,300)
\put(20,80){\includegraphics[width=7cm]{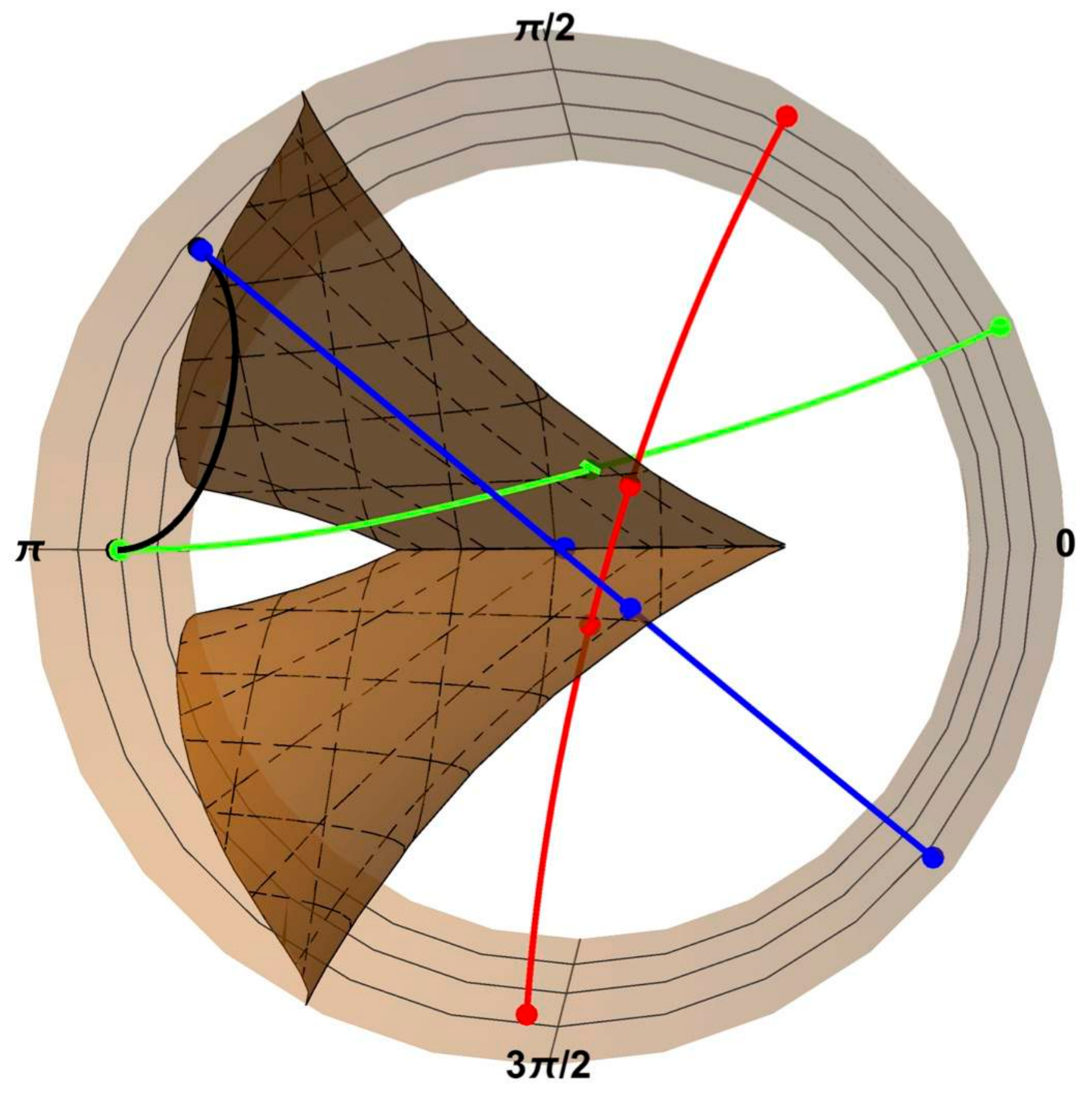}}$\,\,\,$\,\,\,\,\,\,
\put(30,240){$a=(t_a,\phi _a)$}
\put(10,166){$b=(t_b,\phi _b)$}
\put(194,220){$d=a^{\#\#}$}
\put(95,70){$b^*$}
\put(150,270){$d^*=a^{\#}$}
\put(180,110){$b^{**}$}
\put(123,193){$o_2$}
\put(110,206){$o_1$}
\put(120,156){$o_2^{*}$}
\put(107,150){$o_1^{*}$}
\put(210,50){B}
\end{picture}
\caption{Double winding geodesic configuration connecting points $a$ and $b$ for certain boost $\xi$ and mass parameter $\alpha$. The parameter $\alpha=5.4$. Black curve is basic geodesic and wound geodesic consists of three parts. The length of this geodesic is calculated as $l(a,b)=l(a,o_2^{*})+l(o_2,o_1^*)+l(o_1,b)$.}
\label{Fig:tw1}
\end{figure}

\subsubsection{Heavy moving massive particle}\label{Sec:heavy}

If $\alpha>\pi$ the situation differs from the "light" case again. Here we get additional geodesic configurations contributing to the two-point function. The basic geodesic contribution is always present. For simplicity we consider here $\alpha={3\pi}/{2}$. Writing down \eqref{two-point-6} explicitly we get the expression for the correlator
 \bea\label{corH*}\nn
 \mathfrak{G}_{\alpha,\xi}(\phi_a,t_a,\phi_b,t_b)&=&G_{\Delta,AdS}{(\phi_a,t_a,\phi_b,t_b)}\,\Theta_{0}(\phi_a,t_a;\phi_b,t_b;\alpha,\xi)+\\ \nn&+&
G_{\Delta,ren,1}{(\phi_a,t_a,\phi_b^{*},t_b^*)}
\,\Theta_{cr}(t_{a},\phi_{a};t_b^{*},\phi_b^{*};\alpha,\xi)\\\nn&+&
G_{\Delta,ren,1}{(\phi_a,t_a,\phi_b^\#,t_b^\#)}
\,\Theta_{cr}(t_{a},\phi_{a};t_b^\#,\phi_b^\#;\alpha,\xi)\\\nn
&+&G_{\Delta,ren,2}{(\phi_a,t_a,\phi_b^{*2},t_b^{*2})}
\,\Theta_{cr}(\phi_a,t_a,\phi_b^{*2},t_b^{*2};\alpha,\xi)\\\nn
&+&G_{\Delta,ren,2}{(\phi_a,t_a,\phi_b^{\#2},t_b^{\#2})}
\,\Theta_{cr}(\phi_a,t_a,\phi_b^{\#2},t_b^{\#2};\alpha,\xi)
 \eea

 The first term in  \eqref{corH*} corresponds to the "basic geodesic", the second and third terms to the geodesic winding once, as in light particle case and the last two terms correspond to double winding geodesics. The  last term contributes to the two-point function as can be seen in Fig.\ref{Fig:tw1}.

 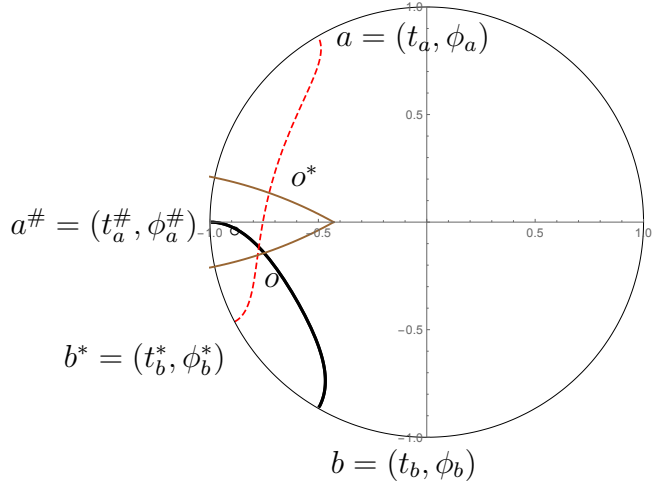
\begin{figure}[h]\label{2diso}
   \centering
\begin{picture}(160,160)
\put(0,0){\includegraphics[width=6cm]{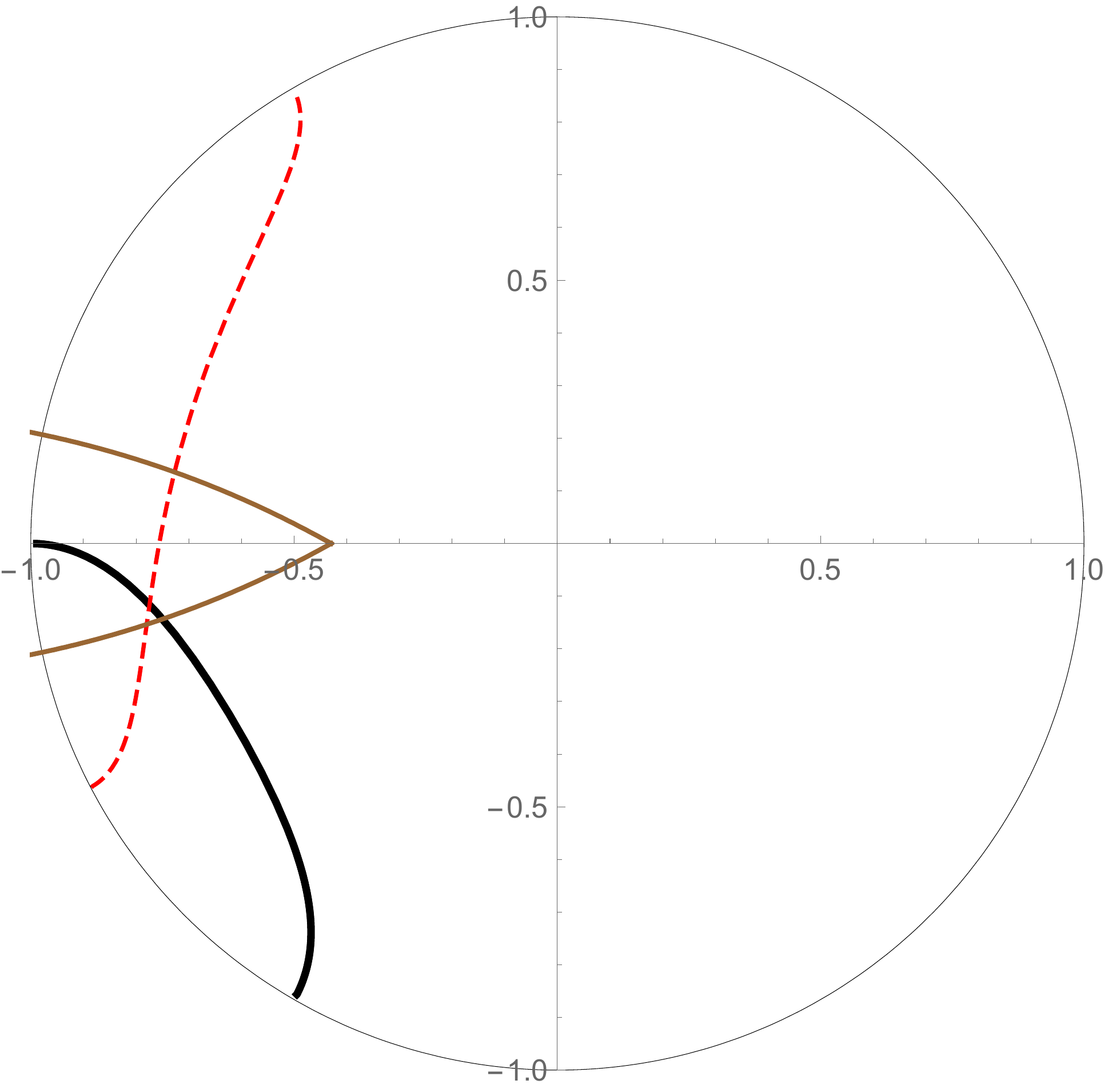}}
\put(52,151){$a=(t_a,\phi _a)$}
\put(50,-10){$b=(t_b,\phi _b)$}
\put(-70,80){$a^\#=(t^\#_a,\phi ^\#_a)$}
\put(-50,30){$b^*=(t^*_b,\phi ^*_b)$}
\put(14,81){\circle{3}}
\put(25,60){$o$}
\put(35,98){$o^*$}
\end{picture}\\
\caption{Geodesic connecting two boundary points following the points $(b,o,o^*,a)$. Black curve is original geodesic and red dashed curve is geodesic between isometry points.} \label{cross}
\end{figure}
\subsection{Renormalization}\label{renorm}
\subsubsection{Spacelike geodesics}

Now we consider the problem of finding the renormalized length of the image geodesic between two points on the boundary.
Consider the geodesic between two near the boundary points $(\phi_a,t_a)$,$(\phi_b,t_b)$ that passes through the wedge. It can be represented as a geodesic consisting of two parts (see Fig.\ref{cross}) whose lengths are  $l_{a,o^*}$ and $l_{o,b}$. Here point $o^*$ is the image of the point $o$ under the isometry (\ref{isom}).

The isometry should respects the geodesic length between two points in the bulk i.e. $l_{o,b}=l_{o^*,b^*}$ therefore the length between points $a$ and $b$ must satisfy:
\bea\label{eq}
{\mathcal L}_{reg}(a,b)={\mathcal L}_{reg}(a,b^*)={\mathcal L}_{reg}(a^{\#},b).
\eea
Here "reg" means the regularized length. The regularization means, as has been explained above, that we consider points $a$, $b$, $a^{\#}$ and $b^{*}$ near the boundary. Expressions for renormalized lengths between points $(a,b^*)$ and $(a^{\#},b)$  are:
\bea\label{wrong-ren}
{\mathcal L}_{ren}(a,b^*)&=&\ln 2[\cos(t_a-t^*_b)-\cos(\phi_a-\phi^*_b)],\nn\\
{\mathcal L}_{ren}(a^{\#},b)&=& \ln 2[\cos(t^\#_a-t_b)-\cos(\phi^\#_a-\phi_b)].
\eea

It is obvious that these lengths are not equal. Let us do the calculations from the beginning taking into account the divergent part dependence on $\chi$ accurately. We define
\bea\label{abs}
{\mathcal L}_{ren}(a,b^*)={\mathcal L}_{reg}(a,b^*)-(\chi_a+\chi_{b^*}),\\\nn
{\mathcal L}_{ren}(a^{\#},b)={\mathcal L}_{reg}(a^{\#},b)-(\chi_{a^{\#}}+\chi_b).
\eea
The renormalized geodesic length between points $a$ and $b$ also is:
\bea
{\mathcal L}_{ren}(a,b)&=&{\mathcal L}_{reg}(a,b)-(\chi_a+\chi_b).\nn
\eea
From (\ref{eq}) we can write:
\bea\label{subs}
{\mathcal L}_{ren}(a,b)={\mathcal L}_{reg}(a^{\#},b)-(\chi_a+\chi_b)={\mathcal L}_{reg}(a,b^*)-(\chi_a+\chi_b).\nn
\eea
Substituting (\ref{wrong-ren}) in (\ref{abs}) we have:
\bea\label{a.b}
{\mathcal L}_{ren}(a^{\#},b)&=&{\mathcal L}_{ren}(a,b)-(\chi_{a^{\#}}+\chi_b)+\chi_a+\chi_b={\mathcal L}_{ren}(a,b)+\chi_a-\chi_{a^{\#}},\\\nn
{\mathcal L}_{ren}(a,b^*)&=&{\mathcal L}_{ren}(a,b)-(\chi_a+\chi_{b^*})+\chi_a+\chi_b={\mathcal L}_{ren}(a,b)+\chi_b-\chi_{b^*},
\eea
and we obtain:
\bea
{\mathcal L}_{ren}(a^{\#},b)={\mathcal L}_{ren}(a,b^*)+\chi_a-\chi_{a^{\#}}-\chi_b+\chi_{b^*}.\nn
\eea
According to \eqref{expchi} for the large $\chi$ we have:
    \bea\nn
\chi_{b^*}&=&\chi_b +\frac{1}{2}\ln C_{b^*}\label{chibs},\,\,
C_{b^*}=\left(\mathcal{B}_{\xi}(\alpha) \cos \phi_b+\sin t_b (1+2\sinh^2\xi\sin^2\frac{\alpha}{2})\right)^2+ \cos ^2t_b,\nn\\
  \chi_{a^\#}&=&\chi_a +\frac{1}{2}\ln C_{a^\#},\nn\,\,
C_{a^\#}= \left(\mathcal{B}_{\xi}(-\alpha)\cos \phi_a+\sin t_a (1+2\sinh^2\xi\sin^2\frac{\alpha}{2})\right)^2+ \cos ^2t_a,
 \nn\eea
 where $\mathcal{B}_{\xi}(\alpha)$ is given by \eqref{B-F}.
Finally, using (\ref{a.b})  we obtain the renormalized image geodesic length in the case of the removed regularization (for points $a$ and $b$ on the boundary):
\bea\nn
{\mathcal L}_{ren}(a,b)&=& \ln [2\left(\cos(t^\#_a-t_b)-\cos(\phi^\#_a-\phi_b)\right)C_{a^\#}^{1/2}]\\
&=&\ln [2\left(\cos(t_a-t^*_b)-\cos(\phi_a-\phi^*_b)\right)C_{b^*}^{1/2}].
\eea

Let us consider the case when the geodesic passes few times through the faces of the wedge and then reach the boundary point forming multiple winding geodesics configuration. This is the case when massive particle deforming the $AdS$ is heavy enough, i.e. $\alpha>\pi$.
 In Fig.\ref{Fig:tw1} we plot this situation. The geodesics of our interest consists of the pieces $(b,o_1)$, $(o_1^*,o_2)$ and $(o_2^*,a)$.
From the previous section we see, that each image of the point to be renormalized (let's assume point $a$) adds the factor $C_{a^{\#}}^{-1/2}$ for the $\#$ image and $C_{a^*}^{-1/2}$ for $*$.
Thus for the two point function for the geodesic connecting $a$ and $b$ as it is show in Fig.\ref{Fig:tw1}.B. we get the representation for factors $Z_n$ and $\bar{Z}_n$ for multiple imaging geodesics:

\bea\label{two-point-three-ren}
Z_n( t^{\#}_{a,n},\phi^{\#}_{a,n};t_b,\phi_b)&=&C_{a^{\#n}}^{-1/2}=C_{a^{\#(n-1)}}^{-1/2}C_{b^{*}}^{-1/2}=...=C_{b^{*n}}^{-1/2}\\\nn
\bar{Z}_n( t^{*}_{a,n},\phi^{*}_{a,n};t_b,\phi_b)&=&C_{a^{*}}^{-1/2}=C_{a^{*(n-1)}}^{-1/2}C_{b^{\#}}^{-1/2}=...=C_{b^{\#n}}^{-1/2}.
\eea

\subsubsection{Quasigeodesics}
\begin{figure}[!h]
\begin{center}
\includegraphics[width=7cm]{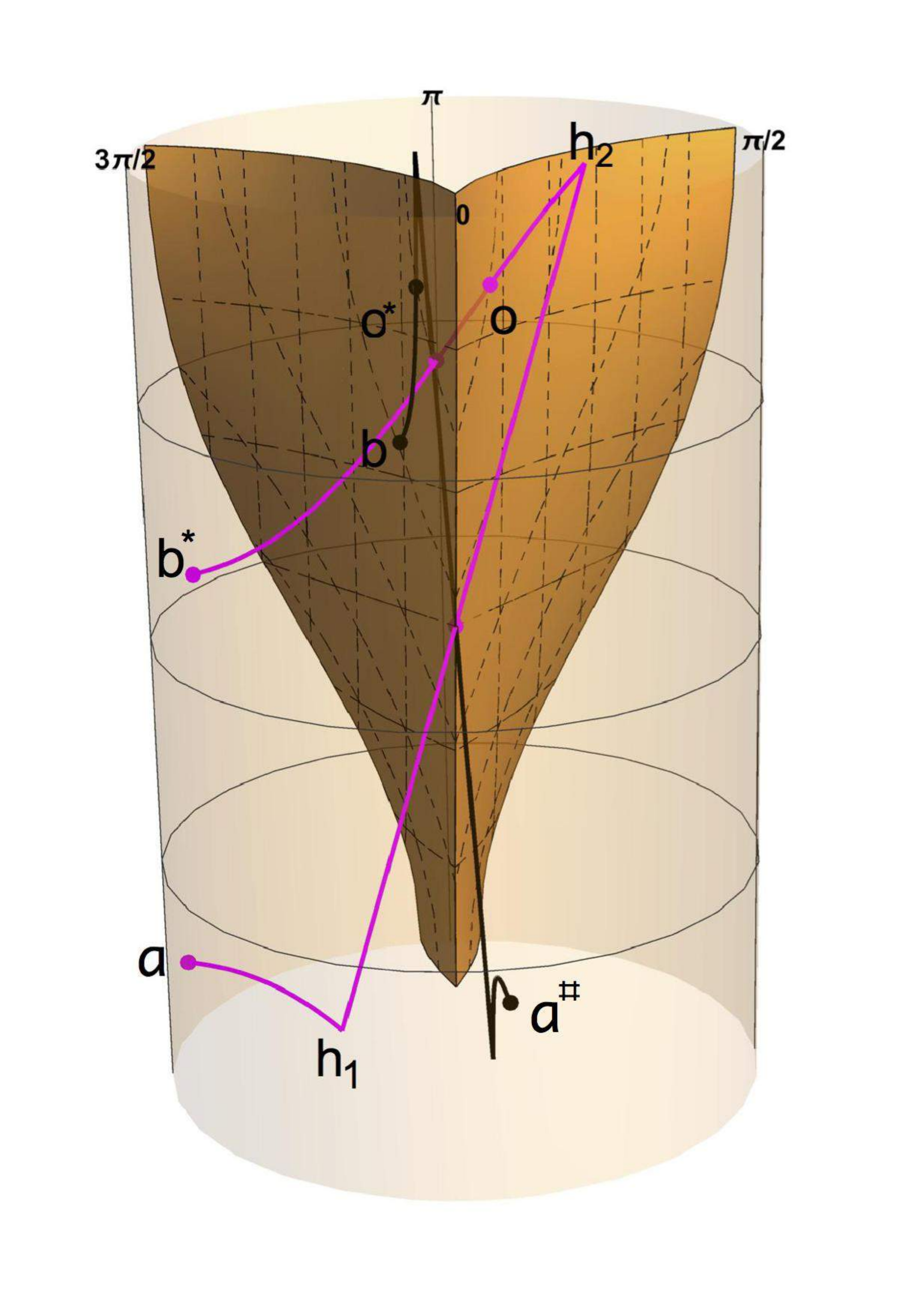}
\caption{ The plot of the quasigeodesic connecting $a$ and $b$ consists from the black curve  between points $a^{\#}$ and $b$ and the magenta curve  between points $a$ and $b^*$. Here we take parameter values: $\xi=1$ and $\alpha=1$. }\label{renorm-quasi}
\end{center}
\end{figure}

In this subsection we will find the renormalized length between timelike separated points ($a$ and $b$). The length between points in the bulk can also be found, as in the previous subsection, using (\ref{Lads}). In accordance to Fig.\ref{renorm-quasi} the length consists of three parts and can be written as:

\be\nn
{\mathcal L}(a,b) = {\mathcal L} (a,h_1) + {\mathcal L}(h_2,o^*) + {\mathcal L} (o,b).
\ee
Taking in to account that  the length is invariant under the isometry and also that points $o$ and $o^*$ are identical we get:
\be\nn
{\mathcal L}(o,b)={\mathcal L}(o^*,b^*),\,\,\,\, {\mathcal L}(a,b) = {\mathcal L}(a,h_1) + {\mathcal L}(h_2,b^*).
\ee

According to (\ref{quasi-half}) we can rewrite the lengths with divergent parts:
\bea\nn
{\mathcal L}(a,h_1) = {\mathcal L}_{ren}(a,h_1) + \chi_a,\,\,\,\,
{\mathcal L}(b^*,h_2) = {\mathcal L}_{ren}(b^*,h_2) + \chi_{b^*}.
\eea
Taking into account expression (\ref{quasiren}) for renormalized length we can write:
\be\label{length1}
{\mathcal L}(a,b)={\mathcal L}(a,b^*) = {\mathcal L}(a,h_1) + {\mathcal L}(b^*, h_2) = {\mathcal L}_{ren}(a,b^*) + \chi_a + \chi_{b^{*}}-2\ln2.
\ee
On the other hand the formula for length between points $a$ and $b$ has a form:
\be\label{length2}
{\mathcal L}(a,b)={\mathcal L}_{ren}(a,b) + \chi_a + \chi_b-2\ln2.
\ee
From (\ref{length1}) and (\ref{length2}) we get the renormalized length for timelike separated points:
\bea\nn
{\mathcal L}_{ren}(a,b) &=&  {\mathcal L}_{ren}(a,b^*) - (\chi_{b} - \chi_{b^{*}}) = {\mathcal L}_{ren}(a,b^*) + \frac{1}{2}\ln C_{b^*}\\\nn
&=& \ln [2|\cos(t_a-t_{b^*})-\cos(\phi_a-\phi_{b^*})|C_{b^*}^{1/2}].
\eea
Also by the same way the formula for the renormalized geodesic length can be calculated using $a^{\#}$ and $b$ points:
\bea\nn
{\mathcal L}_{ren}(a,b) &=& {\mathcal L}_{ren}(a^{\#},b) - (\chi_{a} - \chi_{a^{\#}}) = {\mathcal L}_{ren}(a^{\#},b) + \frac{1}{2}\ln C_{a^{\#}}\\\nn
&=&  \ln [2|\cos(t_{a^{\#}}-t_b)-\cos(\phi_{a^{\#}}-\phi_b)|C_{a^\#}^{1/2}].
\eea
Therefore the formula for renormalization for quasigeodesics is \eqref{two-point-three-ren} again.
\section{Zone structure of correlators}
\subsection{Light particle}
\begin{figure}[h]
    \centering
        \includegraphics[width=4.5cm]{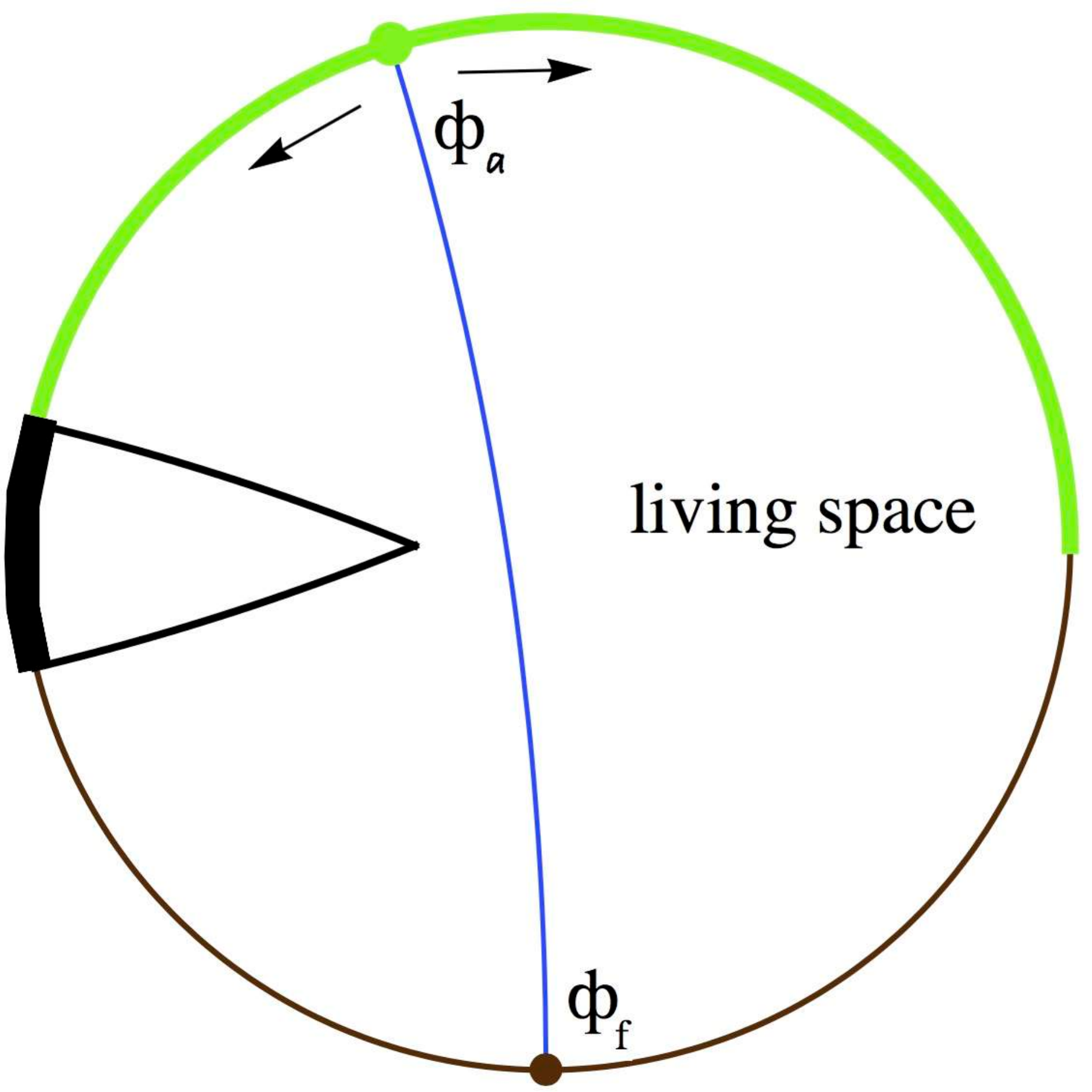}
   \caption{The schematic plot  of  locations of the moving massive light particle and two points that correlators depends on. The circle is the boundary of the constant time section of the $AdS_3$. The point $\phi_f$ is fixed on the brown part of the circle,  $\phi_a$
   varies in the green part of the circle. The removed arc is shown by the thick curve. The living space is indicated by the green and brown curves.}\label{Fig:intro1}
    \end{figure}
Let us consider the light moving particle and  the case when points $a$ and $b$ are taken on the opposite sides of the boundary of $AdS_3$ (the opposite side means the opposite with respect to the massive particle worldline, see Fig.\ref{Fig:intro1} for the schematic plot).
  \begin{figure}[h!]
    \centering
    \centering
\begin{picture}(90,140)
\put(-160,0){\includegraphics[width=8cm]{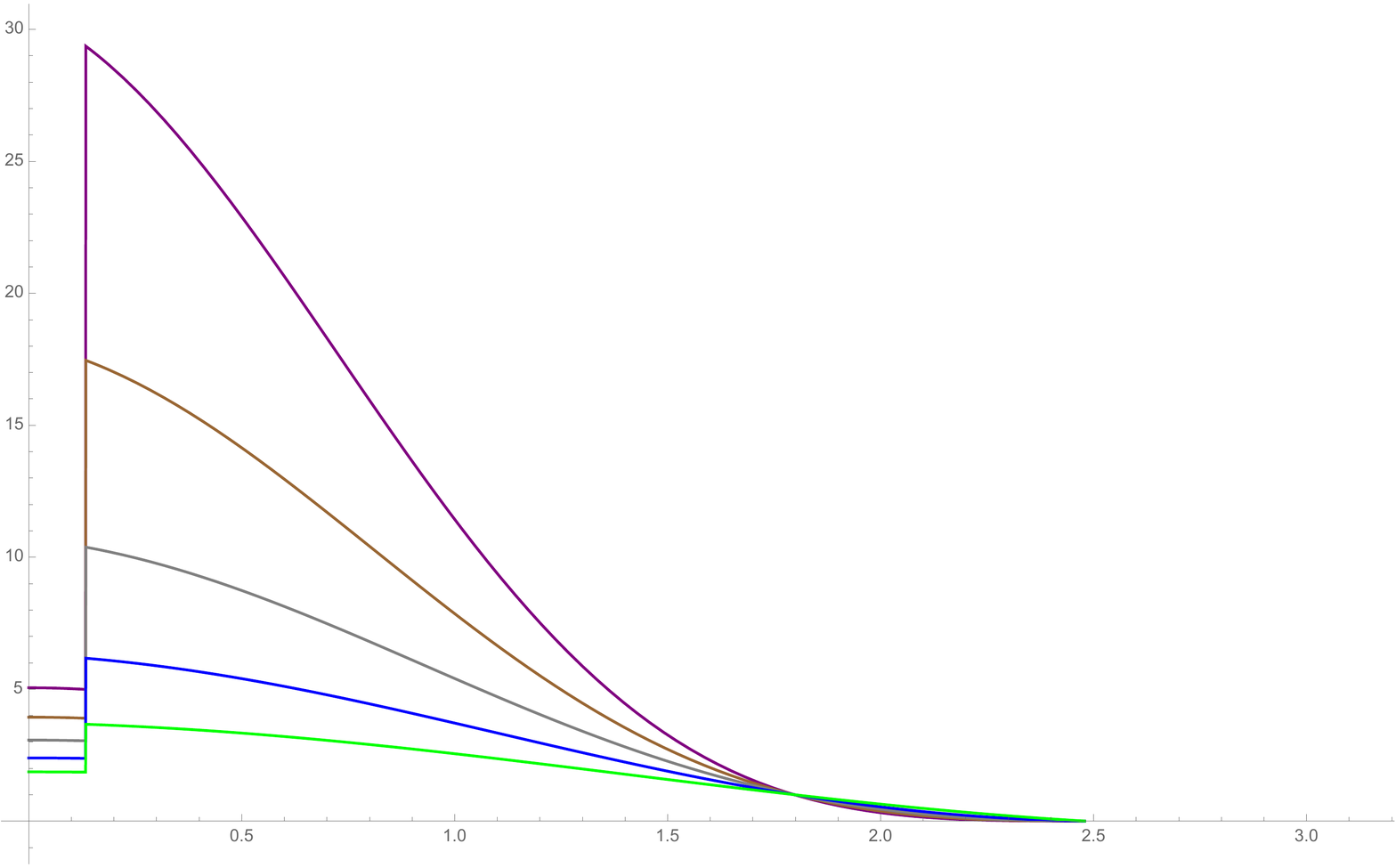}}
\put(90,0){\includegraphics[width=7cm]{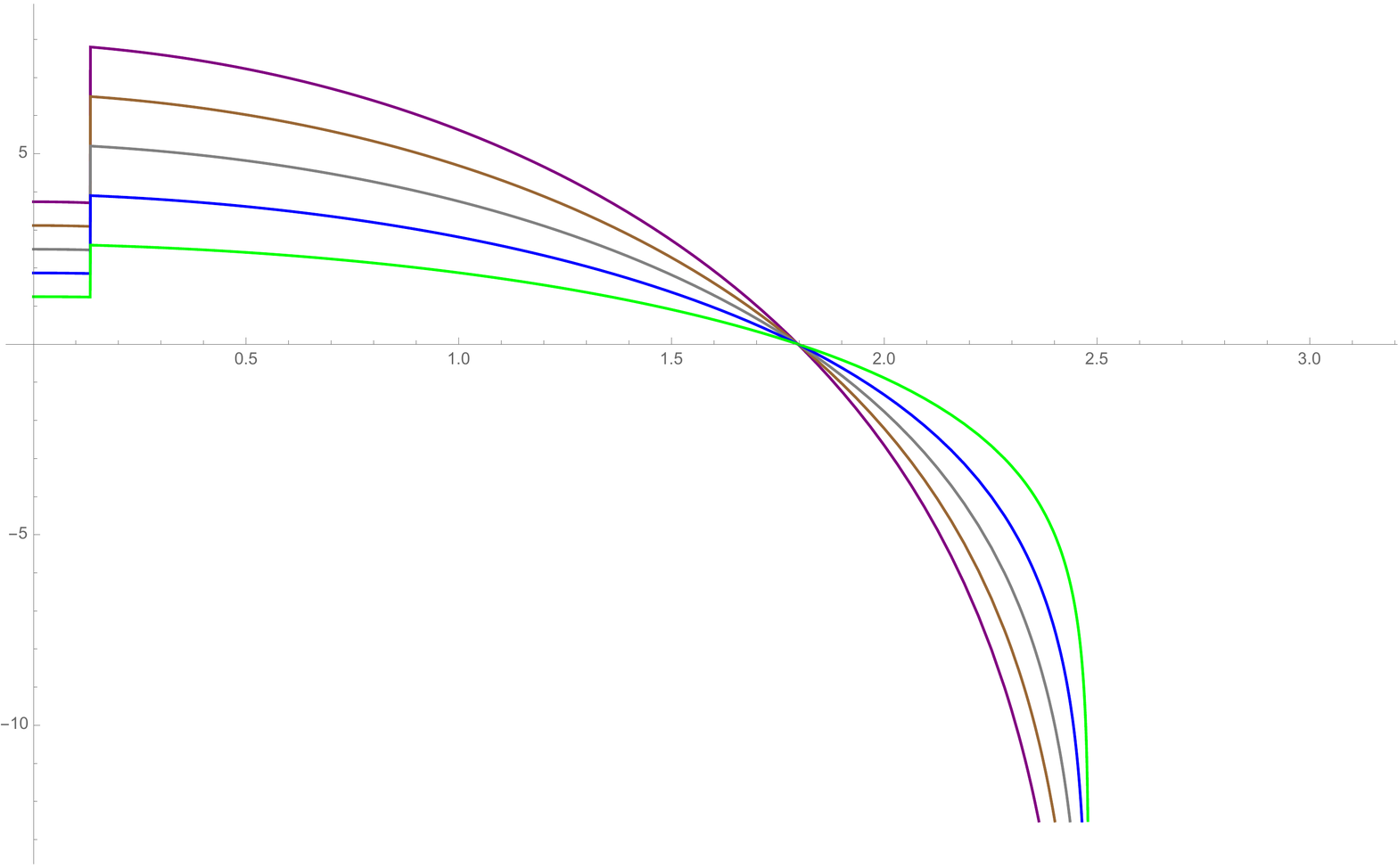}}
\put(70,-11){A}
\put(249,-11){B}
\end{picture}\\
 \caption{A. Plots of $\mathcal{G}_{\alpha,\xi,\phi_f,t_f}(\phi_a,t_a)$ given by \eqref{Gpl} as function of $\phi_a$
   for fixed $t=0.5$,  $\phi_f=\frac{3\pi}{2}$ and $t_f=0$, and parameters  $\alpha=0.3$, $\xi=0.4$ for different values of $\Delta$:  $\Delta=2.6, 2.2,1.8,1.4, 1$ (purple, brown, grey, blue and green lines, respectively). B. Plot of  $\ln \mathcal{G}_{\alpha,\xi,\phi_f,t_f}(\phi_a,t_a)$ for the same parameters
   and $\Delta=6,5,4,3,2$ (purple, brown, grey, blue  and green lines, respectively).  Thick vertical lines show the boundaries of the living space. }
    \label{Fig:2d}
\end{figure}
  \begin{figure}[h!]
    \centering
        \includegraphics[width=12cm]{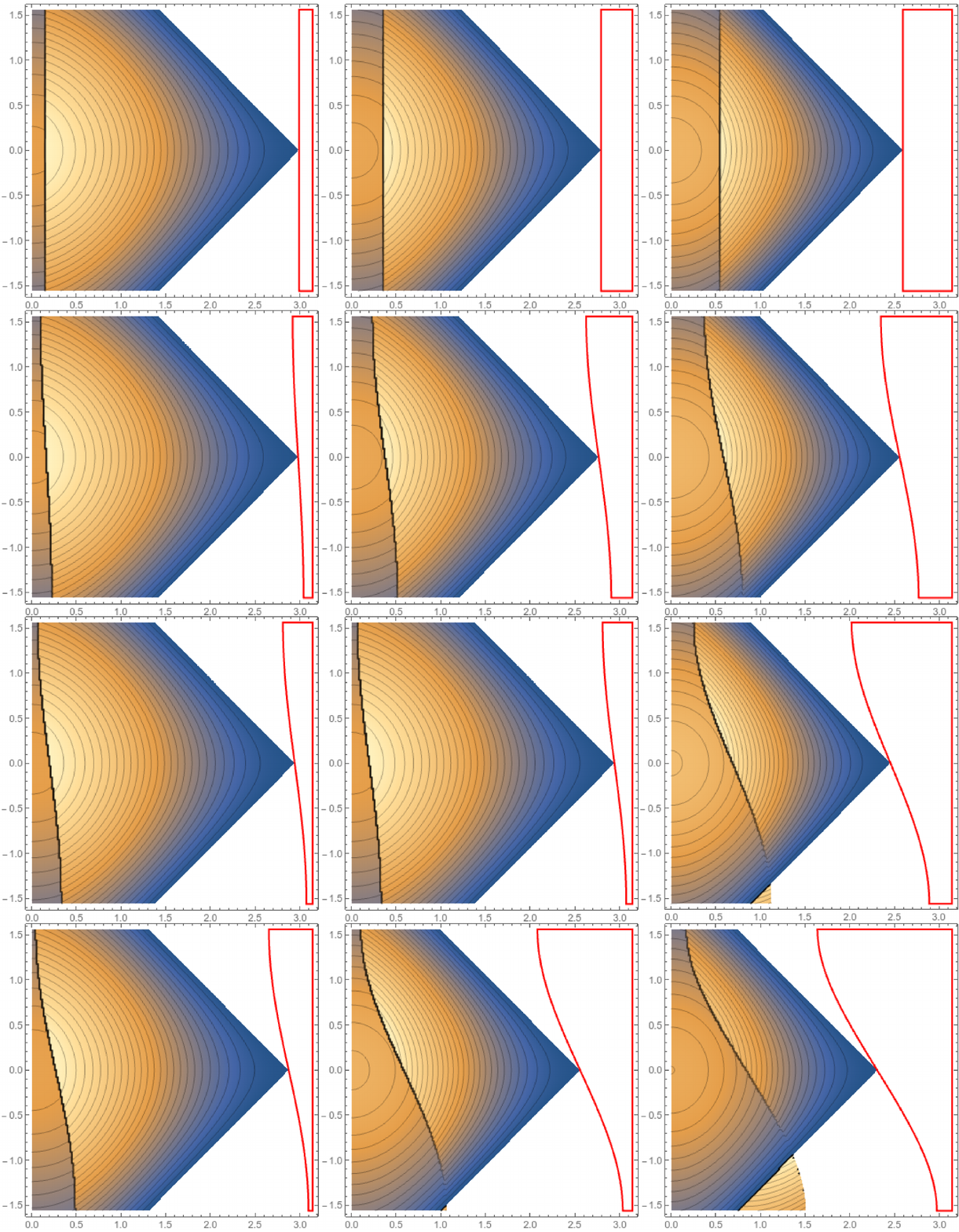}
 \caption{The density plots of  function $\mathcal{G}_{\alpha,\xi,\phi_f,t_f}(\phi_a,t_a)$
 given by \eqref{Gpl} for different values of $\alpha$ and $\xi$. The parameter $\alpha$ increases from the left to right, corresponding to $\alpha=0.3, 0.7, 1.1$.  The parameter $\xi$ increases from top to down, corresponding to $\xi=0, 0.4, 0.8, 1.2$. On each plot $\phi_a$ corresponds to x-axis and $t_a$ to y-axis, $\phi_f={3\pi}/{2}$ and $t_f=0$.
 The red thick curves correspond to the boundaries of the living spaces. }
    \label{Fig:TPO}
\end{figure}

As we have seen in the previous sections the deformation of the $AdS_3$ by moving particles produces changes of correlation functions of the boundary theory. To visualize these effect we depict the density plot of the inverse correlation function $\mathcal{G}_{\alpha,\xi,\phi_f,t_f}(\phi_a,t_a)$, as a function of coordinates of a point $a$ and fixed coordinates $(\phi_f,t_f)$ and $a$ variety of parameters $\alpha$ and $\xi$:

\bea\label{Gpl}
\mathcal{G}_{\alpha,\xi,\phi_f,t_f}(\phi_a,t_a)&=&G^{-1}_{\alpha,\xi}(\phi_a,t_a,\phi_f,t_f),\\
\phi_{max}&<&\phi_a<\pi,\,\,\,\,\,\,\,\,\,\,
2\pi -\phi_{max}<\pi<\phi_f<2\pi,\,\,\,\,\,\,\,\,\,\,\,
\alpha<\pi.\nn
\eea
When  $a$ and $f$ points are located on the opposite halves, the contribution coming from the image geodesic can appear and we can see effects related with the presence of the particle in the bulk (compare with discussion in \cite{Balasubramanian:1999zv}).

 In Fig.\ref{Fig:2d} we plot the  function $\mathcal{G}_{\alpha,\xi,\phi_f,t_f}$ for fixed values of $t_a$, $\phi_f$, $t_f$, $\alpha$ and $\xi$, and various $\Delta$.  In Fig.\ref{Fig:TPO} we present the density plots of the function $\mathcal{G}_{\alpha,\xi,\phi_f,t_f}$ for certain values of $\phi_f$, $t_f$, $\alpha$, $\xi$ and $\Delta=1$. The red curves correspond to the boundary of the removed zone, black curves  indicate locations of discontinuities separating different zones.

In each plot in Fig.\ref{Fig:TPO} there is a zone, where the basic geodesic contribute only, i.e. the correlation function remains unchanged, and there is a zone (next to the $\phi=0$) where the winding geodesic contributes. These regions are separated by discontinuities. The white zone appears when the points on the boundary are timelike separated.

\subsection{Heavy particle}
For the heavy particle we study the similar correlator as for the light particle in the previous section, but in the different region
\bea\label{GplH}
\mathcal{G}_{\alpha,\xi,\phi_f,t_f}(\phi_a,t_a)&=&G^{-1}_{\alpha,\xi}(\phi_a,t_a,\phi_f,t_f)\\\nn
0&<&\phi_a<2\pi,\,\,\,\,\,\,
\frac{\pi}{2}<\phi_f<\frac{3\pi}{2},\,\,\,\,\,\,\,\,
\alpha=\frac{3\pi}{2}>\pi.
\eea
\begin{figure}[h!]
    \centering
        \includegraphics[width=6cm]{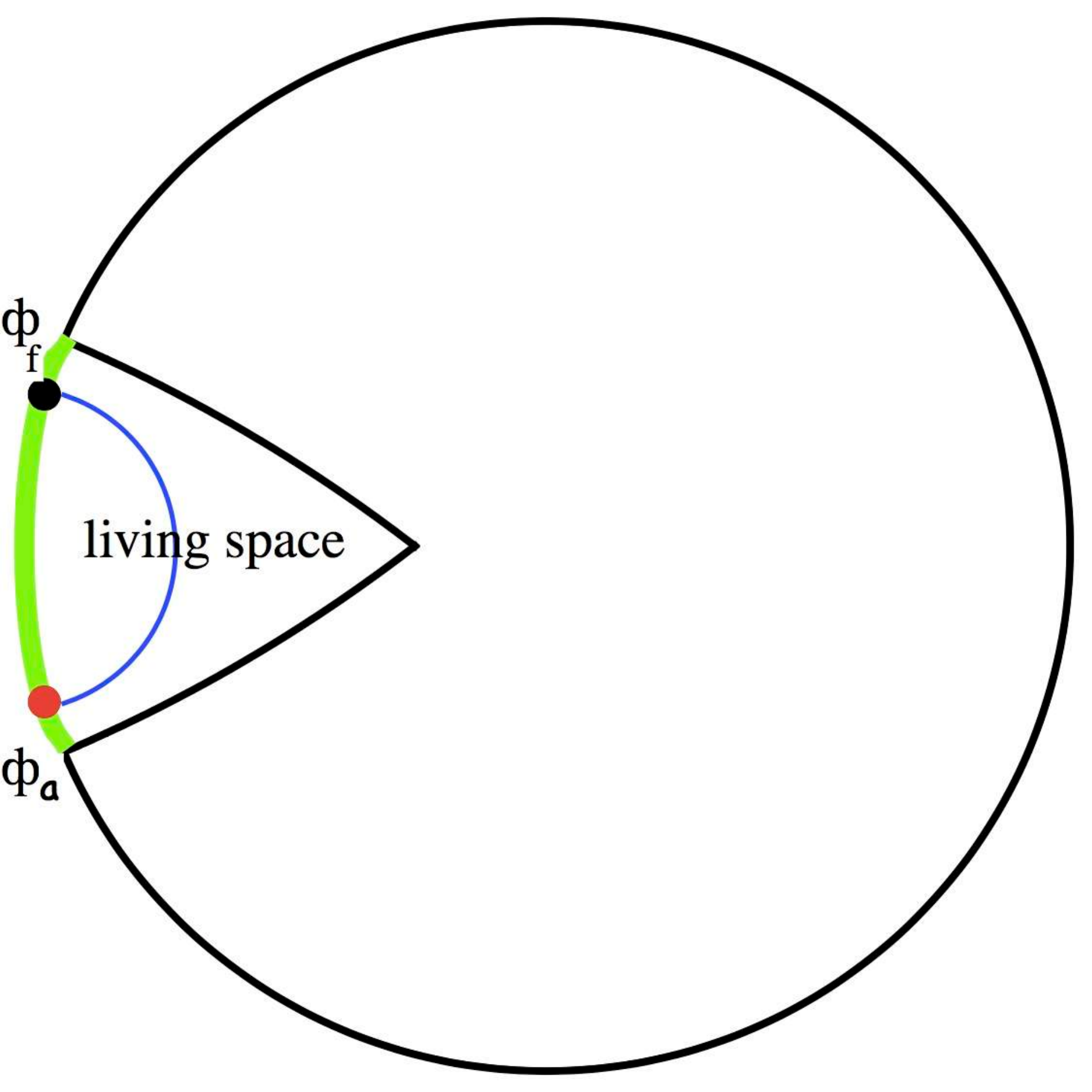}
 \caption{The schematic plot of  locations of the moving massive heavy particle and two points of the correlator. The circle is the boundary of the $AdS_3$  at a constant time section, the point $\phi_f$ is fixed in the living space, the green part of the $AdS_3$ boundary. The point $\phi_a$ belongs to the living area too.}\label{Fig:intro2}
    \end{figure}

\begin{figure}[h!]
    \centering
        \includegraphics[width=6cm]{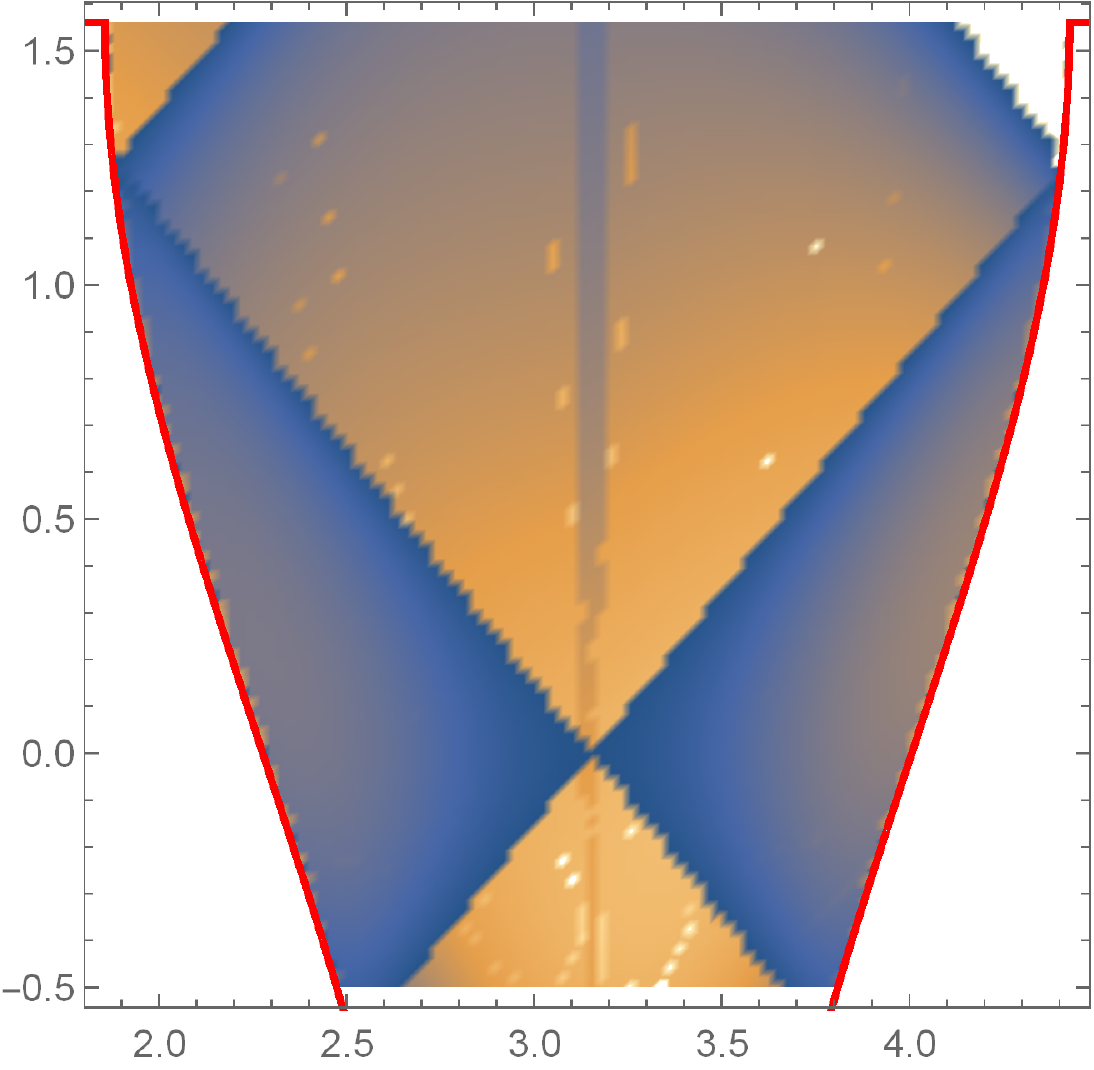}
            \caption{The density plot of  the function $\mathcal{G}_{\alpha,\xi,\phi_f,t_f}(\phi_a,t_a)$ given by \eqref{GplH}. Here $\phi_f=\pi$ and $t_f=0$ and $\alpha=\frac{3\pi}{2}$, and $\xi=0.6$, $\phi_a$ corresponds to x-axis and $t_a$ to y-axis,
    the red thick curves show the boundaries of the removed areas. }
    \label{Fig:total}
\end{figure}

\begin{figure}[h!]
    \centering
    \centering
\begin{picture}(90,140)
\put(-160,0){\includegraphics[width=7cm]{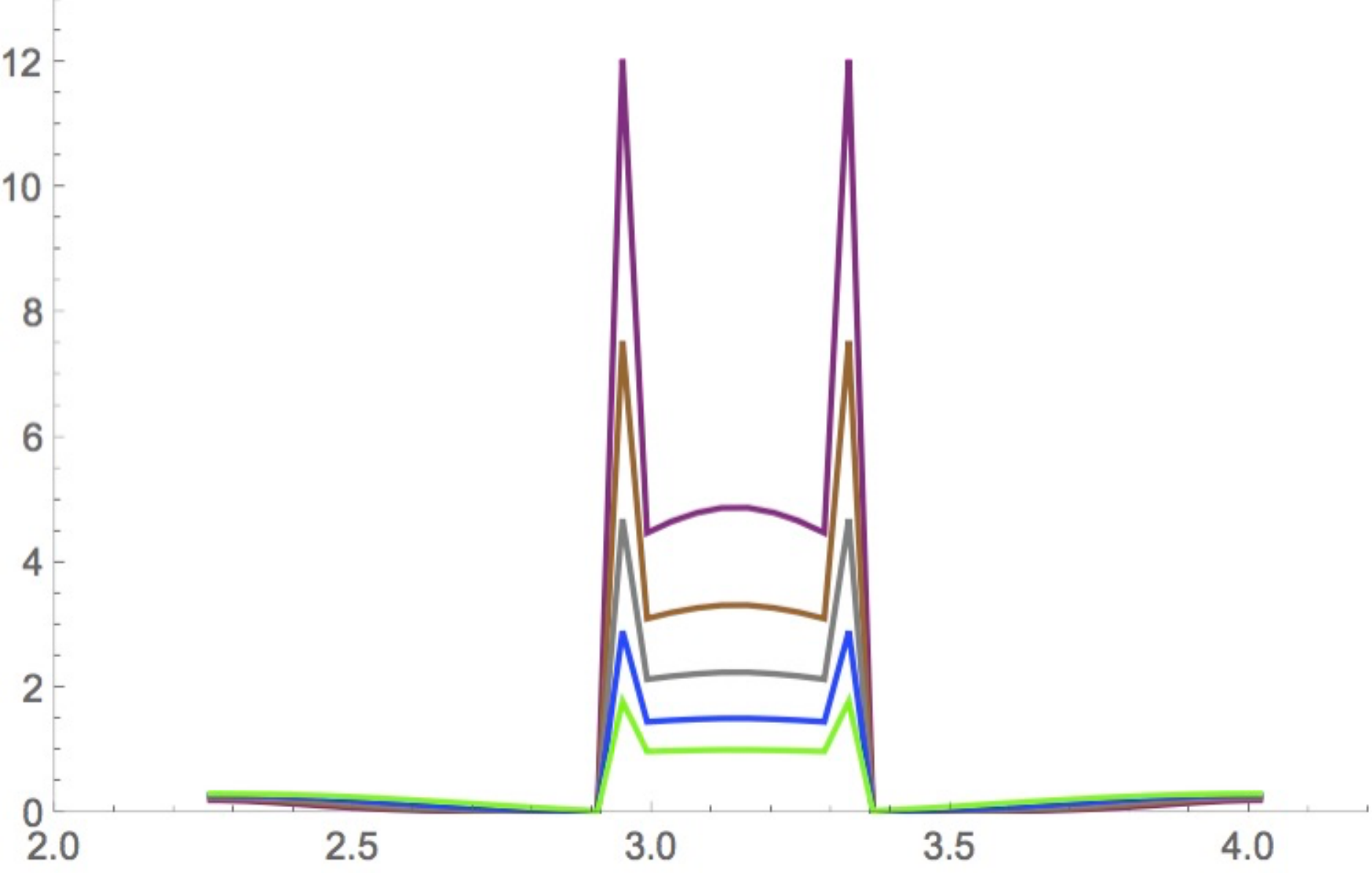}}
\put(50,0){\includegraphics[width=7cm]{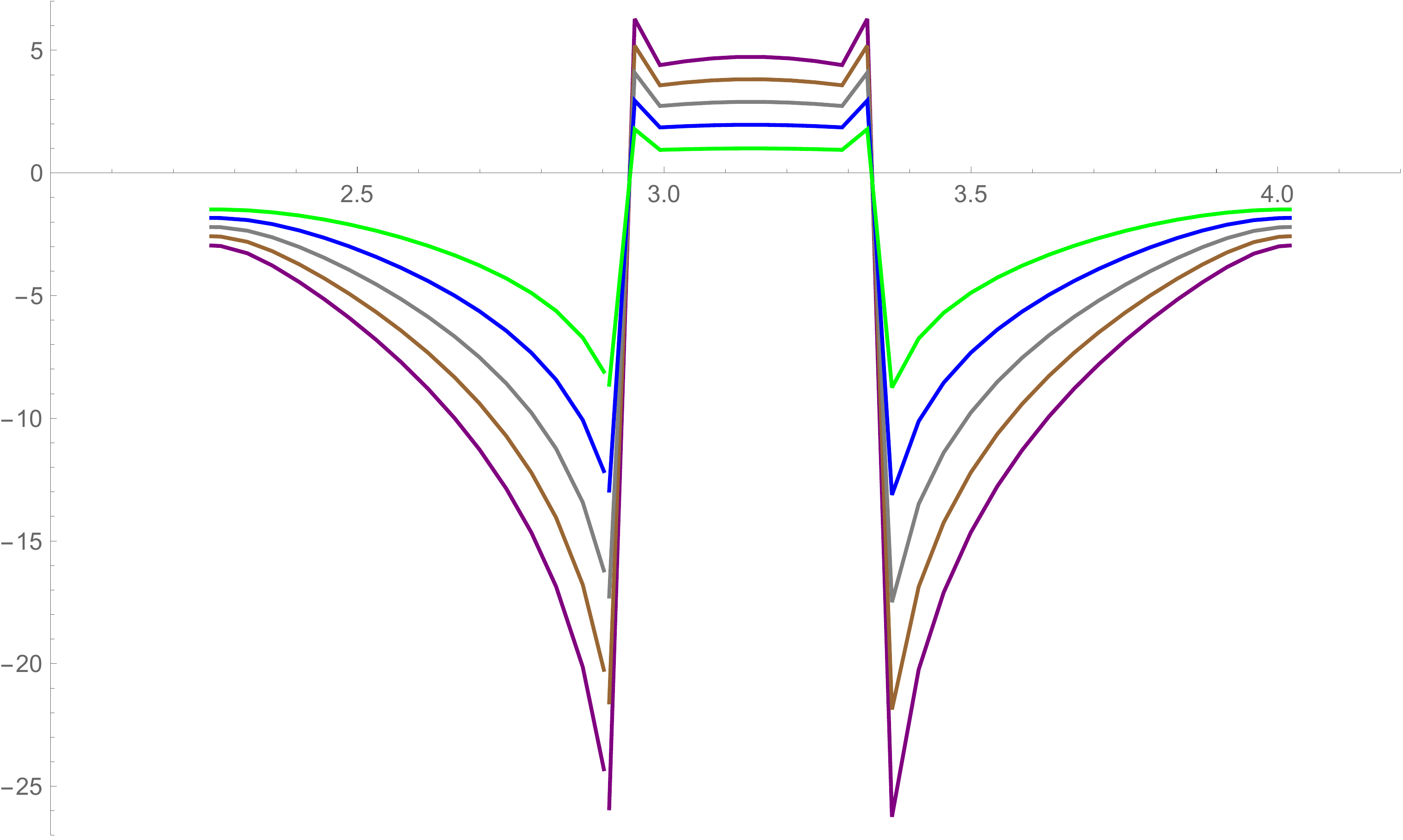}}
\put(-130,9){\line(0,1){110}}
\put(23.6,9){\line(0,1){110}}
\put(233.5,4){\line(0,1){115}}
\put(79.7,4){\line(0,1){115}}
\put(240,-4){B}
\put(30,-4){A}
\end{picture}\\
 \caption{A. Plots of $\mathcal{G}_{\alpha,\xi,\phi_f,t_f}(\phi_a,t_a)$ given by \eqref{GplH}  as function of $\phi_a$
   for fixed $t_a=0.2$,  $\phi_f=\pi$ and $t_f=0$, and parameters  $\alpha=3\pi/2$, $\xi=0.4$ for different values of $\Delta$:  $\Delta=2.6, 2.2,1.8,1.4, 1$ (purple, brown, grey, blue and green lines, respectively). B. Plot of  $\ln \mathcal{G}_{\alpha,\xi,\phi_f,t_f}(\phi_a,t_a)$ for the same parameters
   and $\Delta=6,5,4,3,2$ (purple, brown, grey, blue  and green lines, respectively).  Thick vertical lines show the boundaries of the living space. }
    \label{Fig:2d-heavy}
\end{figure}
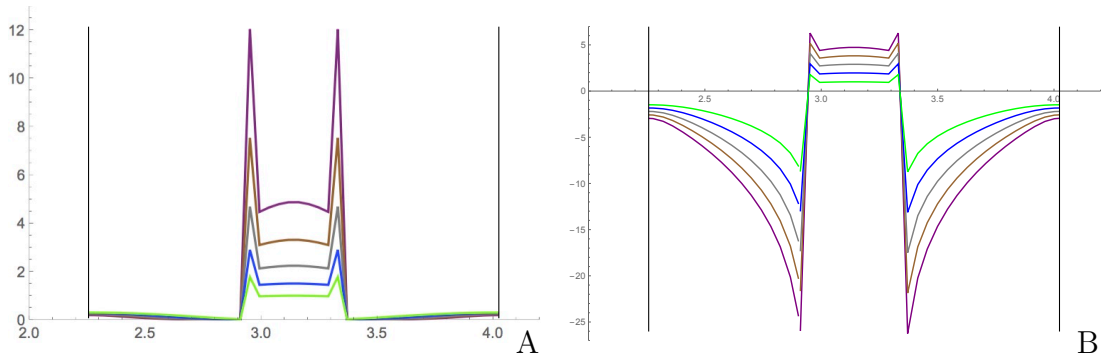

\begin{figure}[!h]
    \centering
        \includegraphics[width=7cm]{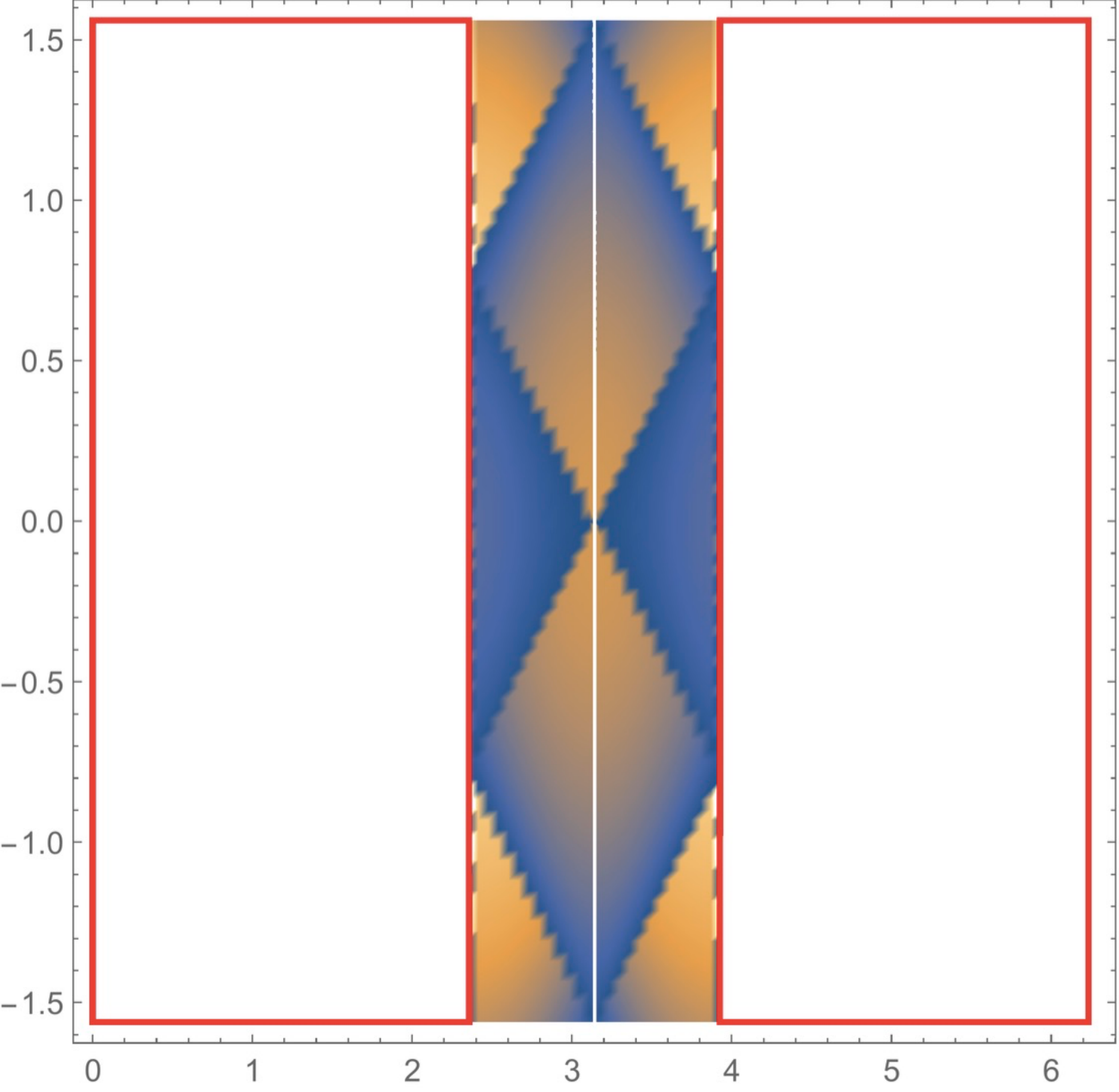}
            \caption{The density plot of the function $\mathcal{G}(a,b)$ on the living space in case of the static heavy particle. In the plot $\alpha=3\pi/2$, $\phi_f=\pi$ and $t_f=0$. Here $\phi_a$ corresponds to x-axis and $t_a$ to y-axis. Note, that the scales on x-axis and y-axis are different.
    The red thick rectangle show the removed  parts of the AdS boundary.}
    \label{Fig:static}
\end{figure}

\begin{figure}[!h]
   \centering
\begin{picture}(90,290)
\put(-110,-58){\includegraphics[width=10cm]{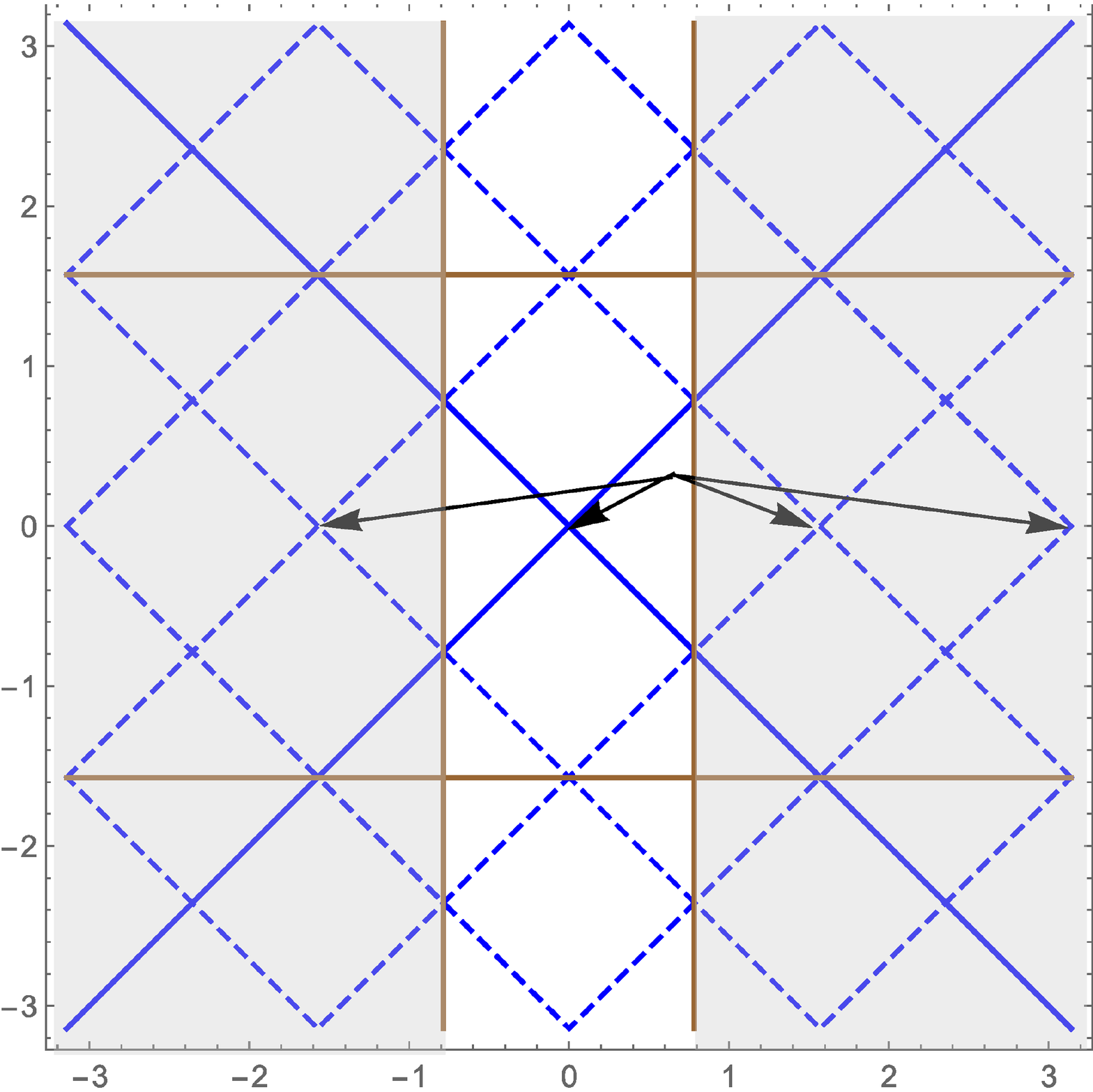}}
\put(63,165){$a$}
\put(25,145){$b$}
\put(100,130){$b^*$}
\put(165,130){$b^{**}$}
\put(-35,155){$b^\#$}
\put(-143,140){$(0,0)$}
\put(25,130){$(\pi,0)$}
\put(177,140){$(2\pi,0)$}
\put(177,280){$(2\pi,\pi)$}
\end{picture}\\
\caption{The schematic plot of different zones of the 2-point correlation function. Here the scales along x-axis and y-axis are the same.
} \label{Zones2}\end{figure}

    The  zone structure of the 2-point correlator on the boundary of $AdS_3$ with a heavy moving particle is presented in Fig.\ref{Fig:total} and Fig.\ref{Fig:2d-heavy}. In these plots we see, that there are several different zones. These zones are typical for heavy particle deformations, and the origin of these zones can be explained first on the static particle, see Fig.\ref{Fig:static} and Fig.\ref{Zones2}.

 Let us take the point $a$ in the darkest zone, see Fig.\ref{Zones2}. The points $a$ and $b$ are spacelike separated points. The point $a$ can be connected by geodesics not only with the point $b$, with coordinates $\phi_b=\phi_f=\pi$, $t_b=t_f=0$, but also with the image points $b^{*}$, $b^{**}$ and $b^{\#}$, and there are several contributions to the propagator. For the case presented in Fig.\ref{Zones2}, there are contributions from 4 terms. 


\begin{figure}[h!]
    \centering
        \includegraphics[width=9cm]{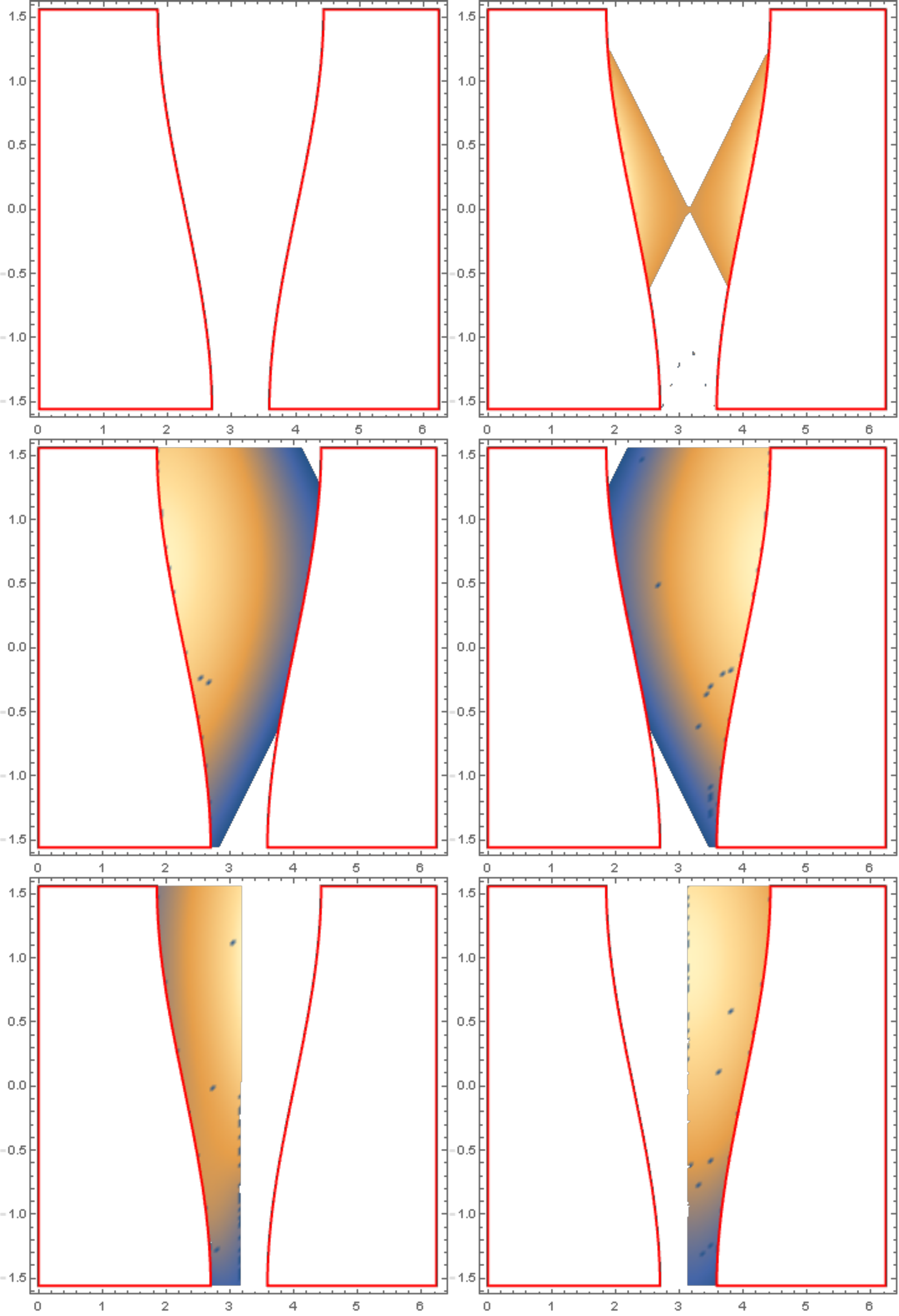}
            \caption{The density plots of separate contributions to the function \eqref{GplH}. For each plot $\phi_f=\pi$, $t_f=0$, $\alpha=\frac{3\pi}{2}$ and $\xi=0.6$. On each plot $\phi_a$ corresponds to x-axis and $t_a$ to y-axis,
    the red thick curves correspond to the boundaries of the removed areas.  The bottom left plot shows the boundary of the removed area, the bottom right shows the contribution from the basic geodesic,
    the left and right plots in the second row show  contributions of double winding geodesics coming from different terms in \eqref{GplH}, the third row shows  the single winding geodesic.}
    \label{Fig:TPO2}
\end{figure}

Let us consider the moving heavy particle located at the initial time as shown in Fig.\ref{Fig:intro2}. Now the living space may be located only in the interval (${\pi}/{2},
{3\pi}/{2}$). In Fig.\ref{Fig:TPO2} we plot contributions for
 different geodesics configurations. In this figure we see that the  basic spacelike geodesics contribution is bounded by lightcone, the single winding geodesic contributes almost everywhere, contributions from different double winding geodesic configurations form zones near the boundary, but the total  double winding geodesics contribution covers all the living space. In Fig.\ref{Fig:total} we show the sum of all
 contributions presented separately in Fig.\ref{Fig:TPO2}. In the Fig.{\ref{Fig:2d-heavy}} the two dimensional plot of the inverse correlation function for the several values of $\Delta$ and for the fixed parameters $\alpha$ and $\xi$ is presented for the case of massive particle. Remind that we do not consider the geodesics between timelike separated points.    

From the plot  in Fig.\ref{Fig:total}  we see that for the heavy particle there is no any "shadow" like in the light particle case.

\section{Conclusion}
In this paper we have investigated  the correlation functions of conformal operators in the theory dual to the $AdS_3$ deformed by moving massive particles. Our calculations are based on the geodesic approximation. This approximation works well for operators with large conformal dimension $\Delta$. However, we have considered how this approximation works starting from $\Delta=1$.
  We find, that the 2-point correlation function gets additional contributions  due to the nontrivial geodesic structure of the deformed spacetime. The presence of these additional geodesics does not depend on the conformal dimension.
The additional geodesics are found  via the renormalized image method. In the work we did not take into account the contribution of geodesics between timelike separated points.
  
  We get two different pictures of behaviour of the 2-point correlators on the boundary of $AdS_3$ deformed by moving/static particles.
The first case, is when the particle deforming the $AdS_3$ is light. In this case additional contributions mentioned above give us to the following picture: we have two different zones separated by discontinuity. One of the zones corresponds to the original correlator of the conformal field on the cylinder. Another zone corresponds to the deformed theory, i.e. constant level lines of the inverse correlator are slightly deformed. 
 The second case is the case of the heavy particle. In this case 2-point correlator differs qualitatively: it is deformed in a whole space and there are many different contributions from different multiple winding geodesics. The number of winding depends on the ratio $2\pi/\bar{\alpha}$, where $\bar\alpha$ is the angle of the living space. 

It is interesting to compare the results presented in the paper with correlators obtained using the scalar field in the bulk via the GKPW prescription \cite{GKP,Witten}. This  is a subject of paper \cite{AK}. The image method for timelike separated points and a continuation of correlators to the entire boundary have been considered in the paper \cite{AKT}

\section*{Acknowlegement}
We would like to thank  Andrey Bagrov, Dmitry Bykov, 	Xian Otero Camanho, Mikhail Khramtsov, Andrey Mikhailov, Giuseppe Policastro and Igor Volovich for useful discussions. This work is supported
by the RFBR grant 14-01-00707  and grant MK-2510.2014.1 (D.S.A.)  of the President of Russia Grant Council. I.Ya. A.  thanks the Galileo Galilei Institute for Theoretical Physics for the
hospitality and the INFN for partial support during the preparation of
this work.

  \end{document}